\documentclass[11pt,a4paper]{article}

\usepackage[a4paper,margin=0.975in]{geometry} 
\usepackage[T1]{fontenc}
\usepackage[utf8]{inputenc}
\usepackage{newpxtext} 
\usepackage[final]{microtype}
\usepackage{setspace}

\usepackage{amsmath,amsthm,mathtools,amsfonts,amssymb} 
\usepackage{bm}
\usepackage{bbm}
\numberwithin{equation}{section}
\allowdisplaybreaks

\usepackage{newpxmath} 

\usepackage{multirow}
\usepackage{graphicx}
\usepackage{subcaption}
\usepackage{booktabs}
\usepackage{threeparttable}
\usepackage{adjustbox}
\usepackage{makecell}
\usepackage{siunitx}
\usepackage{placeins}
\graphicspath{{plots/}}

\usepackage{enumitem}
\setlist{nosep}

\usepackage[authoryear]{natbib}

\usepackage[dvipsnames]{xcolor}
\usepackage[
  colorlinks=true,
  linkcolor=black, 
  citecolor=MidnightBlue, 
  urlcolor=RoyalBlue 
]{hyperref}
\DeclareRobustCommand{\email}[1]{\href{mailto:#1}{\texttt{#1}}}

\theoremstyle{plain}

\theoremstyle{definition}

\theoremstyle{remark}

\title{Bayesian Adaptive Tensor Smoothing for Multiway Demographic Counts}
\author{%
  Gregor Zens\\
  International Institute for Applied Systems Analysis (IIASA)\\
  \email{zens@iiasa.ac.at}}%

\date{October 2026\vspace{-1em}}

\newcommand{\keywords}[1]{\par\smallskip\noindent\textbf{Keywords:} #1\par}

\begin{document}
\maketitle

\begin{abstract}
Demographic data are often collected across multiple dimensions such as time, age, subnational region, and sex. Modeling such data is challenging due to the interplay of high dimensionality, heterogeneity between groups, and stochastic variability within small groups. This paper proposes a Bayesian adaptive tensor smoothing model for multiway demographic counts that represents the latent log-rate array as a sum of rank-one components. Global-local shrinkage priors control the effective rank of the decomposition, the number of modes along which each component varies, and the smoothness that Gaussian process priors impose along ordered modes such as age and time. Posterior inference relies on a data augmentation Markov chain Monte Carlo algorithm that interweaves centered and noncentered parameterizations in every layer of the model to keep mixing satisfactory in large and sparse count arrays. Applied to Swedish county-level and sex-specific mortality from 1969 to 2024, the model recovers stable demographic surfaces and life table quantities from noisy small-population data. In cross-validation exercises, the method outperforms both canonical demographic benchmarks and existing count tensor factorizations.
\end{abstract}

\vspace{0.5em}
\keywords{Bayesian Tensor Factorization; Statistical Demography; Global-Local Shrinkage; Non-Gaussian Data; Markov Chain Monte Carlo; Count Tensor Decomposition}

\section{Introduction}
\label{sec:introduction}

Demographic processes are frequently measured across multiple dimensions, including age, sex, time, and further stratifications including geographical units or educational groups, among others.\footnote{An earlier version of this article was circulated under the title ``Bayesian Matrix Factor Models for Demographic Analysis Across Age and Time''.} For the three primary demographic components -- fertility, mortality, and migration -- such data are almost exclusively records of event counts, namely births, deaths, and moves. Modeling such data has important applications, ranging from providing inputs to subnational population projections and regional infrastructure planning to healthcare resource allocation and assessments of pension system sustainability.

Designing suitable models for such multidimensional data is challenging due to the interplay of high dimensionality, heterogeneity across groups, and stochastic variation in small groups. Every additional dimension along which the data are stratified divides the underlying events across more cells. As a result, counts per cell shrink and stochastic variation grows relative to the demographic signal. A model for such data must therefore separate signal from noise without smoothing away genuine differences between groups. In addition, it must do so at a computational cost that remains feasible for arrays with hundreds of thousands of cells.

\begin{figure}[!t]
    \centering
    \begin{subfigure}[b]{0.325\textwidth}
        \centering
        \includegraphics[width=\textwidth]{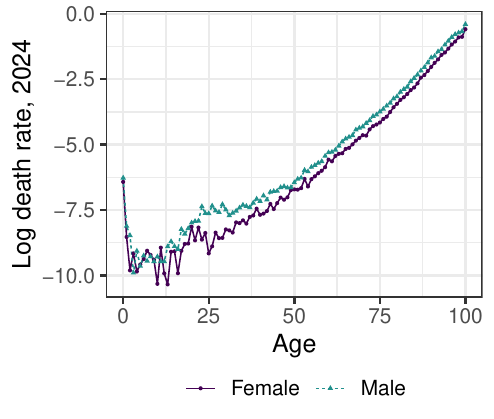}
    \end{subfigure}
    \hfill
    \begin{subfigure}[b]{0.325\textwidth}
        \centering
        \includegraphics[width=\textwidth]{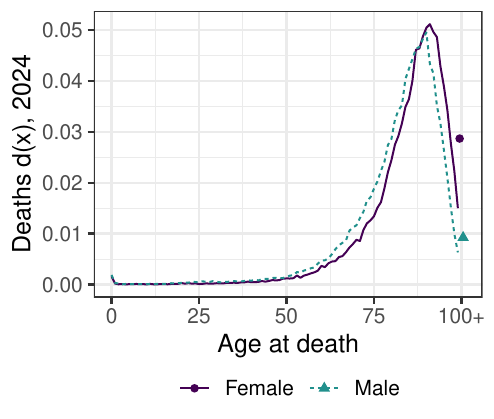}
    \end{subfigure}
    \hfill
    \begin{subfigure}[b]{0.325\textwidth}
        \centering
        \includegraphics[width=\textwidth]{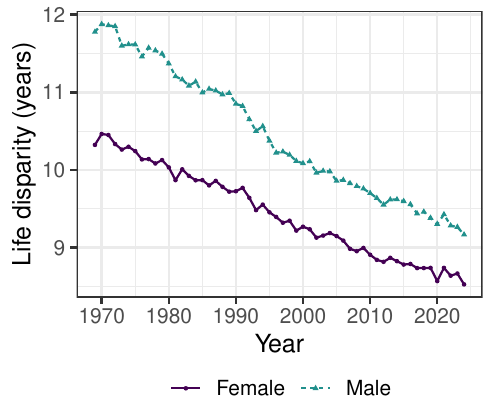}
    \end{subfigure}

    \begin{subfigure}[b]{0.325\textwidth}
        \centering
        \includegraphics[width=\textwidth]{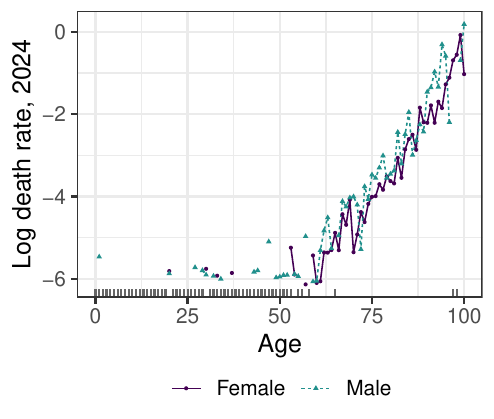}
    \end{subfigure}
    \hfill
    \begin{subfigure}[b]{0.325\textwidth}
        \centering
        \includegraphics[width=\textwidth]{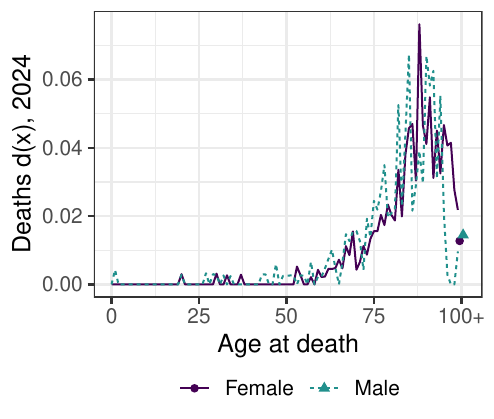}
    \end{subfigure}
    \hfill
    \begin{subfigure}[b]{0.325\textwidth}
        \centering
        \includegraphics[width=\textwidth]{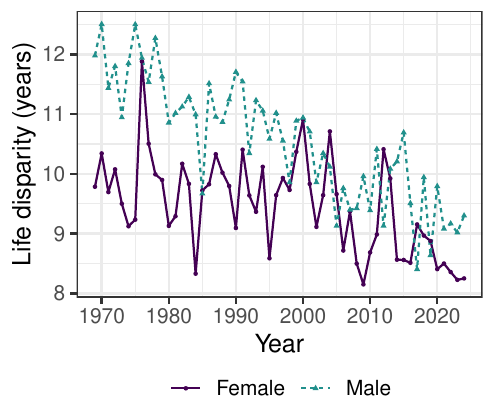}
    \end{subfigure}
    \caption{Observed mortality in Sweden (top row) and in Gotland, its least populous county (bottom row), by sex. Left: log mortality rates by age in 2024. Ages with zero observed deaths are marked by rug ticks. Middle: life table distribution of deaths by age in 2024, computed from the raw rates, with the open age group as a detached point at age 100. Right: life disparity $e^{\dagger}$, the average remaining life expectancy at the age of death, by year. Details on life table quantities are given in the supplementary materials, Sec.~S5.}
    \label{fig:rawdata}
\end{figure}

As a motivating example, Fig.~\ref{fig:rawdata} shows mortality data for Sweden as a whole and for its least populous county, Gotland; see Sec.~\ref{sec:theproblem} for details. The figure shows log mortality rates and the age distribution of deaths in 2024, as well as life disparity, a standard measure of lifespan variation, over the period 1969 to 2024. These are typical examples of quantities routinely used to describe mortality patterns and their change over time.

For Sweden as a whole, these quantities show familiar demographic regularities. Death rates rise almost log-linearly with age after childhood, deaths concentrate around a modal age in the late eighties, and life disparity declines steadily. For Gotland, such patterns are much harder to identify in the raw data. The age-specific rates are sparse and vary strongly between adjacent ages, the distribution of deaths by age is irregular, and life disparity fluctuates far more than any plausible change in mortality would imply. The reason is Gotland’s small population, which makes it difficult to separate demographic signal from stochastic variability. Conclusions drawn directly from the raw data can therefore be misleading, and reliable insights into the underlying process typically require model-based approaches.

One family of models that has proven useful in this context relies on low-rank approximations, based, for example, on singular value decompositions or parametric factor models. Because demographic processes typically exhibit strong regularities (e.g., across age and time), large datasets can often be described accurately using a small number of latent factors (\citealp{lee1992modeling, alexander2017flexible, clark2019general, czado2005bayesian, wisniowski2015bayesian, zens2024flexible}). Recent advances in the statistical and machine learning literature on matrix and tensor factorizations offer a promising direction for generalizing these ideas to high-dimensional multi-way contexts (\citealp{wang2019factor, hoff2011separable, hoff2015multilinear, schein2015bayesian, billio2023bayesian, qin2025bayesian}). However, existing matrix and tensor factorization methods rarely incorporate the domain-specific knowledge and requirements central to demographic modeling. They are typically not designed for overdispersed counts with known exposures and encode no prior knowledge of demographic processes, such as smoothness across age and time.

This article addresses this gap by developing a Bayesian adaptive tensor smoothing model for multiway demographic count data. The model represents the latent log-rate array as a sum of rank-one components, that is, as a canonical polyadic (CP) decomposition \citep{kolda2009tensor}. A global-local shrinkage prior induces parsimony at three levels. First, it shrinks the effective rank of the decomposition. Second, it pulls mode vectors toward constants, so that each component varies only along the dimensions where the data require it. Third, it shrinks the mode vectors of ordered modes such as age and time toward smooth profiles via Gaussian process priors, while unordered modes such as sex or region remain exchangeable. The model thus favors parsimonious and smooth factorizations. Embedding the model in a Bayesian framework provides fully probabilistic uncertainty quantification.

We demonstrate the model on annual age- and sex-specific mortality across Sweden's counties from 1969 to 2024 where the method recovers stable demographic surfaces and life table quantities from a large and sparse demographic tensor. The smoothed tensor reveals that regional inequality in life expectancy declined until the late 1990s and has risen since, a pattern that sampling noise conceals in the raw data. In cross-validation exercises, the method outperforms canonical demographic smoothers by a wide margin and improves on existing count tensor factorizations in predictive accuracy and calibration. A replication on Austrian district-level outmigration data supports these findings beyond the mortality setting. 

This work thus makes three main contributions to the literature on tensor models and high-dimensional statistics for demographic applications. From a methodological point of view, we construct a shrinkage prior for CP-type models that simultaneously learns the effective tensor rank and the modes along which each component varies, and, through Gaussian process priors on ordered modes, how smoothly the components vary. In terms of applications, we provide a single framework for smoothing multiway demographic rate arrays across all dimensions at once, which contains classical structures such as additive age and period effects and Lee--Carter-type factor models as special cases. From a computational perspective, we develop a data augmentation Markov chain Monte Carlo (MCMC) algorithm for large and sparse Poisson lognormal arrays, where MCMC samplers typically mix slowly. Since cells with many events and cells with few or none call for different parameterizations, the algorithm interweaves centered and noncentered updates in every layer of the model \citep{yu2011center} and exploits partially collapsed Gibbs steps \citep{vandyk2008partially} where possible.

The remainder of the article is structured as follows. Sec.~\ref{sec:theproblem} presents the motivating dataset and practical modeling problem. Sec.~\ref{sec:literature} reviews low-rank factorization and smoothing approaches in the demographic and statistical literature. Sec.~\ref{sec:model} introduces the proposed tensor model, prior specification, and posterior simulation algorithm. Sec.~\ref{sec:prior_demo} demonstrates the effect of the prior on a small subset of the data. Sec.~\ref{sec:cv} presents a comparative cross-validation exercise. Sec.~\ref{sec:app} details the real-data application. Sec.~\ref{sec:conclusion} provides concluding remarks and discusses extensions and future research directions.

\section{Motivating Application and Data}
\label{sec:theproblem}

The motivating application concerns mortality data, a natural demonstration case for a multiway demographic model for two reasons. First, canonical smoothing methods exist for mortality, which allows a rigorous comparative validation. Second, smoothed mortality surfaces feed directly into life tables, which are central to demography, actuarial science, and public health studies. The methodology itself is, however, general and applies to any multiway count array, including fertility or migration counts, and, with a Gaussian instead of a Poisson observation layer, to continuous array-valued data.

In particular, we consider the problem of modeling a comprehensive dataset on Swedish mortality, comprising county-specific death counts by single year of age ($x = 0, \dots, 99$, plus an open age group 100 and above, $A = 101$) and sex across 20 counties from 1969 to 2024 ($T = 56$). The data, sourced from the Swedish national statistical office (\textit{Statistics Sweden}), combine register-based death counts with population stocks recorded on December 31 of each year. The counties of Uppsala and V\"astmanland are merged into one unit throughout, because a municipality was transferred between them in 2007. Person-year exposures are constructed following the conventions of the Human Mortality Database as the average of the two adjacent December stocks (\citealp{wilmoth2007methods}). Rare cells at the highest ages that record deaths with zero averaged exposure receive an exposure floor of one half person-year, because an observed death implies partial-year residence. The full array contains $56 \times 101 \times 2 \times 20 = 226{,}240$ cells. In 2024, county populations range from about 2.5 million residents in Stockholm to 61{,}000 in Gotland, so the level of stochastic variation differs widely across units. At the same time, the data span more than half a century, both sexes, and regions ranging from metropolitan areas to sparsely populated rural counties, so that genuine heterogeneity in demographic patterns is to be expected.

The array records about 5.1 million deaths over 502 million person-years. Counts per cell are highly skewed, with a median of 5 deaths, a 90th percentile of 60, and a maximum of 455. A fifth of all cells record no deaths and half record at most five. In Gotland, these shares rise to two fifths and three quarters. A model for these data must therefore pool information across the array to stabilize estimates in the many sparse cells, without smoothing away the detail that the cells with large counts carry.

\section{Related Literature}
\label{sec:literature}

Low-rank models and statistical smoothing have a long history in statistical demography. Their success rests on the strong regularities that demographic processes exhibit, most notably across age and time. Regularities across age stem from the close link of mortality and fertility to biological processes, and of migration to life-course events. Regularities across time stem from co-movements such as the historical decline of mortality risk at all ages and in all subnational units of a country. Such common variation can often be summarized effectively by a small number of smooth latent factors.

Classical factorizations such as the Lee--Carter model and its extensions extract latent time factors \citep{lee1992modeling, li2005coherent, czado2005bayesian, wisniowski2015bayesian}. Alternative approaches instead extract common age profiles \citep{hyndman2007robust, hyndman2013coherent, alexander2017flexible, clark2019general, dharamshi2025jointly, zens2024flexible}. Typically, only a single margin of the array is factorized and the loadings form an array over all remaining dimensions. Hierarchical or smoothing priors on the loadings are then widely used to address the resulting overfitting concerns \citep{alexander2017flexible, susmann2022temporal, dharamshi2025jointly}. However, the number of loadings grows with the product of the sizes of all non-factorized dimensions, which makes such approaches impractical for arrays with several dimensions.

A complementary demographic tradition models schedules and surfaces directly rather than through latent factors. Its oldest members impose pre-specified functional forms on age schedules, as in the Gompertz law of mortality \citep{gompertz1825nature} and the model migration schedules of \citet{rogers1981model}. Modern smoothers relax the functional form. A canonical benchmark for mortality is the penalized spline model of \citet{currie2004smoothing}, see also \citet{camarda2012mortalitysmooth}. Related approaches such as TOPALS spline-adjust a standard schedule \citep{debeer2012smoothing} and have been embedded in Bayesian small-area estimation \citep{schmertmann2018bayesian}. Whether parametric or not, these methods typically fit one age curve or one age--time surface at a time. When the data carry additional dimensions, they are applied per subpopulation or linked through further hierarchical layers.

Neither tradition shares information across all dimensions of a multiway array. Thus, a natural way forward is to factorize and smooth all margins simultaneously. The literature on modern tensor models provides important building blocks. Models imposing multi-way low-rank structure are well established, for instance through array-normal models with separable covariances \citep{hoff2011separable, hoff2011hierarchical, hoff2015multilinear}, tensor-train structures \citep{oseledets2011tensor}, and general low-rank tensor estimation methods \citep{han2022optimal}. Shrinkage priors that adapt the effective rank have been developed for Gaussian and binary tensors \citep{rai2014scalable, zhao2015bayesian, stolf2025bayesian}, building on similar priors for factor models \citep{bhattacharya2011sparse, legramanti2020bayesian}. Smoothness of the factors may be modeled through penalties or Gaussian process priors \citep{yokota2016smooth, lei2025scalable}, or latent time series processes in dynamic factor models \citep{calder2007dynamic, wang2019factor, chen2022factor, billio2023bayesian, qin2025bayesian}. For count data, Poisson factorizations with nonnegative factors range from classical nonnegative matrix factorization to Bayesian Poisson tensor factorization \citep{lee1999learning, hu2015scalable, gopalan2015scalable, schein2015bayesian}.

A small number of demographic contributions have taken steps in this direction. \citet{russolillo2011extending} apply a Tucker component analysis to an array of log mortality rates over age, year, and country. \citet{fosdick2014separable} place a separable covariance with factor-analytic structure in every mode on log mortality rates indexed by age, sex, country, and year. \citet{zhang2024poisson} combine Poisson regression with a tensor-train decomposition for cause-specific mortality counts. \citet{martin2026modeling} extend penalized spline smoothing to age--time--space arrays to estimate small-area mortality. \citet{boonen2026lowrank} model multiway mortality arrays with low-rank tensor autoregressions for forecasting. Together, these works show that multiway low-rank structure is a natural description of demographic arrays.

What remains open is how to turn these ideas into a general model for multiway demographic arrays. The Swedish data in Sec.~\ref{sec:theproblem} show the typical features of such arrays, and together with the literature reviewed above they suggest three requirements. First, the model should respect the count nature of the data, with exposures entering as offsets and a mechanism that accommodates overdispersion, so that every cell is weighted by the information it carries and zero counts are informative observations rather than undefined log rates. Second, it should share information across all dimensions of the array at once, as common structure exists across age, time, and further stratifications alike, while preserving genuine differences between units. Third, it should encourage smoothness along ordered dimensions such as age and time, where adjacent levels typically carry similar demographic rates.

These requirements guide the model development in Sec.~\ref{sec:model}. A Poisson lognormal likelihood with exposure offsets handles the counts on their natural scale and accommodates overdispersion. A latent CP decomposition shares information across all modes. A shrinkage prior adapts both the number of components and the modes along which each component varies, so that parsimonious factorizations are favored but genuine differences between units are retained where the data support them. Gaussian process priors on ordered modes encode smoothness, with its degree learned per component and mode. Sec.~\ref{sec:cv} shows that this combination outperforms both canonical demographic benchmarks and existing count tensor factorizations in predictive accuracy and calibration. 

\section{Tensor Factor Models for Multiway Demographic Analysis}
\label{sec:model}

We consider modeling count-valued observations organized in a $D$-way array $\mathbf{Y}$ with mode sizes $n_1, \dots, n_D$ and cells indexed by $c = (c_1, \dots, c_D)$, where $c_d \in \{1, \dots, n_d\}$. In the motivating application, the modes are calendar year, age, sex, and county, so $D = 4$ and a cell is one year, age, sex, and county combination. The counts \( y_{c} \) are assumed to arise from conditionally independent Poisson distributions,
\begin{equation}
\label{eq:poisson}
y_{c} \sim \mathcal{P}\bigl(O_{c}\, e^{z_{c}}\bigr),
\end{equation}
where \( O_{c} \) is a fixed offset or exposure term representing an underlying known population at risk, so that \( z_c \) is the latent log rate of cell \( c \). We decompose the latent log rates into a structured surface and an error term,
\begin{equation}
\label{eq:state}
z_c = \eta_c + \varepsilon_c, \qquad \varepsilon_{c} \sim \mathcal{N}\big(0, \sigma^2_{c_s}\big),
\end{equation}
where the errors are independent across cells and account for measurement error, stochastic variation, and overdispersion. This renders the model \eqref{eq:poisson}--\eqref{eq:state} effectively a Poisson lognormal model (\citealp{aitchison1989multivariate}). The error variance is indexed by the levels of one designated mode \( s \), so heteroskedastic dispersion is permitted along that mode. Beyond accounting for overdispersion, a major benefit of the setup in \eqref{eq:state} is that, conditional on \( \mathbf{z} \), the model is Gaussian, so the conditional posteriors of all structural parameters remain tractable under the Gaussian process and global-local shrinkage priors introduced in Sec.~\ref{sec:prior}, which have no conjugate counterpart in Poisson--gamma factorizations. 

For the structured surface, we consider a CP-type low-rank tensor decomposition
\begin{equation}
\label{eq:main}
\eta_c = \mu + \sum_{r=1}^{R} g_r \prod_{d=1}^{D} m_{d,r}(c_d),
\end{equation}
where $\mu$ is a global level, $g_r$ is the weight of component $r$, $\mathbf{m}_{d,r} \in \mathbb{R}^{n_d}$ is its mode-$d$ vector, and $m_{d,r}(c_d)$ is entry $c_d$ of $\mathbf{m}_{d,r}$. Each component is thus a rank-one array, the outer product of one vector per mode. Every array can be represented exactly in this form for sufficiently large $R$ \citep{ kolda2009tensor}. Here $R$ is an upper bound on the number of components. The prior determines how many remain active and how complex each is, see Sec.~\ref{sec:prior}.

The proposed factorization extracts common variation across all data dimensions into low-dimensional mode vectors, which yields a parsimonious representation of the data. In the motivating application, each component requires \( 56 + 101 + 2 + 20 = 179 \) mode vector elements plus one weight, so at the upper bound of \( R = 20 \) components used below, the structural parameterization comprises about \( 3{,}600 \) parameters. This compares favorably with \( 226{,}240 \) free values for an unstructured surface, about \( 10{,}000 \) coefficients for separate Lee--Carter models by sex and subnational unit, about \( 7{,}000 \) parameters for an age factorization in the spirit of \citet{alexander2017flexible} with three common profiles and loadings for every year--sex--county combination, and about \( 7{,}000 \) covariance parameters for an array normal model with unstructured mode covariances, in addition to its mean model. The CP multiway decomposition is thus more parsimonious than any of these alternatives even at its upper bound. Importantly, its parameter count grows additively in the mode sizes when further modes are added, whereas the count of many factorization alternatives grows with the product of the sizes of the dimensions they do not factorize, and that of the array normal model grows with the squares of the mode sizes.

\subsection{Prior Specification}
\label{sec:prior}

For the latent mode vectors, we use priors that reflect demographic domain knowledge and let the data decide how much structure each component carries. Every mode vector is modeled as a common level plus deviations from it,
\begin{equation}
\label{eq:mode}
\mathbf{m}_{d,r} = \frac{a_{d,r}\mathbf{1}_{n_d} + \kappa_{d,r}\,\mathbf{v}_{d,r}}{\sqrt{1 + \kappa_{d,r}^2}}, \qquad a_{d,r} \sim \mathcal{N}(0,1), \qquad \mathbf{v}_{d,r} \sim \mathcal{N}\big(\mathbf{0}, \mathbf{C}_d(\phi_{d,r})\big),
\end{equation}
so that marginally \( \mathbf{m}_{d,r} \sim \mathcal{N}\big(\mathbf{0}, \mathbf{K}_d(\kappa_{d,r}, \phi_{d,r})\big) \)\footnote{Given \( \kappa_{d,r} \) and \( \phi_{d,r} \), \( \mathbf{m}_{d,r} \) is a linear combination of the independent zero-mean Gaussian variables \( a_{d,r} \) and \( \mathbf{v}_{d,r} \). It is therefore Gaussian with mean zero and covariance \( \{\mathrm{Var}(a_{d,r})\,\mathbf{1}\mathbf{1}^{\top} + \kappa_{d,r}^2\,\mathrm{Cov}(\mathbf{v}_{d,r})\}/(1+\kappa_{d,r}^2) = \{\mathbf{1}\mathbf{1}^{\top} + \kappa_{d,r}^2\,\mathbf{C}_d(\phi_{d,r})\}/(1+\kappa_{d,r}^2) \), where cross terms vanish by independence.} with
\begin{equation}
\label{eq:kernel}
\mathbf{K}_d(\kappa, \phi) = \frac{\mathbf{1}\mathbf{1}^{\top} + \kappa^2\, \mathbf{C}_d(\phi)}{1 + \kappa^2}.
\end{equation}
The parameter $\kappa_{d,r}$ thus controls how far the vector departs from the constant $a_{d,r} \mathbf{1}$ along mode $d$, and the correlation matrix $\mathbf{C}_d$ controls how.

Each mode is declared \textit{smooth} or \textit{exchangeable} a priori. Smooth modes are ordered dimensions such as age and time, whose neighboring levels typically carry similar outcomes and for which smoothness is well supported by demographic theory and evidence (\citealp{camarda2012mortalitysmooth}; \citealp{hyndman2013coherent}; \citealp{pavone2024learning}). For smooth modes, $\mathbf{C}_d(\phi)$ is a Mat{\'e}rn correlation matrix with smoothness parameter $3/2$ and length scale $L_d/\phi$ \citep{rasmussen2006gaussian}, evaluated at the coordinates of the mode, such as calendar years or single ages, where $L_d$ is the coordinate range of the mode. Hence, $\phi$ is an inverse length scale in units of the range. For exchangeable modes, such as sex and county, \( \mathbf{C}_d = \mathbf{I}_{n_d} \). Where a smooth mode contains a known discontinuity, \( \mathbf{C}_d \) can be made block-diagonal while the common level remains shared. In the application below, we will separate infant mortality at age 0 from the remaining ages.

This construction has three properties that the priors below build on. First, as \( \kappa_{d,r} \to 0 \), the mode vector becomes constant and component \( r \) does not vary along mode \( d \). As \( \kappa_{d,r} \to \infty \), only the deviations remain. Second, for smooth modes, \( \phi \to \infty \) makes the deviations independent across levels and recovers the exchangeable kernel, whereas small values of \( \phi \) make them vary smoothly over the whole range. The prior thus moves continuously between a constant, a smooth, and an exchangeable mode vector. Spline smoothers do not span this range, as their basis cannot resolve variation below the knot spacing. Third, since $\mathbf{C}_d$ is a correlation matrix in all cases, $\mathbf{K}_d$ has unit diagonal. Hence, every entry of a mode vector has prior variance equal to one. The scale of the mode vectors is therefore fixed a priori, and the amplitude of component \( r \) is carried by the weight \( g_r \) alone.

The first two properties motivate the priors on $\kappa_{d,r}$ and $\phi_{d,r}$. The share of the variance of a mode vector that is carried by its deviations is $\kappa^2/(1+\kappa^2)$. Under a standard half-Cauchy prior $\kappa \sim \mathcal{C}^{+}(0,1)$, this share follows a $\mathrm{Beta}(1/2, 1/2)$ distribution, the law of the shrinkage factor of the horseshoe prior \citep{car-etal:hor}, so the prior concentrates near a constant and near a fully varying mode vector while remaining proper. We use this prior with an additional global scale per mode, and the same construction for the inverse length scales of smooth modes,
\begin{equation}
\label{eq:kappa}
\kappa_{d,r} \mid \tau_{\kappa,d} \sim \mathcal{C}^{+}(0, \tau_{\kappa,d}), \qquad \phi_{d,r} \mid \tau_{\phi,d} \sim \mathcal{C}^{+}(0, \tau_{\phi,d}), \qquad \tau_{\kappa,d},\, \tau_{\phi,d} \sim \mathcal{C}^{+}(0,1),
\end{equation}
so that the amount of variation and its smoothness are learned per component and mode, while the global scales \( \tau_{\kappa,d} \) and \( \tau_{\phi,d} \) pool this information across the components of a mode.

The third property makes a shrinkage prior on the weights meaningful, because each weight alone carries the amplitude of its component. Again, we use a horseshoe prior,
\begin{equation}
\label{eq:hs}
g_r \mid \tau, \lambda_r \sim \mathcal{N}\big(0, \tau^2 \lambda_r^2\big), \qquad
\lambda_r \sim \mathcal{C}^{+}(0,1), \qquad
\tau \sim \mathcal{C}^{+}(0,1).
\end{equation}

The model thus contains two sparsity mechanisms. Because all mode vectors have fixed scale, the global-local shrinkage in \eqref{eq:hs} acts on the single quantity that carries the component amplitude and shrinks the effective number of components. If the data carry little information on a component, the global-local prior pushes its weight \( g_r \) toward zero, and the component contributes nothing to \( \boldsymbol{\eta} \) while its mode vectors simply follow their fixed-scale priors. This is the analogue of shrinkage priors used for complexity control in Bayesian factor models \citep{bhattacharya2011sparse, legramanti2020bayesian} and in Bayesian CP factorizations \citep{zhao2015bayesian, rai2014scalable}. The prior \eqref{eq:kappa} in turn shrinks every active component toward a simpler one, because a component with a constant mode vector in mode \( d \) is a lower-order interaction that does not involve that mode. 

We choose the remaining priors to be at most weakly informative. The global level receives the prior \( \mu \sim \mathcal{N}(0, s_0) \) with $s_0=100$, and the error variances receive conjugate priors \( \sigma^2_j \sim \mathcal{IG}(c_0, C_0) \) with $c_0=1$, $C_0=0.01$.

\subsection{Induced Prior Dependence}
\label{sec:covariance}

Write \( s_r^2 = \tau^2 \lambda_r^2 \) and let \( \mathbf{K}_{d,r} = \mathbf{K}_d(\kappa_{d,r}, \phi_{d,r}) \) denote the prior covariance of \( \mathbf{m}_{d,r} \). Conditional on $\mu$, the kernel parameters and the weight scales, components are independent, and within a component the weight and the mode vectors are independent with mean zero. Hence \( \mathbb{E}(\eta_c) = \mu \) and
\begin{equation}
\mathrm{Cov}\big(\eta_c, \eta_{c'}\big)
= \sum_{r=1}^{R} s_r^2 \prod_{d=1}^{D} \mathbf{K}_{d,r}\big(c_d, c_d'\big),
\label{eq:cov}
\end{equation}
so that every cell has prior variance \( \sum_r s_r^2 \) because all kernels have unit diagonal, and \( \mathrm{Cov}(z_c, z_{c'}) = \mathrm{Cov}(\eta_c, \eta_{c'}) + \sigma^2_{c_s} \mathbf{1}\{c = c'\} \).

The kernel \eqref{eq:kernel} also determines the implied prior correlation between two levels of a mode vector. The shared level \( a_{d,r} \) contributes a correlation of one with weight \( 1/(1+\kappa^2) \), and the deviations \( \mathbf{v}_{d,r} \) contribute their own correlation with weight \( \kappa^2/(1+\kappa^2) \). For an exchangeable mode, the deviations are independent across levels, so any two levels have correlation \( 1/(1+\kappa^2) \). Under \( \kappa \sim \mathcal{C}^{+}(0,1) \), that is, for \( \tau_{\kappa,d} = 1 \), this correlation follows a \( \mathrm{Beta}(1/2,1/2) \) distribution, which places most of its mass near zero and near one. For a smooth mode, two levels at distance \( \delta \) have correlation
\begin{equation}
\label{eq:adjcor}
\rho(\delta) = \frac{1 + \kappa^2 c_{\phi}(\delta)}{1+\kappa^2}, \qquad c_{\phi}(\delta) = \Big(1 + \frac{\sqrt{3}\,\phi\,\delta}{L_d}\Big)\exp\Big(-\frac{\sqrt{3}\,\phi\,\delta}{L_d}\Big),
\end{equation}
where \( c_{\phi} \) is the Mat\'ern correlation of the deviations. The correlation between neighboring ages or years is therefore at least \( 1/(1+\kappa^2) \). It approaches one when \( \kappa \) is small, so that the component barely varies along the mode, or when \( \phi \) is small, so that it varies smoothly. 


Three comparison cases are instructive. For a classical CP decomposition with independent standard Gaussian priors on all factor entries, \eqref{eq:cov} collapses to $\sum_r s_r^2 \mathbf{1}\{c = c'\}$, so distinct cells are a priori uncorrelated, including neighboring ages or years. Similarly, single-margin models such as the Lee--Carter model imply independence across populations when fitted separately, and they induce covariance along age and time only where priors on their age loadings or period index specify it directly. The array normal and separable factor models of \citet{hoff2011separable} and \citet{fosdick2014separable}, in turn, place a separable covariance $\boldsymbol{\Sigma}_1 \otimes \cdots \otimes \boldsymbol{\Sigma}_D$ on the array. In such models, dependence across modes factorizes. For instance, the rate at which temporal dependence decays does not depend on the age difference between observations. Equation \eqref{eq}, by contrast, sums $R$ separable terms with component-specific kernels, allowing different degrees of smoothness to coexist and dependence across modes to interact.

\subsection{Nested Demographic Model Structures}
\label{sec:nesting}

Two observations show how familiar model structures are nested inside the decomposition. First, as noted in Sec.~\ref{sec:prior}, a small value of $\kappa_{d,r}$ makes a mode vector effectively constant while the component itself remains active. Second, a component varying in a single mode acts as a main effect for that mode. Similarly, a component varying in two modes contributes a rank-one interaction between them, and so on for higher orders. 

Several established mortality models arise as special cases in this way. $D$ components that each vary in a single mode reduce the decomposition to the classical main-effects model $\mu + \sum_d a_d(c_d)$, and selected or all pairwise interactions, as well as interactions of higher order, arise in the same way. The estimation layer of the Lee--Carter model \citep{lee1992modeling} is an age main effect plus one rank-one age--time interaction, that is, one component varying in age only and one varying in age and time. Functional demographic models in the tradition of \citet{hyndman2007robust} add further smooth age--time components of this kind. Models with population-specific terms are nested as well, because a component whose vectors in the population modes, such as sex and county, are indicators of a single population acts within that population only. Separate Lee--Carter models per population are therefore contained at the cost of one set of components per population, and so is the augmented common factor model of \citet{li2005coherent}, which combines a common age--time factor with population-specific age levels and age--time deviations. The same holds for principal component representations of age schedules, as used in Bayesian small-area estimation \citep{alexander2017flexible}, which express every population schedule as a combination of a few common age profiles with population-specific loadings. The prior, however, favors structure that is shared across populations and uses population-specific components only where the data require them.

\subsection{Identification}
\label{sec:ident}

Latent factorizations such as \eqref{eq:main} are subject to several identification issues \citep{conti2014bayesian, assmann2016bayesian, fruhwirth2024sparse, qin2025bayesian}. Components can be permuted, signs can be flipped jointly within a component, and the split of scale between $g_r$ and the mode vectors is fixed only by the unit prior variance of the mode vector entries in \eqref{eq:kernel}. For $D \geq 3$, a CP decomposition is nevertheless essentially unique when $\sum_{d=1}^{D} k_d \geq 2R + D - 1$, where $k_d$, the Kruskal rank of the mode-$d$ factor matrix $[\mathbf{m}_{d,1}, \dots, \mathbf{m}_{d,R}]$, is the largest $k$ such that every set of $k$ columns is linearly independent \citep{kruskal1977three, sidiropoulos2000uniqueness}. In an overfitted model, this condition fails easily. Components that are shrunk toward constant vectors in the same mode have nearly collinear columns, which lowers $k_d$, and weak components can split or merge their structure without changing the implied surface. Individual components are therefore typically not interpretable on their own.

We thus report results exclusively through identified quantities, that is, functionals of the surface $\boldsymbol{\eta}$, and use its Gram matrices for this purpose. For mode $d$, unfold the surface into the $n_d \times (N/n_d)$ matrix $\tilde{\boldsymbol{\eta}}_{(d)}$, where $N$ is the number of cells. Its rows are the levels of mode $d$, its columns the combinations of the levels of the other modes, and each column is centered to mean zero. The Gram matrix $\boldsymbol{\Gamma}_d = \tilde{\boldsymbol{\eta}}_{(d)}\tilde{\boldsymbol{\eta}}_{(d)}^{\top}/N$ is its crossproduct. Its trace is the variation of the log rates along mode $d$ per cell, and its eigendecomposition splits this variation into orthogonal directions $\mathbf{u}_j$ with eigenvalues $\lambda_j$, so that $\lambda_j/\mathrm{tr}(\boldsymbol{\Gamma}_d)$ is the share carried by direction $j$. The eigenvectors $\mathbf{u}_j$ of the Gram matrix have unit length, so their entries carry no scale. We multiply them by $\sqrt{n_d\lambda_j}$, which gives a direction whose mean square over the levels equals $\lambda_j$, the variation it carries, so that its entries are typical deviations of the levels in log-rate units. These scaled directions are a multiway analogue of the age and time patterns of the Lee--Carter model. All of these quantities are computed per posterior draw, see Sec.~S2 for details.

\subsection{Posterior Simulation}
\label{sec:mcmc}

Posterior simulation relies on the observation that, conditional on \( \mathbf{z} \), the remaining model is a Gaussian multilinear model, offering closed-form conditional posteriors for all structural parameters of interest. This makes data augmentation, where the algorithm iterates between (i) imputing the latent \( z_c \) conditional on the remaining parameters and (ii) updating the remaining parameters conditional on \( \mathbf{z} \), a natural candidate for posterior simulation (\citealp{tanner1987calculation}). To update \( z_{c} \), we use an independence Metropolis--Hastings step whose proposal is the Laplace approximation of its full conditional. As the conditionals are typically close to Gaussian, acceptance rates are close to one. Cells with zero exposure or missing counts contribute no likelihood term, so their latent rates are drawn exactly from the Gaussian conditional law implied by \eqref{eq:state}.

Conditional on $\mathbf{z}$, the mode vectors $\mathbf{M}_d = [\mathbf{m}_{d,1}, \dots, \mathbf{m}_{d,R}]$ of every mode $d$ and the block $(\mu, \mathbf{g})$ of the global level and the weights have Gaussian full conditionals whose precisions and canonical means are per-mode contractions of the type used in alternating least squares for CP decompositions (\citealp{kolda2009tensor}). Each $\mathbf{M}_d$ is drawn jointly, and so is $(\mu, \mathbf{g})$. Two further blocks are updated with one of these Gaussian blocks integrated out, which removes a strong posterior coupling in each case and constitutes a partially collapsed Gibbs step \citep{vandyk2008partially}. The kernel parameters $(\kappa_{d,r}, \phi_{d,r})$ are moved by random-walk Metropolis steps on the log scale with their mode vector $\mathbf{m}_{d,r}$ integrated out analytically, after which $\mathbf{m}_{d,r}$ is re-drawn exactly given the other columns of $\mathbf{M}_d$. The horseshoe scales $(\lambda_1, \dots, \lambda_R, \tau)$ are moved in the same way with $(\mu, \mathbf{g})$ integrated out, which is possible because the surface is linear in $(\mu, \mathbf{g})$ given the mode vectors, after which $(\mu, \mathbf{g})$ is re-drawn exactly. The remaining parameters of the half-Cauchy hierarchies, the global scales $\tau_{\kappa,d}$ and $\tau_{\phi,d}$ and the auxiliary variables, have closed-form conditional updates via an auxiliary variable representation (\citealp{makalic2015simple}). The error variances $\sigma^2_j$ of the levels of mode $s$ have conjugate inverse gamma full conditionals.

Data augmentation samplers for large and sparse count arrays are, however, notoriously slow to mix. In a cell with few or no events, the latent log rate $z_c$ is determined mostly by its Gaussian layer $\mathcal{N}(\eta_c, \sigma^2_{c_s})$, so $z_c$ follows $\eta_c$ closely and the structural parameters move more freely in a noncentered parameterization that holds the residual $z_c - \eta_c$ fixed. On the contrary, in a cell with many events, $z_c$ is pinned down by the data, and the structural parameters are best updated conditional on $z_c$, a centered parameterization. Demographic arrays typically contain both kinds of cells, so no single parameterization serves all of them. We therefore use, in every layer of the model, the ancillarity--sufficiency interweaving strategy (ASIS) of \citet{yu2011center}, which has proven effective in other latent variable model contexts \citep{kastner2014ancillarity, kastner2017efficient}. This applies to three blocks, the mode vectors $\mathbf{M}_d$ of each mode, the global level and weights $(\mu, \mathbf{g})$, and the error variances $\sigma^2_j$ of the levels of mode $s$. On every second sweep, after its centered update, each block is moved once more in the noncentered parameterization. For $\mathbf{M}_d$ and $(\mu, \mathbf{g})$ this is a Metropolis--Hastings step with a Newton-type proposal (\citealp{gamerman1997sampling}), and for the $\sigma^2_j$ a random-walk step on the log scale. These steps leave the posterior invariant and add a fixed fraction to the cost of a sweep. In test runs on subsets of the data, they raised the effective sample sizes of the identified quantities per unit of computing time by one to two orders of magnitude, and with them convergence of all quantities of interest is satisfactory (see Sec.~\ref{sec:app}). Full details of the MCMC algorithm are given in Sec.~S1, and Sec.~S3 reports two simulation studies. The first shows that the algorithm recovers the surface, its leading directions, and the error variances of a known four-way truth well. The second is a small factorial experiment that varies the rank, noise level, and exposure of the true surface. It shows that the shrinkage prior removes the spurious structure that an unshrunk CP model fits to sampling noise, and that the smoothing kernels add accuracy, most of all at low exposure.

\section{Demonstrating the Prior}
\label{sec:prior_demo}

To illustrate the two sparsity mechanisms described in Sec.~\ref{sec:prior}, we fit the model to a subset of the Swedish data, restricted to the four least populated counties (Gotland, J\"amtland, Blekinge, and Kronoberg) in the years 2021 to 2024. The subset comprises \( 3{,}232 \) cells with \( 22{,}952 \) deaths, so likelihood information is moderate and the effect of the prior is clearly visible. Full panel results are discussed in Sec.~\ref{sec:app}. We compare the proposed Bayesian adaptive tensor smoothing (BATS) model to the same model with independent standard Gaussian priors on all mode vector entries and \( \mathcal{N}(0, 10) \) priors on the weights, which we refer to as the CP Poisson lognormal model (CP-PLN). Both models use \( R = 20 \) and the posterior sampling settings given in Sec.~\ref{sec:app}.

\begin{figure}[!t]
    \centering
    \includegraphics[width=\textwidth]{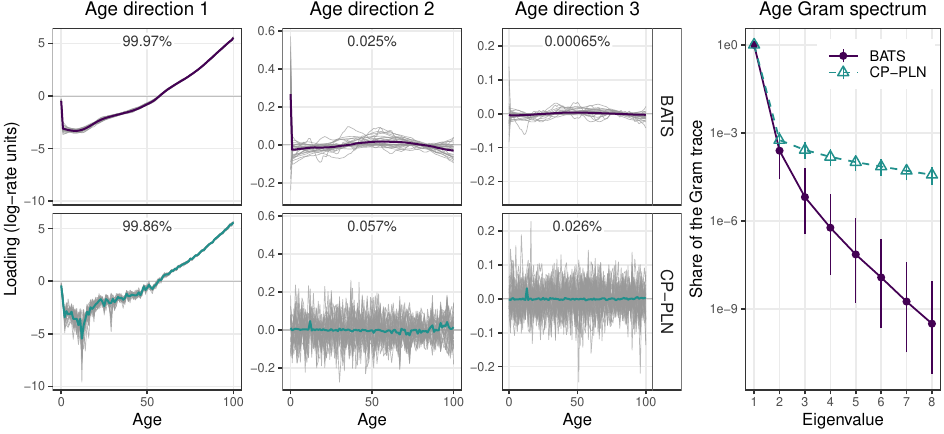}
    \caption{Effect of the prior on the subset of the four least populous Swedish counties, 2021--2024. Columns 1--3: the first three directions of the age mode, \( \sqrt{n_d\lambda_j}\,\mathbf{u}_j \) in log-rate units, under BATS (top) and CP-PLN (bottom), with 25 posterior draws (thin grey lines), the posterior median (thick line), and the median share of the variation along age. Right: posterior medians and 90 percent intervals of the eigenvalues of the age Gram matrix as shares of its trace, on a log scale.}
    \label{fig:prior}
\end{figure}

Fig.~\ref{fig:prior} shows the first three directions of the age mode, computed per posterior draw from the age Gram matrix as described in Sec.~\ref{sec:ident}, together with the spectrum of its eigenvalues. Both models agree on the leading direction, the typical age schedule of mortality, which carries more than \( 99.8 \) percent of the variation along age. Under CP-PLN, however, this direction is jagged below age 30, where most cells of the subset are empty, and the second and third directions are noise of up to about \( 0.2 \) on the log scale without any structure across ages, whose eigenvalues level off between \( 10^{-5} \) and \( 10^{-3} \) of the total variation. Under BATS, all estimated patterns are smooth, the second and third directions are shrunk towards zero, with shares of about \( 2.5 \times 10^{-4} \) and \( 10^{-5} \), and the eigenvalues continue to decay by about an order of magnitude per direction. Note that the somewhat larger uncertainty at age 0 in directions 2 and 3 under BATS results from excluding this age from the Mat\'ern kernel to capture the typically non-smooth infant mortality spike.

The prior thus removes the spurious interactions that the unshrunk model fits to sampling noise, and it does so via two channels at once. It reduces the effective number of components via the prior on $g_r$, and it simplifies the components themselves, because mode vectors that the data do not require to vary are pulled toward constants. BATS thus learns the effective tensor rank and the modes along which each component varies simultaneously, a stronger form of regularization than shrinkage of the number of components alone.

The supplementary materials (Sec.~S4) report the full comparison for all four modes and the fitted age schedules, together with a third model that retains both shrinkage mechanisms but replaces the Mat\'ern kernels by exchangeable ones, which isolates the contribution of smoothing.

\section{Comparative Cross-Validation Exercise}
\label{sec:cv}

To assess the usefulness of the model for reconstructing demographic patterns, we implement a systematic cross-validation exercise using a random cell holdout design. The \( 226{,}240 \) cells of the Swedish mortality data are split once into twenty folds by a random permutation, so each fold holds out a uniformly random five percent of the cells (roughly \( 11{,}300 \) cells). Because individual cells are held out at random, the neighboring ages and years of every held-out cell typically remain observed, and even a model without any smoothing can relatively easily impute a held-out cell from its neighbors. This favors methods without explicit smoothing mechanisms, so the gains of BATS reported below are likely conservative.

We compare BATS to eight alternative specifications. The first four are established demographic point estimators, representing different degrees of information pooling. A one-dimensional Poisson P-spline smoother fitted to every single age schedule (\citealp{eilers1996flexible}; \citealp{camarda2012mortalitysmooth}) pools no information across years or units. A two-dimensional Poisson P-spline age--time surface per county--sex unit, as in \citet{currie2004smoothing} and \citet{camarda2012mortalitysmooth}, pools information across ages and years within a unit. For both, penalties are selected using the Bayesian information criterion within each training fold. A Poisson Lee--Carter model per unit (\citealp{brouhns2002poisson}) imposes a rank-one age--time structure within a county--sex unit without any smoothness. A TOPALS relational model (\citealp{debeer2012smoothing}) with a sex-specific standard schedule pooled from the training cells of the fold and a penalized linear-spline deviation per age schedule pools information through the standard schedule only.

The remaining four specifications are Bayesian models. An additive main-effects model, \( \eta_c = \mu + \sum_d a_d(c_d) \), with the Poisson lognormal observation layer of the proposed model, serves as the interaction-free baseline. CP-PLN, as introduced in Sec.~\ref{sec:prior_demo}, keeps the PLN count layer with CP structure, but without shrinkage or smoothing mechanisms. BATS and CP-PLN use \( R = 20 \). The variance \( \sigma^2_{c_s} \) is indexed by age, as in Sec.~\ref{sec:app}. Finally, the Bayesian Poisson tensor factorization (BPTF) of \citet{schein2015bayesian} is considered with \( R = 10 \) and \( R = 20 \) components. This is a Poisson--gamma CP model in which the expected count is the exposure times a sum of products of nonnegative gamma-distributed factor entries, with no overdispersion layer, no explicit component weights, and no smoothing priors. All Bayesian models are fitted by MCMC with \( 20{,}000 \) warm-up and \( 20{,}000 \) sampling sweeps, and every fourth draw is retained, so \( 5{,}000 \) posterior draws enter all summaries. Further implementation details are given in the supplementary materials (Sec.~S6).

For evaluation, we collect log predictive scores (LPS), computed cell-wise by adaptive Gauss--Hermite quadrature over the Gaussian error layer, or by the exact Poisson mixture over the rate draws for BPTF. Point predictions are assessed by the mean Poisson deviance of the held-out counts against the posterior mean expected counts of the Bayesian models and the fitted counts of the point estimators. Calibration is assessed through randomized probability integral transform (PIT) ranks of the held-out counts among the stored predictive draws, by the empirical coverage of central predictive intervals and by the Kolmogorov--Smirnov distance of the PIT values from uniformity. All comparisons are based on \( 226{,}174 \) held-out cells with positive exposure.

\begin{table}[!t]
\renewcommand\baselinestretch{1.15}
  \centering
  \small
  \begin{threeparttable}
    \caption{Cross-validation performance over all held-out cells}
    \label{tab:cv}
      \begin{tabular}{lccccc}
        \toprule
        Model & LPS & Deviance & Cov.\ 90 & Cov.\ 95 & PIT KS \\
        \midrule
        1-D P-spline (per age schedule)   & --       & 1.813 & --    & --    & --    \\
        2-D P-spline (per unit)           & --       & 1.087 & --    & --    & --    \\
        Lee--Carter (per unit)            & --       & 1.130 & --    & --    & --    \\
        TOPALS                            & --       & 1.197 & --    & --    & --    \\
        Additive main effects             & $-2.279$ & 1.345 & 0.900 & 0.950 & 0.007 \\
        BPTF, $R = 10$                    & $-2.193$ & 1.070 & 0.890 & 0.943 & 0.008 \\
        BPTF, $R = 20$                    & $-2.192$ & 1.069 & 0.891 & 0.943 & 0.008 \\
        CP-PLN                            & $-2.191$ & 1.068 & 0.900 & 0.949 & $\mathbf{0.002}$ \\
        BATS                              & $\mathbf{-2.186}$ & $\mathbf{1.058}$ & 0.900 & 0.950 & $\mathbf{0.002}$ \\
        \bottomrule
      \end{tabular}
    \begin{tablenotes}[flushleft]
      \footnotesize
      \item\textit{Notes}: LPS = mean log predictive score per cell. Deviance = mean Poisson deviance per cell of the posterior mean expected count (Bayesian models) or of the fitted count (point estimators). Cov.\ 90 and Cov.\ 95 report empirical central coverage of nominal 90 and 95 percent predictive intervals based on randomized PIT ranks, and PIT KS the Kolmogorov--Smirnov distance of the PIT values from the uniform distribution. Results are averaged across all $226{,}174$ scored held-out cells with positive exposure, belonging to $2{,}240$ age schedules. Column-wise best performance is highlighted in \textbf{bold}. The first four methods are point estimators, so probabilistic scores are not defined for them.
    \end{tablenotes}
  \end{threeparttable}
\end{table}

Tab.~\ref{tab:cv} reports the results. BATS attains the best log predictive score and the smallest deviance, its predictive intervals are calibrated to within a tenth of a percentage point at both nominal levels considered, and its PIT values are practically indistinguishable from uniform draws. All differences between BATS and its competitors in both log predictive score and deviance are at least \( 5 \) standard errors from zero (Sec.~S5).

Among the demographic benchmarks, the ordering follows the degree of pooling. The schedule-wise P-spline and TOPALS, which pool little or no information across years, perform worst. The P-spline surface gives better point estimates, with a deviance three percent above that of BATS. Fitting separate Lee--Carter models to subpopulations appears to be of limited use for subnational data, as it is outperformed both by a per-unit smoother and by every model that pools information across units.

Two conclusions follow from comparing the results of the Bayesian competitors. First, most of the distance to the demographic benchmarks is covered by combining a multiway low-rank structure with a count likelihood. The four count tensor models are clearly ahead of every other method. The additive model shows the price of removing the interactions, with a lower log score and a deviance higher by more than a quarter. Second, the priors introduced in this article improve on the CP-PLN and BPTF specifications in point prediction and in predictive distribution alike. The differences are consistent across unit sizes, and the gains in log predictive score are on average larger in the two smaller quartiles of unit size than in the two larger ones (Sec.~S5). Both models with the overdispersion layer attain nominal coverage to within a tenth of a percentage point, whereas BPTF undercovers at both nominal levels, since a pure Poisson likelihood cannot absorb the extra-Poisson variation of the rates. A visual comparison of the fitted schedules of all nine methods for the least populous unit, which illustrates the mechanisms behind these scores, is given in Sec.~S6.

In order to explore whether these findings hold beyond the motivating mortality application, the cross-validation exercise is reproduced in Sec.~S7 of the supplementary materials on a second data set describing Austrian district-level outmigration by single year of age, sex, district, and year, with a total of \( 376{,}376 \) cells. The count tensor models are again ahead of all other methods, BATS again attains the best log predictive score and deviance with coverage within a tenth of a percentage point of nominal, and BPTF undercovers more strongly. The lead of BATS over CP-PLN in log predictive score and in deviance is again substantial. Against BPTF with $R = 20$, BATS is well ahead in log predictive score and slightly ahead in deviance (see Sec.~S7 for detailed results).

\section{Application to Swedish Mortality Data}
\label{sec:app}

The model is fitted to the full four-way array of Swedish death counts described in Sec.~\ref{sec:theproblem}, with person-year exposures as offsets. The time and age modes are declared smooth, with kernels evaluated at calendar years and at single ages. The age kernel separates age 0 from the ages above by a break point, because infant mortality is driven by perinatal conditions and changes discontinuously into childhood mortality. The sex and county modes are treated as exchangeable. We use \( R = 20 \) as the upper bound on the number of components. The variance $\sigma^2_{c_s}$ is indexed by age, so that the error variance of the log rates can vary over the life course. Because the error layer absorbs what the low-rank surface does not capture, this allows for larger residual variation at ages where the schedule is more finely structured and the decomposition fits less closely.

\begin{figure}[!t]
    \centering
    \begin{subfigure}[b]{0.325\textwidth}
        \centering
        \includegraphics[width=\textwidth]{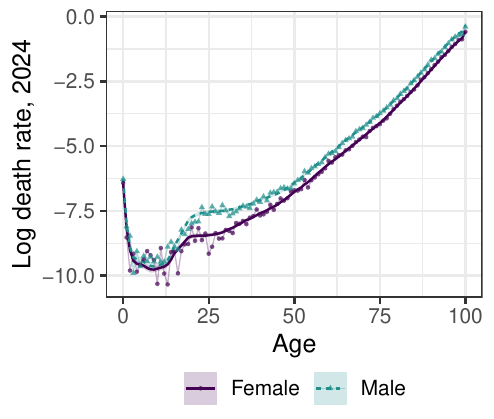}
    \end{subfigure}
    \hfill
    \begin{subfigure}[b]{0.325\textwidth}
        \centering
        \includegraphics[width=\textwidth]{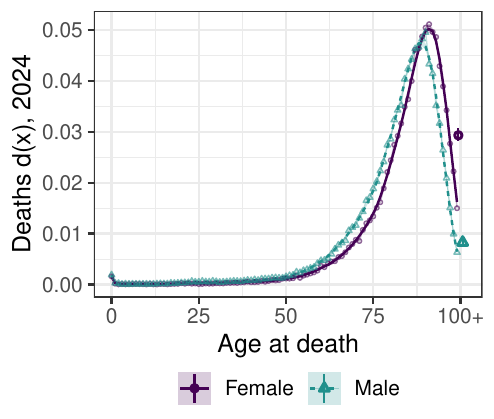}
    \end{subfigure}
    \hfill
    \begin{subfigure}[b]{0.325\textwidth}
        \centering
        \includegraphics[width=\textwidth]{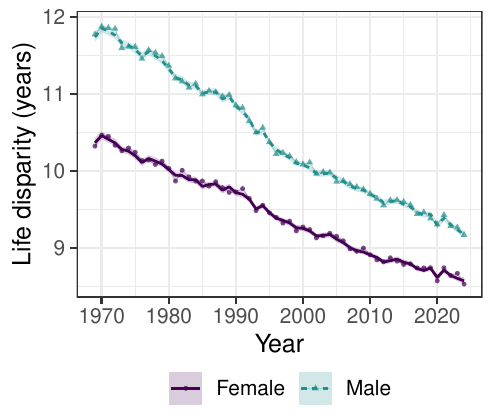}
    \end{subfigure}

    \begin{subfigure}[b]{0.325\textwidth}
        \centering
        \includegraphics[width=\textwidth]{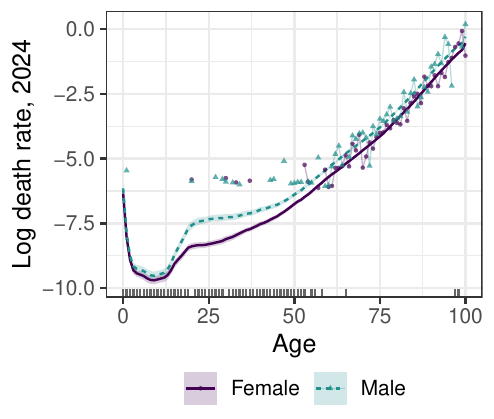}
    \end{subfigure}
    \hfill
    \begin{subfigure}[b]{0.325\textwidth}
        \centering
        \includegraphics[width=\textwidth]{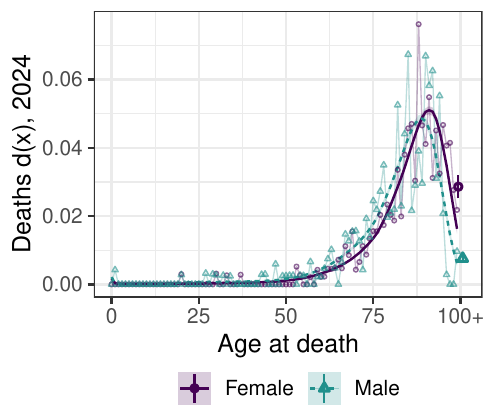}
    \end{subfigure}
    \hfill
    \begin{subfigure}[b]{0.325\textwidth}
        \centering
        \includegraphics[width=\textwidth]{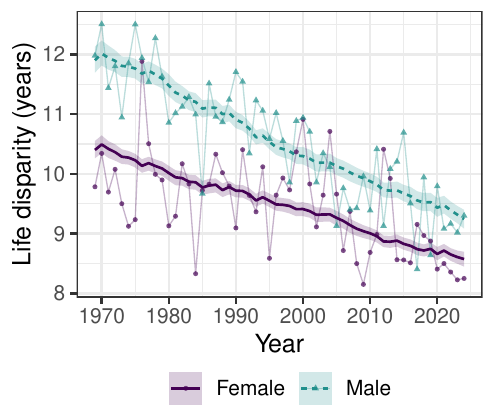}
    \end{subfigure}
\caption{Posterior distributions for the quantities of Fig.~\ref{fig:rawdata}, for Sweden (top row) and Gotland (bottom row), by sex. Lines and bands are posterior medians and 90 percent credible intervals, points are the raw values. Left: log mortality rates by age in 2024. Ages with zero observed deaths are marked by rug ticks. Middle: life table distribution of deaths by age in 2024, with the open age group as a detached point at age 100. Right: life disparity $e^{\dagger}$ by year.}
    \label{fig:fit}
\end{figure}

We run $20{,}000$ warm-up sweeps followed by $20{,}000$ sampling sweeps and retain every fourth draw, so that $5{,}000$ posterior draws are used for inference. On the application array with $226{,}240$ cells and $R = 20$, one sweep takes about one second on a single core, so that the main fit takes about eleven hours. This is about the cost of the Poisson tensor factorization of \citet{schein2015bayesian} with $R = 20$ under full MCMC. Convergence is satisfactory, with split-$\hat{R}$ values at or below $1.02$ for all monitored quantities. Computation times for all methods and convergence diagnostics are given in Secs.~S1 and~S5 of the supplementary materials.

\begin{figure}[!t]
    \centering
    \begin{subfigure}[b]{0.49\textwidth}
        \centering
        \includegraphics[width=\textwidth]{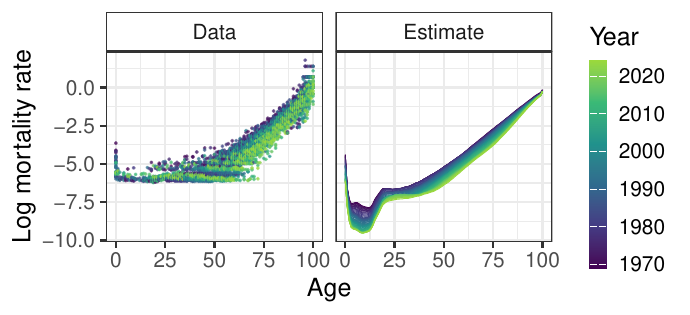}
    \end{subfigure}
    \hfill
    \begin{subfigure}[b]{0.49\textwidth}
        \centering
        \includegraphics[width=\textwidth]{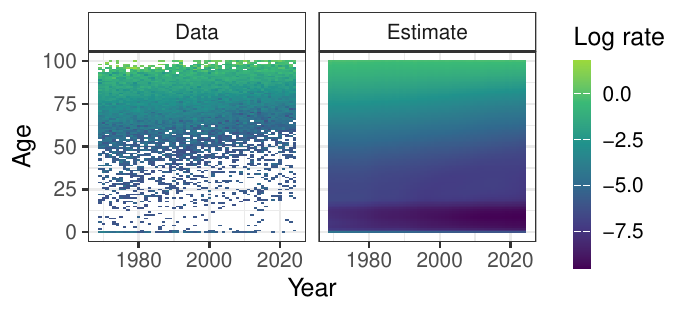}
    \end{subfigure}
    \caption{Observed and fitted log mortality rates for men in Gotland, a sparsely populated island county with substantial stochastic variation in the raw data, as age schedules colored by year (left) and as age--year tiles (right). Within each panel, the left facet shows the observed rates and the right facet the posterior mean of the smoothed surface \( \boldsymbol{\eta} \). Cells with zero counts or zero exposure are dropped on the data side.}
    \label{fig:true_vs_fitted}
\end{figure}

\paragraph*{Smoothing behavior} Fig.~\ref{fig:fit} repeats the panels of Fig.~\ref{fig:rawdata} with the posterior distributions of the corresponding quantities overlaid. The posterior for Sweden as a whole is not fitted separately but aggregated per draw from the county cell estimates with exposure weights (Sec.~S5). For Sweden as a whole, where the data are dense, the posterior follows the raw values closely and its credible bands are narrow. For Gotland, the posterior age schedules are smooth while retaining the infant drop at age 0, the age-at-death distributions concentrate around a well-determined mode, and life disparity declines steadily over the study period. Patterns that are invisible in the raw data also emerge, such as clearly separated schedules for men and women with the well-known young-adult mortality hump of men in the late teens and twenties, typically stemming largely from accidents and other external causes. Because the entire life table is a deterministic function of the age schedule, every posterior draw of the surface yields one draw of every life table quantity, and derived quantities are therefore smoothed and equipped with uncertainty in one step, without any additional modeling. Details of the life table computations are given in the supplementary materials (Sec.~S5). Here and below, posterior summaries of rates and life table quantities are computed per draw from the smoothed surface $\exp(\boldsymbol{\eta})$. They reflect uncertainty about the smoothed rates and exclude the cell-level error $\varepsilon_c$ and Poisson variation.

Fig.~\ref{fig:true_vs_fitted} shows observed and fitted rate schedules for men in Gotland over all years, as age curves colored by year and as age--year tiles. The tile representation makes the structure of the missing information visible. A large share of the age--year cells carries zero deaths, and a few at the highest ages zero exposure, so that their raw log rates are undefined and the cells are blank in the data panel. The fitted surface fills these cells with model-based rates consistent with the neighboring ages and years of the same unit and with the other units of the panel, borrowing information about demographic patterns from other subpopulations.  As a more systematic assessment beyond this illustrative case, the supplementary materials also rank all \( 2{,}240 \) age schedules by the held-out log predictive score they received in the cross-validation exercise of Sec.~\ref{sec:cv}, and display the in-sample fits of the ten best, the ten worst, and ten randomly chosen age schedules (Sec.~S5). We find that even the worst-ranked schedules are fitted with smooth and demographically plausible age profiles that are not distorted by the sampling noise of sparse cells. For the most informative schedules of large units, however, the low-rank surface is somewhat too structured to reproduce year-specific irregularities in the age profile, which is why these schedules tend to rank lowest. A larger upper bound $R$ would likely resolve this, but at a computational cost that is not warranted here.

\begin{figure}[!t]
    \centering
    \includegraphics[width=\textwidth]{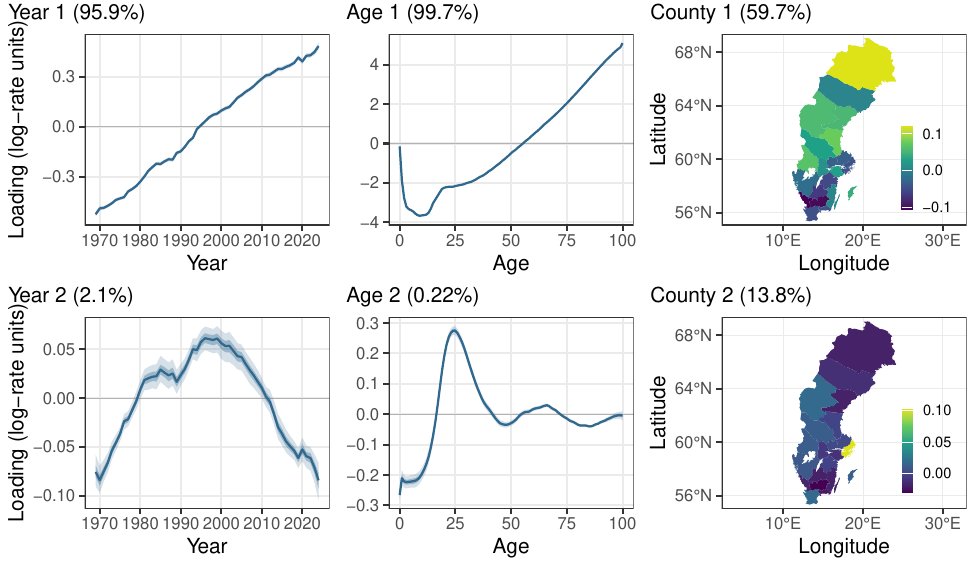}
    \caption{Leading directions of the main fit. The first (top) and second (bottom) directions of the time, age, and county modes, \( \sqrt{n_d\lambda_j}\,\mathbf{u}_j \) in log-rate units. The county directions are shown as maps of posterior medians in longitude and latitude. Lines are posterior medians, bands show 50 and 90 percent credible intervals, and the panel titles give the median share of the variation along the mode. The sign of every direction is arbitrary.}
    \label{fig:directions}
\end{figure}

\paragraph*{Leading directions} To gain structural insights into broad patterns across all data dimensions, we inspect the identified directions based on the eigendecompositions of the Gram matrices, as discussed in Sec.~\ref{sec:ident}. Fig.~\ref{fig:directions} shows the two leading directions of the time, age, and county modes, each in log-rate units with credible bands. The leading time direction, with \( 96 \) percent of the variation along time, is the secular decline of mortality, which proceeds almost linearly over the study period. The second time direction, with $2$ percent, rises until the late 1990s and falls thereafter. Had all combinations of age, sex, and county followed the same course over time, differing only in the strength of their decline, the time mode would show a single direction. The second direction thus captures how their trajectories differed and implies some declined relatively faster before the late 1990s, others afterwards. The leading age direction, with nearly all of the variation along age, is the age schedule of mortality with the infant drop, the minimum around age ten, and the near-linear increase thereafter. The second age direction peaks between ages 20 and 30, the ages of young-adult excess mortality, which typically differs between men and women and changes over time. The leading county direction, with \( 60 \) percent of the variation along counties, is a north--south gradient, with the northernmost counties at one end and the southernmost at the other. It is worth noting that the model recovers this pattern although it treats counties as exchangeable a priori. The second county direction, with \( 14 \) percent, contrasts the Stockholm metropolitan region with the remaining counties. The sex mode has a single direction, which separates men and women by about half a log unit on average.

\begin{figure}[!t]
    \centering
    \begin{subfigure}[b]{0.49\textwidth}
        \centering
        \includegraphics[width=\textwidth]{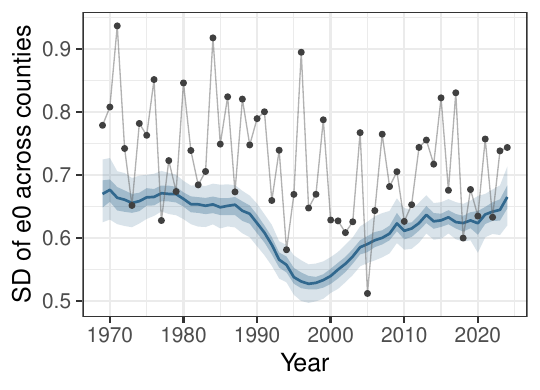}
    \end{subfigure}
    \hfill
    \begin{subfigure}[b]{0.49\textwidth}
        \centering
        \includegraphics[width=\textwidth]{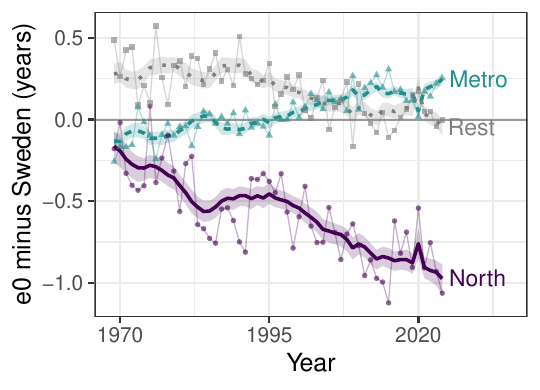}
    \end{subfigure}
    \caption{Regional inequality in life expectancy at birth, men. Left: standard deviation of life expectancy across the 20 counties by year, computed from the raw rates (points) and per posterior draw of the smoothed surface (line with 50 and 90 percent credible bands). Right: life expectancy of the northern counties (G\"avleborg, V\"asternorrland, J\"amtland, V\"asterbotten, Norrbotten), the metropolitan counties (Stockholm, Uppsala--V\"astmanland, Sk\aa ne, V\"astra G\"otaland), and the remaining counties, each expressed as the difference from the life expectancy of Sweden as a whole in the same year and sex, from pooled raw rates (points) and per posterior draw of the smoothed surface (lines with 90 percent credible bands). National and group values are exposure-weighted aggregates of the county rates (Sec.~S5).}
    \label{fig:regional}
\end{figure}

\paragraph*{Regional inequality in life expectancy} It is worth highlighting that the smoothed tensor also supports substantive analyses that the raw data cannot. Consider tracking the standard deviation of life expectancy at birth across counties, a natural summary of regional inequality in mortality. Computed from the raw mortality rates, this quantity is inflated by sampling noise in the small counties. For men, the standard deviation of life expectancy across counties averages $0.72$ years over the study period when computed from the raw rates, but only $0.62$ years under the smoothed posterior. Moreover, the raw standard deviation fluctuates considerably from one year to the next, concealing any temporal trend (Fig.~\ref{fig:regional}, left). The posterior of this inequality measure, by contrast, shows a clear pattern. It is stable at about \( 0.65 \) years until the late 1980s, declines to a minimum of \( 0.53 \) years in 1997, and rises again to \( 0.67 \) years in 2024. A decomposition by region (Fig.~\ref{fig:regional}, right) points to geographic divergence as one underlying reason. Until the late 1990s, the counties outside the metropolitan areas and the North lost their earlier advantage while the metropolitan counties of Stockholm, Uppsala--V\"astmanland, Sk\aa ne, and V\"astra G\"otaland improved, and both groups had converged by about 2000, while the five northern counties remained about half a year below the national level. Since 2000, the metropolitan counties have moved ahead, to \( 0.25 \) years above the national level in 2024, and the North has fallen further behind, to almost one year below it. Results for women, with a similar but weaker pattern, are given in the supplementary materials (Sec.~S5). Which underlying causes of death and age groups drive this divergence, and why, requires further investigation beyond the scope of this article.

\section{Discussion and Concluding Remarks}
\label{sec:conclusion}

We propose a Bayesian adaptive tensor smoothing model for count arrays, particularly demographic data stratified by age, time, and other characteristics. The prior adapts the effective number of components, the modes along which each component varies, and its smoothness to the data. A data augmentation MCMC algorithm with ancillarity--sufficiency interweaving is used for posterior simulation. In an application to Swedish county-level mortality, the model denoises the sparse schedules of small populations, yields interpretable directions of variation, propagates posterior uncertainty to life table functionals in one step, and reveals a decline and subsequent rise of regional inequality in life expectancy that the raw data do not show. In cross-validation exercises the proposed method outperforms both canonical demographic benchmarks and existing count tensor factorizations in predictive accuracy and calibration.

Several avenues for future research emerge. First, forecasting is the most prominent extension. The time mode of the present model carries a correlation kernel. This is appropriate for smoothing, but under extrapolation implies mean reversion, which does not suit non-stationary demographic processes such as mortality. Three routes may lead to principled forecasts. First, the kernel on the time mode can be replaced by an explicit time series process. Such a prior rarely has a fixed marginal variance, however, so the weight $g_r$ would no longer carry the amplitude of its component alone, and the shrinkage prior on the weights would lose its scale. Components may also require different dynamics, stationary for some and nonstationary for others, complicating a general specification. Second, in line with current practice in functional demographic forecasting (\citealp{hyndman2007robust}; \citealp{hyndman2013coherent}), time series models for the time directions may be fitted ex-post, ideally by posterior draw, and used for extrapolation. This resembles a ``cut'' posterior in the sense of \citet{plummer2015cuts}. It separates smoothing from forecasting and propagates uncertainty from the smoothed surface into forecasts, but does not let the forecasting model inform smoothing. Third, the model could be formulated directly for latent log mortality improvements -- the first differences of log rates over time -- which are plausibly stationary. Smoothing and forecasting would then take place within one model. This route, however, requires non-trivial changes to the model, since the level of the surface enters through cumulated improvements and the fixed-scale construction of \eqref{eq:mode} no longer applies, and in particular to the posterior simulation algorithm, whose Gaussian regression and interweaving steps rely on the multilinear form of $\boldsymbol{\eta}$.

A second direction relates to computational scalability. As the number and size of modes grow, decomposition formats that represent the interaction structure between subsets of modes more compactly than a sum of rank-one terms and faster estimation routes such as variational or maximum a posteriori estimation appear attractive.

A third point concerns the priors on the mode vectors. Any correlation matrix with unit diagonal is admissible as \( \mathbf{C}_d \) in \eqref{eq:kernel}, so the exchangeable kernel on the unordered modes can be replaced by structured alternatives. For regional units, a spatial kernel on centroid distances or a standardized intrinsic autoregressive prior on the adjacency graph would let neighboring units borrow strength a priori, with an inverse length scale learned per component as for age and time. More generally, the mode correlation matrices themselves can be learned more flexibly, either unrestricted or with the factor-analytic structure of \citet{fosdick2014separable}.

\subsection*{Data Availability Statement}

The data that support the findings of this study are openly available in the public statistical database of \textit{Statistics Sweden} at \url{https://www.statistikdatabasen.scb.se} (tables BE0101D8 and BE0101N1) and, for the Austrian migration panel of the supplementary materials, in the \textit{StatCube} database of \textit{Statistik Austria} at \url{https://statcube.at}. 

\singlespacing
\bibliographystyle{apalike}
\bibliography{Bibliography-MM-MC}

\begin{thebibliography}{}

\bibitem[Aitchison and Ho, 1989]{aitchison1989multivariate}
Aitchison, J. and Ho, C. (1989).
\newblock {The multivariate {P}oisson-log normal distribution}.
\newblock {\em Biometrika}, 76(4):643--653.

\bibitem[Alexander et~al., 2017]{alexander2017flexible}
Alexander, M., Zagheni, E., and Barbieri, M. (2017).
\newblock {A flexible Bayesian model for estimating subnational mortality}.
\newblock {\em Demography}, 54(6):2025--2041.

\bibitem[Andreev and Kingkade, 2015]{andreev2015average}
Andreev, E.~M. and Kingkade, W.~W. (2015).
\newblock {Average age at death in infancy and infant mortality level:
  reconsidering the Coale--Demeny formulas at current levels of low mortality}.
\newblock {\em Demographic Research}, 33:363--390.

\bibitem[A{\ss}mann et~al., 2016]{assmann2016bayesian}
A{\ss}mann, C., Boysen-Hogrefe, J., and Pape, M. (2016).
\newblock {Bayesian analysis of static and dynamic factor models: An ex-post
  approach towards the rotation problem}.
\newblock {\em Journal of Econometrics}, 192(1):190--206.

\bibitem[Bhattacharya and Dunson, 2011]{bhattacharya2011sparse}
Bhattacharya, A. and Dunson, D.~B. (2011).
\newblock {Sparse {Bayesian} infinite factor models}.
\newblock {\em Biometrika}, 98(2):291--306.

\bibitem[Billio et~al., 2023]{billio2023bayesian}
Billio, M., Casarin, R., Iacopini, M., and Kaufmann, S. (2023).
\newblock {Bayesian dynamic tensor regression}.
\newblock {\em Journal of Business \& Economic Statistics}, 41(2):429--439.

\bibitem[Boonen and Chen, 2026]{boonen2026lowrank}
Boonen, T.~J. and Chen, Y. (2026).
\newblock Low-rank tensor autoregressive models for mortality modelling.
\newblock {\em Journal of the Royal Statistical Society Series A: Statistics in
  Society}.
\newblock qnag048.

\bibitem[Brouhns et~al., 2002]{brouhns2002poisson}
Brouhns, N., Denuit, M., and Vermunt, J.~K. (2002).
\newblock {A Poisson log-bilinear regression approach to the construction of
  projected lifetables}.
\newblock {\em Insurance: Mathematics and Economics}, 31(3):373--393.

\bibitem[Calder, 2007]{calder2007dynamic}
Calder, C.~A. (2007).
\newblock {Dynamic factor process convolution models for multivariate
  space--time data with application to air quality assessment}.
\newblock {\em Environmental and Ecological Statistics}, 14(3):229--247.

\bibitem[Camarda, 2012]{camarda2012mortalitysmooth}
Camarda, C.~G. (2012).
\newblock {MortalitySmooth: An R package for smoothing Poisson counts with
  P-splines}.
\newblock {\em {Journal of Statistical Software}}, 50:1--24.

\bibitem[Carvalho et~al., 2010]{car-etal:hor}
Carvalho, C.~M., Polson, N.~G., and Scott, J.~G. (2010).
\newblock {The horseshoe estimator for sparse signals}.
\newblock {\em Biometrika}, 97(2):465--480.

\bibitem[Chen et~al., 2022]{chen2022factor}
Chen, R., Yang, D., and Zhang, C.-H. (2022).
\newblock {Factor models for high-dimensional tensor time series}.
\newblock {\em Journal of the American Statistical Association},
  117(537):94--116.

\bibitem[Clark, 2019]{clark2019general}
Clark, S.~J. (2019).
\newblock {A general age-specific mortality model with an example indexed by
  child mortality or both child and adult mortality}.
\newblock {\em Demography}, 56(3):1131--1159.

\bibitem[Conti et~al., 2014]{conti2014bayesian}
Conti, G., Fr{\"u}hwirth-Schnatter, S., Heckman, J.~J., and Piatek, R. (2014).
\newblock {Bayesian exploratory factor analysis}.
\newblock {\em Journal of Econometrics}, 183(1):31--57.

\bibitem[Currie et~al., 2004]{currie2004smoothing}
Currie, I.~D., Durban, M., and Eilers, P.~H. (2004).
\newblock {Smoothing and forecasting mortality rates}.
\newblock {\em Statistical Modelling}, 4(4):279--298.

\bibitem[Czado et~al., 2005]{czado2005bayesian}
Czado, C., Delwarde, A., and Denuit, M. (2005).
\newblock {Bayesian Poisson log-bilinear mortality projections}.
\newblock {\em Insurance: Mathematics and Economics}, 36(3):260--284.

\bibitem[de~Beer, 2012]{debeer2012smoothing}
de~Beer, J. (2012).
\newblock {Smoothing and projecting age-specific probabilities of death by
  TOPALS}.
\newblock {\em Demographic Research}, 27:543--592.

\bibitem[Dharamshi et~al., 2025]{dharamshi2025jointly}
Dharamshi, A., Alexander, M., Winant, C., and Barbieri, M. (2025).
\newblock {Jointly estimating subnational mortality for multiple populations}.
\newblock {\em Demographic Research}, 52:71--110.

\bibitem[Eilers and Marx, 1996]{eilers1996flexible}
Eilers, P.~H. and Marx, B.~D. (1996).
\newblock {Flexible smoothing with B-splines and penalties}.
\newblock {\em Statistical Science}, 11(2):89--121.

\bibitem[Fosdick and Hoff, 2014]{fosdick2014separable}
Fosdick, B.~K. and Hoff, P.~D. (2014).
\newblock {Separable factor analysis with applications to mortality data}.
\newblock {\em The Annals of Applied Statistics}, 8(1):120--147.

\bibitem[Fr{\"u}hwirth-Schnatter et~al., 2024]{fruhwirth2024sparse}
Fr{\"u}hwirth-Schnatter, S., Hosszejni, D., and Lopes, H.~F. (2024).
\newblock {Sparse {B}ayesian factor analysis when the number of factors is
  unknown}.
\newblock {\em Bayesian Analysis}.
\newblock Advance publication.

\bibitem[Gamerman, 1997]{gamerman1997sampling}
Gamerman, D. (1997).
\newblock {Sampling from the posterior distribution in generalized linear mixed
  models}.
\newblock {\em Statistics and Computing}, 7(1):57--68.

\bibitem[Gompertz, 1825]{gompertz1825nature}
Gompertz, B. (1825).
\newblock On the nature of the function expressive of the law of human
  mortality, and on a new mode of determining the value of life contingencies.
\newblock {\em Philosophical Transactions of the Royal Society of London},
  115:513--583.

\bibitem[Gopalan et~al., 2015]{gopalan2015scalable}
Gopalan, P., Hofman, J.~M., and Blei, D.~M. (2015).
\newblock {Scalable recommendation with hierarchical Poisson factorization}.
\newblock In {\em Proceedings of the Thirty-First Conference on Uncertainty in
  Artificial Intelligence}, pages 326--335.

\bibitem[Han et~al., 2022]{han2022optimal}
Han, R., Willett, R., and Zhang, A.~R. (2022).
\newblock {An optimal statistical and computational framework for generalized
  tensor estimation}.
\newblock {\em The Annals of Statistics}, 50(1):1--29.

\bibitem[Hoff, 2011a]{hoff2011hierarchical}
Hoff, P.~D. (2011a).
\newblock {Hierarchical multilinear models for multiway data}.
\newblock {\em Computational Statistics \& Data Analysis}, 55(1):530--543.

\bibitem[Hoff, 2011b]{hoff2011separable}
Hoff, P.~D. (2011b).
\newblock {Separable covariance arrays via the Tucker product, with
  applications to multivariate relational data}.
\newblock {\em Bayesian Analysis}, 6(2):179--196.

\bibitem[Hoff, 2015]{hoff2015multilinear}
Hoff, P.~D. (2015).
\newblock {Multilinear tensor regression for longitudinal relational data}.
\newblock {\em The Annals of Applied Statistics}, 9(3):1169--1193.

\bibitem[Hu et~al., 2015]{hu2015scalable}
Hu, C., Rai, P., Chen, C., Harding, M., and Carin, L. (2015).
\newblock {Scalable Bayesian non-negative tensor factorization for massive
  count data}.
\newblock In {\em Joint European Conference on Machine Learning and Knowledge
  Discovery in Databases}, pages 53--70. Springer.

\bibitem[Hyndman et~al., 2013]{hyndman2013coherent}
Hyndman, R.~J., Booth, H., and Yasmeen, F. (2013).
\newblock {Coherent mortality forecasting: the product-ratio method with
  functional time series models}.
\newblock {\em Demography}, 50(1):261--283.

\bibitem[Hyndman and Ullah, 2007]{hyndman2007robust}
Hyndman, R.~J. and Ullah, M.~S. (2007).
\newblock {Robust forecasting of mortality and fertility rates: A functional
  data approach}.
\newblock {\em Computational Statistics \& Data Analysis}, 51(10):4942--4956.

\bibitem[Kastner and Fr{\"u}hwirth-Schnatter, 2014]{kastner2014ancillarity}
Kastner, G. and Fr{\"u}hwirth-Schnatter, S. (2014).
\newblock Ancillarity-sufficiency interweaving strategy ({ASIS}) for boosting
  {MCMC} estimation of stochastic volatility models.
\newblock {\em Computational Statistics \& Data Analysis}, 76:408--423.

\bibitem[Kastner et~al., 2017]{kastner2017efficient}
Kastner, G., Fr{\"u}hwirth-Schnatter, S., and Lopes, H.~F. (2017).
\newblock {Efficient Bayesian inference for multivariate factor stochastic
  volatility models}.
\newblock {\em Journal of Computational and Graphical Statistics},
  26(4):905--917.

\bibitem[Kolda and Bader, 2009]{kolda2009tensor}
Kolda, T.~G. and Bader, B.~W. (2009).
\newblock {Tensor decompositions and applications}.
\newblock {\em SIAM Review}, 51(3):455--500.

\bibitem[Kruskal, 1977]{kruskal1977three}
Kruskal, J.~B. (1977).
\newblock {Three-way arrays: rank and uniqueness of trilinear decompositions,
  with application to arithmetic complexity and statistics}.
\newblock {\em Linear Algebra and its Applications}, 18(2):95--138.

\bibitem[Lee and Seung, 1999]{lee1999learning}
Lee, D.~D. and Seung, H.~S. (1999).
\newblock Learning the parts of objects by non-negative matrix factorization.
\newblock {\em Nature}, 401(6755):788--791.

\bibitem[Lee and Carter, 1992]{lee1992modeling}
Lee, R.~D. and Carter, L.~R. (1992).
\newblock {Modeling and forecasting {US} mortality}.
\newblock {\em {Journal of the American Statistical Association}},
  87(419):659--671.

\bibitem[Legramanti et~al., 2020]{legramanti2020bayesian}
Legramanti, S., Durante, D., and Dunson, D.~B. (2020).
\newblock {Bayesian cumulative shrinkage for infinite factorizations}.
\newblock {\em Biometrika}, 107(3):745--752.

\bibitem[Lei et~al., 2025]{lei2025scalable}
Lei, M., Labbe, A., and Sun, L. (2025).
\newblock {Scalable spatiotemporally varying coefficient modeling with Bayesian
  kernelized tensor regression}.
\newblock {\em Bayesian Analysis}, 20(3):919--947.

\bibitem[Li and Lee, 2005]{li2005coherent}
Li, N. and Lee, R. (2005).
\newblock {Coherent mortality forecasts for a group of populations: An
  extension of the Lee-Carter method}.
\newblock {\em Demography}, 42(3):575--594.

\bibitem[Makalic and Schmidt, 2016]{makalic2015simple}
Makalic, E. and Schmidt, D.~F. (2016).
\newblock {A simple sampler for the horseshoe estimator}.
\newblock {\em IEEE Signal Processing Letters}, 23(1):179--182.

\bibitem[Martin and Camarda, 2026]{martin2026modeling}
Martin, J. and Camarda, C.~G. (2026).
\newblock Modeling smooth and localized mortality patterns across age, time,
  and space to uncover small-area inequalities.
\newblock {\em Journal of the Royal Statistical Society Series A: Statistics in
  Society}.
\newblock qnag093.

\bibitem[Oseledets, 2011]{oseledets2011tensor}
Oseledets, I.~V. (2011).
\newblock {Tensor-train decomposition}.
\newblock {\em SIAM Journal on Scientific Computing}, 33(5):2295--2317.

\bibitem[Pavone et~al., 2024]{pavone2024learning}
Pavone, F., Legramanti, S., and Durante, D. (2024).
\newblock {Learning and forecasting of age-specific period mortality via
  B-spline processes with locally-adaptive dynamic coefficients}.
\newblock {\em The Annals of Applied Statistics}, 18(3):1965--1987.

\bibitem[Plummer, 2015]{plummer2015cuts}
Plummer, M. (2015).
\newblock Cuts in {B}ayesian graphical models.
\newblock {\em Statistics and Computing}, 25(1):37--43.

\bibitem[Qin et~al., 2025]{qin2025bayesian}
Qin, L., Wang, Y., Zhu, Y., and Shia, B.-C. (2025).
\newblock {Bayesian Dynamic Matrix Factor Models}.
\newblock {\em Journal of Business \& Economic Statistics}, pages 1--13.

\bibitem[Rai et~al., 2014]{rai2014scalable}
Rai, P., Wang, Y., Guo, S., Chen, G., Dunson, D., and Carin, L. (2014).
\newblock {Scalable Bayesian low-rank decomposition of incomplete multiway
  tensors}.
\newblock In {\em Proceedings of the 31st International Conference on Machine
  Learning}, pages 1800--1808. PMLR.

\bibitem[Rasmussen and Williams, 2006]{rasmussen2006gaussian}
Rasmussen, C.~E. and Williams, C.~K. (2006).
\newblock {\em {Gaussian Processes for Machine Learning}}.
\newblock MIT Press, Cambridge, MA.

\bibitem[Rogers and Castro, 1981]{rogers1981model}
Rogers, A. and Castro, L.~J. (1981).
\newblock Model migration schedules.
\newblock Research Report RR-81-30, International Institute for Applied Systems
  Analysis, Laxenburg, Austria.

\bibitem[Russolillo et~al., 2011]{russolillo2011extending}
Russolillo, M., Giordano, G., and Haberman, S. (2011).
\newblock {Extending the Lee--Carter model: a three-way decomposition}.
\newblock {\em Scandinavian Actuarial Journal}, 2011(2):96--117.

\bibitem[Schein et~al., 2015]{schein2015bayesian}
Schein, A., Paisley, J., Blei, D.~M., and Wallach, H. (2015).
\newblock {Bayesian Poisson tensor factorization for inferring multilateral
  relations from sparse dyadic event counts}.
\newblock In {\em Proceedings of the 21st ACM SIGKDD International Conference
  on Knowledge Discovery and Data Mining}, pages 1045--1054.

\bibitem[Schmertmann and Gonzaga, 2018]{schmertmann2018bayesian}
Schmertmann, C.~P. and Gonzaga, M.~R. (2018).
\newblock {Bayesian estimation of age-specific mortality and life expectancy
  for small areas with defective vital records}.
\newblock {\em Demography}, 55(4):1363--1388.

\bibitem[Sidiropoulos and Bro, 2000]{sidiropoulos2000uniqueness}
Sidiropoulos, N.~D. and Bro, R. (2000).
\newblock {On the uniqueness of multilinear decomposition of N-way arrays}.
\newblock {\em Journal of Chemometrics}, 14(3):229--239.

\bibitem[Stolf and Canale, 2026]{stolf2025bayesian}
Stolf, F. and Canale, A. (2026).
\newblock {Bayesian Adaptive Tucker Decompositions for Tensor Factorization}.
\newblock {\em Journal of Computational and Graphical Statistics},
  35(1):119--128.

\bibitem[Susmann et~al., 2022]{susmann2022temporal}
Susmann, H., Alexander, M., and Alkema, L. (2022).
\newblock {Temporal Models for Demographic and Global Health Outcomes in
  Multiple Populations: Introducing a New Framework to Review and Standardise
  Documentation of Model Assumptions and Facilitate Model Comparison}.
\newblock {\em International Statistical Review}, 90(3):437--467.

\bibitem[Tanner and Wong, 1987]{tanner1987calculation}
Tanner, M.~A. and Wong, W.~H. (1987).
\newblock {The calculation of posterior distributions by data augmentation}.
\newblock {\em Journal of the American Statistical Association},
  82(398):528--540.

\bibitem[van Dyk and Park, 2008]{vandyk2008partially}
van Dyk, D.~A. and Park, T. (2008).
\newblock Partially collapsed {G}ibbs samplers: Theory and methods.
\newblock {\em Journal of the American Statistical Association},
  103(482):790--796.

\bibitem[Vaupel and Canudas~Romo, 2003]{vaupel2003decomposing}
Vaupel, J.~W. and Canudas~Romo, V. (2003).
\newblock Decomposing change in life expectancy: A bouquet of formulas in honor
  of {N}athan {K}eyfitz's 90th birthday.
\newblock {\em Demography}, 40(2):201--216.

\bibitem[Wang et~al., 2019]{wang2019factor}
Wang, D., Liu, X., and Chen, R. (2019).
\newblock {Factor models for matrix-valued high-dimensional time series}.
\newblock {\em Journal of Econometrics}, 208(1):231--248.

\bibitem[Wilmoth et~al., 2007]{wilmoth2007methods}
Wilmoth, J.~R., Andreev, K., Jdanov, D., and Glei, D.~A. (2007).
\newblock {Methods protocol for the Human Mortality Database}.
\newblock Technical report, University of California, Berkeley, and Max Planck
  Institute for Demographic Research.

\bibitem[Wi{\'s}niowski et~al., 2015]{wisniowski2015bayesian}
Wi{\'s}niowski, A., Smith, P.~W., Bijak, J., Raymer, J., and Forster, J.~J.
  (2015).
\newblock {Bayesian population forecasting: extending the Lee-Carter method}.
\newblock {\em Demography}, 52(3):1035--1059.

\bibitem[Yokota et~al., 2016]{yokota2016smooth}
Yokota, T., Zhao, Q., and Cichocki, A. (2016).
\newblock {Smooth PARAFAC decomposition for tensor completion}.
\newblock {\em IEEE Transactions on Signal Processing}, 64(20):5423--5436.

\bibitem[Yu and Meng, 2011]{yu2011center}
Yu, Y. and Meng, X.-L. (2011).
\newblock To center or not to center: That is not the question--{A}n
  ancillarity--sufficiency interweaving strategy ({ASIS}) for boosting {MCMC}
  efficiency.
\newblock {\em Journal of Computational and Graphical Statistics},
  20(3):531--570.

\bibitem[Zens, 2025]{zens2024flexible}
Zens, G. (2025).
\newblock {Flexible Bayesian modelling of age-specific counts in many
  demographic subpopulations}.
\newblock {\em Journal of the Royal Statistical Society Series A: Statistics in
  Society}.
\newblock Forthcoming.

\bibitem[Zhang et~al., 2024]{zhang2024poisson}
Zhang, W., Mira, A., and Wit, E.~C. (2024).
\newblock {Bayesian Poisson regression tensor train decomposition model for
  learning mortality pattern changes during COVID-19 pandemic}.
\newblock {\em Journal of Applied Statistics}.
\newblock DOI 10.1080/02664763.2024.2411608.

\bibitem[Zhao et~al., 2015]{zhao2015bayesian}
Zhao, Q., Zhang, L., and Cichocki, A. (2015).
\newblock {Bayesian CP factorization of incomplete tensors with automatic rank
  determination}.
\newblock {\em IEEE Transactions on Pattern Analysis and Machine Intelligence},
  37(9):1751--1763.

\end{thebibliography}

\clearpage
\onehalfspacing
\setcounter{equation}{0}
\setcounter{figure}{0}
\setcounter{section}{0}
\setcounter{table}{0}
\counterwithout{equation}{section}
\renewcommand\thesection{S\arabic{section}}
\renewcommand{\theHsection}{S\arabic{section}}
\renewcommand\theequation{S\arabic{equation}}
\renewcommand\thefigure{S\arabic{figure}}
\renewcommand\thetable{S\arabic{table}}
\renewcommand{\theHequation}{supp.\arabic{equation}}
\renewcommand{\theHfigure}{supp.\arabic{figure}}
\renewcommand{\theHtable}{supp.\arabic{table}}

\makeatletter
\setlength{\@fptop}{0pt plus 1fil}
\setlength{\@fpsep}{2.5cm}
\setlength{\@fpbot}{0pt plus 1fil}
\makeatother

\begin{center}
\vspace*{2em}
{\large\MakeUppercase{Supplementary Material}}\\[1em]
\end{center}

\section{Posterior Simulation}
\label{sec:S1}

\subsection{Notation}
\label{sec:S1not}

Let $N = \prod_d n_d$ be the number of cells and $\mathbf{M}_d = [\mathbf{m}_{d,1}, \dots, \mathbf{m}_{d,R}]$ the $n_d \times R$ matrix of the vectors of mode $d$. Conditional on $\mathbf{z}$, every update of the CP part is a Bayesian linear regression, because the surface is linear in the vectors of any one mode when the others are held fixed. For mode $d$, write
\begin{equation}
x_{d,r}(c) = g_r \prod_{d' \neq d} m_{d',r}(c_{d'}),
\label{eq:reg}
\end{equation}
so that $\eta_c = \mu + \sum_r m_{d,r}(c_d)\, x_{d,r}(c)$. For every level $j$ of mode $d$, the cells with $c_d = j$ follow a linear regression of $z_c - \mu$ on the $R$ known regressors $x_{d,r}(c)$ with coefficients $m_{d,1}(j), \dots, m_{d,R}(j)$. The prior on the mode vectors couples the levels. The regression requires only the weighted crossproducts
\begin{equation}
\mathbf{A}_d(j) = \sum_{c :\, c_d = j} w_c\, \mathbf{x}_d(c)\, \mathbf{x}_d(c)^\top, \qquad
\mathbf{b}_d(j) = \sum_{c :\, c_d = j} w_c\, (z_c - \mu)\, \mathbf{x}_d(c),
\label{eq:stats}
\end{equation}
for every level $j$, where $w_c = 1/\sigma^2_{c_s}$ is the precision of cell $c$. Because the precisions vary along mode $s$ only, $\mathbf{A}_d(j)$ factorizes over the other modes,
\begin{equation}
\mathbf{A}_d(j) = \omega_j\, \mathrm{diag}(\mathbf{g}) \Big( \mathop{\circ}_{d' \neq d} \mathbf{M}_{d'}^\top \boldsymbol{\Omega}_{d'} \mathbf{M}_{d'} \Big) \mathrm{diag}(\mathbf{g}),
\label{eq:Ab}
\end{equation}
where $\circ$ is the elementwise product \citep{kolda2009tensor}. Here $\boldsymbol{\Omega}_{d'} = \mathrm{diag}(w_1, \dots, w_{n_s})$ if $d' = s$ and the identity otherwise, and $\omega_j = w_j$ if $d = s$ and $1$ otherwise. All precisions are thus assembled from $R \times R$ matrices. Only $\mathbf{b}_d(j)$ needs one pass over the cells per mode. Held-out cells and cells with zero exposure carry no Poisson likelihood term and are excluded from all Poisson-based steps (Sections~\ref{sec:S1z}, \ref{sec:S1asis}, \ref{sec:S1g}, and~\ref{sec:S1s2}). Their latent log rates are drawn from the Gaussian layer and enter the Gaussian regression steps and the residual sums of squares like every other cell, which keeps the precisions separable. A break point in a smooth mode makes $\mathbf{C}_d$ block-diagonal and changes nothing else.

\subsection{One sweep}
\label{sec:S1sweep}

Initial values come from a warm start. The latent log rates are set to $\log\{(y_c + 1/2)/O_c\}$, the components are obtained by an alternating least squares algorithm, and the kernel parameters and error variances are initialized from the resulting fit. Starting from these values, one sweep of the sampler performs the following steps.
\begin{enumerate}
\item Draw the latent log rates $z_c$ of all cells (Section~\ref{sec:S1z}).
\item For every mode $d = 1, \dots, D$:
  \begin{enumerate}
  \item draw all vectors of the mode jointly, $\mathbf{M}_d \mid \mathbf{z}, \cdot$ (Section~\ref{sec:S1m});
  \item for every component $r$, update the kernel parameters $(\kappa_{d,r}, \phi_{d,r})$ with $\mathbf{m}_{d,r}$ integrated out, then draw $\mathbf{m}_{d,r} \mid \kappa_{d,r}, \phi_{d,r}, \cdot$ (Section~\ref{sec:S1kappa});
  \item perform an interweaving step for $\mathbf{M}_d$ (Section~\ref{sec:S1asis});
  \item draw the global scales $\tau_{\kappa,d}$ and $\tau_{\phi,d}$ and their auxiliary variables (Section~\ref{sec:S1glob}).
  \end{enumerate}
\item For the global level and the weights:
  \begin{enumerate}
  \item update the horseshoe scales $\lambda_r$ and $\tau$ with $(\mu, \mathbf{g})$ integrated out, then draw their auxiliary variables (Section~\ref{sec:S1hs});
  \item draw $(\mu, \mathbf{g}) \mid \mathbf{z}, \cdot$ jointly (Section~\ref{sec:S1g});
  \item perform an interweaving step for $(\mu, \mathbf{g})$ (Section~\ref{sec:S1g}).
  \end{enumerate}
\item For the error variances:
  \begin{enumerate}
  \item draw $\sigma^2_j$ for every level $j$ of mode $s$ (Section~\ref{sec:S1s2});
  \item perform an interweaving step for them (Section~\ref{sec:S1s2}).
  \end{enumerate}
\end{enumerate}
For exchangeable modes, $\mathbf{C}_d = \mathbf{I}$ has no parameter $\phi$, and all steps involving $\phi$ are skipped. The interweaving steps are carried out on every second sweep. The proposal scales of all Metropolis steps are adapted during warm-up only.

\subsection{Latent log rates}
\label{sec:S1z}

Conditional on all other quantities, the latent log rates are independent across cells with density
\begin{equation}
p(z_c \mid \cdot) \propto \exp\big( y_c z_c - O_c e^{z_c} \big) \, \mathcal{N}\big(z_c ;\, \eta_c, \sigma^2_{c_s}\big).
\end{equation}
For every cell with positive exposure, this density is log-concave with a unique mode $z_c^{*}$, the root of $y_c - O_c e^{z} - (z - \eta_c)/\sigma^2_{c_s} = 0$, which can be written through the Lambert W function as
\begin{equation}
z_c^{*} = \eta_c + \sigma^2_{c_s} y_c - W\big(\sigma^2_{c_s}\, O_c \exp\{\eta_c + \sigma^2_{c_s} y_c\}\big)
\end{equation}
and is evaluated to machine precision by a few Newton steps. A Laplace approximation of the conditional is therefore available for all cells at once, $q(z) = \mathcal{N}\{z;\, z_c^{*},\, (O_c e^{z_c^{*}} + 1/\sigma^2_{c_s})^{-1}\}$, and serves as the proposal of an independence Metropolis--Hastings step. A proposal $z_c'$ is accepted with probability $\min\{1, p(z_c' \mid \cdot)\, q(z_c) / [p(z_c \mid \cdot)\, q(z_c')]\}$. Because the conditionals are typically close to Gaussian, acceptance rates are close to one. For cells with zero exposure or missing counts, the likelihood term vanishes and $z_c$ is drawn exactly from $\mathcal{N}(\eta_c, \sigma^2_{c_s})$.

\subsection{Gaussian regression steps}
\label{sec:S1m}

The centered updates of the mode vectors and the global coefficients use the same Gaussian regression calculation. For a response $\mathbf{u}$, design matrix $\mathbf{X}$, diagonal precision matrix $\mathbf{W}$, and coefficient vector $\boldsymbol{\theta}$ with prior $\mathcal{N}(\mathbf{0},\mathbf{P}_0^{-1})$, the Gaussian conditional is
\begin{equation}
\boldsymbol{\theta} \mid \cdot \sim \mathcal{N}\big(\mathbf{Q}^{-1}\mathbf{h},\ \mathbf{Q}^{-1}\big), \qquad
\mathbf{Q} = \mathbf{P}_0 + \mathbf{X}^\top \mathbf{W}\mathbf{X}, \qquad
\mathbf{h} = \mathbf{X}^\top \mathbf{W}\mathbf{u}.
\label{eq:generic}
\end{equation}
The draw is obtained from a Cholesky factorization of $\mathbf{Q}$. The same calculation applies in the following forms.
\begin{enumerate}
\item \textit{Mode vectors $\mathbf{M}_d$.} Take $\boldsymbol{\theta} = \mathrm{vec}(\mathbf{M}_d)$, $\mathbf{u} = \mathbf{z} - \mu\mathbf{1}$, and $\mathbf{P}_0 = \mathrm{blockdiag}(\mathbf{K}_{d,1}^{-1}, \dots, \mathbf{K}_{d,R}^{-1})$. With the regressors and statistics of \eqref{eq:reg}--\eqref{eq:stats}, the likelihood contribution to the precision and its canonical vector are
\begin{equation*}
\mathbf{X}^\top \mathbf{W}\mathbf{X} = \sum_{j=1}^{n_d} \mathbf{A}_d(j) \otimes \mathbf{E}_{jj}, \qquad
\mathbf{X}^\top \mathbf{W}\mathbf{u} = \mathrm{vec}(\mathbf{B}_d),
\end{equation*}
where $\mathbf{B}_d$ has rows $\mathbf{b}_d(j)^\top$ and $\mathbf{E}_{jj}$ has a one in position $(j,j)$ and zeros elsewhere. For CP-PLN, $\mathbf{K}_{d,r} = \mathbf{I}$, and the joint update reduces to independent row draws with precision $\mathbf{I} + \mathbf{A}_d(j)$ and canonical vector $\mathbf{b}_d(j)$.
\item \textit{One column $\mathbf{m}_{d,r}$.} Holding the other columns of $\mathbf{M}_d$ fixed gives the same regression with one component and the contribution of the other components subtracted from $\mathbf{z} - \mu\mathbf{1}$. Its conditional is \eqref{eq:colcond}, used in the collapsed kernel-parameter update (Section~\ref{sec:S1kappa}).
\item \textit{Global level and weights.} Take $\boldsymbol{\theta} = \boldsymbol{\beta} = (\mu, \mathbf{g}^\top)^\top$, $\mathbf{u} = \mathbf{z}$, $\mathbf{X} = \tilde{\mathbf{X}}$, and $\mathbf{P}_0 = \mathbf{P}_{\beta}$. The $(R + 1)$ regressors and the prior precision are specified in Section~\ref{sec:S1g}. This is the single-level version of \eqref{eq:generic}.
\item \textit{Interweaving steps.} The Poisson IWLS calculations for the mode vectors and global weights use the same weighted-crossproduct structure, with Poisson means $\lambda_c$ in place of $w_c$ and the corresponding working responses (Sections~\ref{sec:S1asis} and~\ref{sec:S1g}). These provide Metropolis proposals rather than exact Gaussian conditional draws.
\end{enumerate}
For the Gaussian updates in instances 1--3, the precisions are separable, and the crossproducts can be assembled using \eqref{eq:Ab} without a pass over the cells. For the Poisson IWLS calculations in instance 4, the means are not separable, so the crossproducts are accumulated over cells.

\subsection{Kernel parameters}
\label{sec:S1kappa}

For every component $r$, the conditional of $\mathbf{m}_{d,r}$ given the other columns of $\mathbf{M}_d$ is instance 2 of Section~\ref{sec:S1m},
\begin{equation}
p(\mathbf{m}_{d,r} \mid \cdot) \propto \exp\big( \mathbf{h}_r^\top \mathbf{m}_{d,r} - \tfrac{1}{2}\, \mathbf{m}_{d,r}^\top \mathbf{D}_r\, \mathbf{m}_{d,r} \big)\, \mathcal{N}\big(\mathbf{m}_{d,r};\, \mathbf{0}, \mathbf{K}_{d,r}\big),
\label{eq:colcond}
\end{equation}
where $\mathbf{D}_r$ is diagonal with entries $\mathbf{A}_d(j)_{rr}$ and $\mathbf{h}_r$ has entries
\begin{equation}
h_r(j) = \mathbf{b}_d(j)_r - \sum_{r' \neq r} \mathbf{A}_d(j)_{rr'}\, m_{d,r'}(j), \qquad j = 1, \dots, n_d.
\end{equation}
Integrating $\mathbf{m}_{d,r}$ out of \eqref{eq:colcond} gives the marginal likelihood of the kernel parameters,
\begin{equation}
L(\kappa_{d,r}, \phi_{d,r}) = \big|\mathbf{I} + \mathbf{D}_r^{1/2}\mathbf{K}_{d,r}\mathbf{D}_r^{1/2}\big|^{-1/2} \exp\big\{\tfrac{1}{2}\, \mathbf{h}_r^\top \big(\mathbf{K}_{d,r}^{-1} + \mathbf{D}_r\big)^{-1} \mathbf{h}_r\big\}.
\label{eq:marglik}
\end{equation}
The parameter $\kappa_{d,r}$ is updated by a random-walk Metropolis step on $\log \kappa_{d,r}$. A proposal $\kappa^*$ is accepted with probability
\begin{equation}
\min\left\{1,\ \frac{L(\kappa^*, \phi_{d,r})\; p(\kappa^* \mid \nu^{\kappa}_{d,r})\; \kappa^*}{L(\kappa_{d,r}, \phi_{d,r})\; p(\kappa_{d,r} \mid \nu^{\kappa}_{d,r})\; \kappa_{d,r}}\right\},
\end{equation}
where $p(\kappa \mid \nu)$ is the density of $\kappa$ implied by $\kappa^2 \mid \nu \sim \mathcal{IG}(1/2, 1/\nu)$ and the factor $\kappa$ is the Jacobian of the log transformation. For smooth modes, $\phi_{d,r}$ is then updated in the same way, with $L(\kappa_{d,r}, \phi^*)$ and the density $p(\phi \mid \nu^{\phi}_{d,r})$. Finally, $\mathbf{m}_{d,r}$ is drawn from \eqref{eq:colcond}, that is, from $\mathcal{N}\{(\mathbf{K}_{d,r}^{-1} + \mathbf{D}_r)^{-1}\mathbf{h}_r,\, (\mathbf{K}_{d,r}^{-1} + \mathbf{D}_r)^{-1}\}$. Updating the kernel parameters with $\mathbf{m}_{d,r}$ integrated out and then $\mathbf{m}_{d,r}$ given them is a partially collapsed Gibbs step, which leaves the posterior invariant \citep{vandyk2008partially}. A break point in a smooth mode makes $\mathbf{C}_d$ block-diagonal and changes nothing else.

\subsection{Global scales of the kernel parameters}
\label{sec:S1glob}

Given the kernel parameters of mode $d$, the auxiliary variables and the global scale of $\kappa$ have the inverse gamma conditionals
\begin{align}
\nu^{\kappa}_{d,r} \mid \cdot &\sim \mathcal{IG}\big(1,\ 1/\kappa_{d,r}^2 + 1/\tau_{\kappa,d}^2\big), \qquad r = 1, \dots, R, \nonumber \\
\tau_{\kappa,d}^2 \mid \cdot &\sim \mathcal{IG}\Big(\tfrac{R+1}{2},\ 1/\xi^{\kappa}_d + \textstyle\sum_r 1/\nu^{\kappa}_{d,r}\Big), \\
\xi^{\kappa}_d \mid \cdot &\sim \mathcal{IG}\big(1,\ 1 + 1/\tau_{\kappa,d}^2\big), \nonumber
\end{align}
which are drawn in this order, and analogously for $\phi$ with $\nu^{\phi}_{d,r}$, $\tau_{\phi,d}$, and $\xi^{\phi}_d$.

\subsection{Interweaving step for the mode vectors}
\label{sec:S1asis}

The updates above condition on $\mathbf{z}$, a centered parameterization in the sense of \citet{yu2011center}. The interweaving step moves $\mathbf{M}_d$ once more in the noncentered parameterization, in which the residuals $e_c = z_c - \eta_c$ are held fixed and the latent log rates move with $\mathbf{M}_d$, $z_c = \eta_c(\mathbf{M}_d) + e_c$. With $\mathbf{x} = \mathrm{vec}(\mathbf{M}_d)$, the target is
\begin{equation}
\pi(\mathbf{x}) \propto \prod_{c \,:\, O_c > 0} \mathrm{Pois}\big\{y_c \mid \lambda_c(\mathbf{x})\big\} \prod_{r=1}^{R} \mathcal{N}\big(\mathbf{m}_{d,r};\, \mathbf{0}, \mathbf{K}_{d,r}\big), \qquad \lambda_c(\mathbf{x}) = O_c \exp\{\eta_c(\mathbf{x}) + e_c\},
\end{equation}
because the Gaussian density of the fixed residuals does not depend on $\mathbf{x}$. Write $\mathbf{P}_0 = \mathrm{blockdiag}(\mathbf{K}_{d,1}^{-1}, \dots, \mathbf{K}_{d,R}^{-1})$, and let $\lambda_c = 0$ at the excluded cells. A Newton step for $\log \pi$ at $\mathbf{x}$, which is one iteratively reweighted least squares (IWLS) step, is instance 4 of Section~\ref{sec:S1m}. It uses the gradient $\nabla(\mathbf{x}) = \mathrm{vec}\{\mathbf{S}(\mathbf{x})\} - \mathbf{P}_0\mathbf{x}$ and the precision $\mathbf{P}(\mathbf{x}) = \mathbf{P}_0 + \sum_{j=1}^{n_d} \mathbf{F}(j) \otimes \mathbf{E}_{jj}$, where
\begin{equation}
\mathbf{S}(\mathbf{x})_{jr} = \sum_{c :\, c_d = j} \{y_c - \lambda_c(\mathbf{x})\}\, x_{d,r}(c), \qquad
\mathbf{F}(j) = \sum_{c :\, c_d = j} \lambda_c(\mathbf{x})\, \mathbf{x}_d(c)\, \mathbf{x}_d(c)^\top,
\end{equation}
which give the Newton point $\hat{\mathbf{x}} = \mathbf{x} + \mathbf{P}(\mathbf{x})^{-1}\nabla(\mathbf{x})$. The Gaussian statistics $\mathbf{b}_d(j)$ and $\mathbf{A}_d(j)$ of the centered step have the same form with $w_c (z_c - \mu)$ and $w_c$ in place of $y_c - \lambda_c(\mathbf{x})$ and $\lambda_c(\mathbf{x})$. The proposal is
\begin{equation}
q(\mathbf{x}^* \mid \mathbf{x}) = \mathcal{N}\big\{\mathbf{x}^*;\ a\,\mathbf{x} + (1 - a)\,\hat{\mathbf{x}},\ (1 - a^2)\, \mathbf{P}(\mathbf{x})^{-1}\big\},
\label{eq:asisprop}
\end{equation}
with a constant $a \in [0, 1)$. For $a = 0$ this is the IWLS proposal of \citet{gamerman1997sampling}, and larger values keep the proposal closer to the current value. During warm-up, $a$ is adapted so that the acceptance rate approaches $0.5$ from above, that is, $a$ is raised only where the IWLS proposal is accepted less than half of the time. In the application, $a$ stays close to zero in every mode, with acceptance rates between $0.6$ and $0.95$. The proposal is accepted with probability
\begin{equation}
\min\left\{1,\ \frac{\pi(\mathbf{x}^*)\, q(\mathbf{x} \mid \mathbf{x}^*)}{\pi(\mathbf{x})\, q(\mathbf{x}^* \mid \mathbf{x})}\right\},
\label{eq:asisratio}
\end{equation}
where $q(\mathbf{x} \mid \mathbf{x}^*)$ uses the Newton step at $\mathbf{x}^*$. On acceptance, $\mathbf{M}_d$ is replaced and $z_c \leftarrow z_c + \eta_c(\mathbf{x}^*) - \eta_c(\mathbf{x})$ for all cells. For CP-PLN, the rows of $\mathbf{M}_d$ are moved separately in the same way.

\subsection{Horseshoe scales}
\label{sec:S1hs}

Given $\mathbf{z}$ and the mode vectors, the surface is linear in $\boldsymbol{\beta} = (\mu, \mathbf{g}^\top)^\top$ (Section~\ref{sec:S1g}), so $\boldsymbol{\beta}$ can be integrated out analytically. With $\mathbf{C} = \tilde{\mathbf{X}}^\top \mathbf{W} \tilde{\mathbf{X}}$, $\mathbf{b} = \tilde{\mathbf{X}}^\top \mathbf{W} \mathbf{z}$, and $\mathbf{D} = \mathrm{diag}\{100,\ \tau^2\lambda_1^2, \dots, \tau^2\lambda_R^2\}$ in the notation of that section, the marginal likelihood of the scales is
\begin{equation}
L(\lambda_1, \dots, \lambda_R, \tau) = |\mathbf{D}|^{-1/2}\, |\mathbf{C} + \mathbf{D}^{-1}|^{-1/2} \exp\big\{\tfrac{1}{2}\, \mathbf{b}^\top (\mathbf{C} + \mathbf{D}^{-1})^{-1} \mathbf{b}\big\},
\end{equation}
which involves only $(R + 1) \times (R + 1)$ matrices. Every $\lambda_r$ in turn and then $\tau$ is updated by a random-walk Metropolis step on the log scale against $L$ times its prior, $\lambda_r^2 \mid \nu_r \sim \mathcal{IG}(1/2, 1/\nu_r)$ and $\tau^2 \mid \xi \sim \mathcal{IG}(1/2, 1/\xi)$, with the Jacobian of the log transformation. The auxiliary variables are then drawn from
\begin{equation}
\nu_r \mid \cdot \sim \mathcal{IG}\!\left(1,\, 1 + \tfrac{1}{\lambda_r^2}\right), \qquad \xi \mid \cdot \sim \mathcal{IG}\!\left(1,\, 1 + \tfrac{1}{\tau^2}\right),
\end{equation}
and $\boldsymbol{\beta}$ is drawn exactly given the new scales. This is a partially collapsed Gibbs step \citep{vandyk2008partially}. It removes the coupling between a weight close to zero and its own scale, under which the weight keeps the scale small and the scale keeps the weight close to zero, so that weak components would switch on and off only slowly.

\subsection{Global level and component weights}
\label{sec:S1g}

With $\tilde{\mathbf{X}}$ the $N \times (R + 1)$ matrix with rows $\tilde{\mathbf{x}}_c^\top = (1,\ \prod_d m_{d,1}(c_d), \dots, \prod_d m_{d,R}(c_d))$, $\boldsymbol{\eta} = \tilde{\mathbf{X}}\boldsymbol{\beta}$ with $\boldsymbol{\beta} = (\mu, \mathbf{g}^\top)^\top$, and the prior is $\boldsymbol{\beta} \sim \mathcal{N}(\mathbf{0}, \mathbf{P}_{\beta}^{-1})$ with $\mathbf{P}_{\beta} = \mathrm{diag}\{1/100,\ 1/(\tau^2\lambda_1^2), \dots, 1/(\tau^2\lambda_R^2)\}$. The full conditional is instance 3 of Section~\ref{sec:S1m}, Gaussian with precision $\tilde{\mathbf{X}}^\top \mathbf{W} \tilde{\mathbf{X}} + \mathbf{P}_{\beta}$ and canonical mean $\tilde{\mathbf{X}}^\top \mathbf{W} \mathbf{z}$, where $\mathbf{W} = \mathrm{diag}(w_c)$. By \eqref{eq:Ab}, the $R \times R$ block of $\tilde{\mathbf{X}}^\top \mathbf{W} \tilde{\mathbf{X}}$ is $\circ_d\, \mathbf{M}_d^\top \boldsymbol{\Omega}_d \mathbf{M}_d$, its cross terms are $\prod_d \mathbf{1}^\top \boldsymbol{\Omega}_d \mathbf{m}_{d,r}$, and its first entry is $\sum_c w_c$.

The full conditional is drawn after the collapsed update of the horseshoe scales (Section~\ref{sec:S1hs}). The interweaving step holds $e_c = z_c - \eta_c$ fixed and targets
\begin{equation}
\pi(\boldsymbol{\beta}) \propto \prod_{c \,:\, O_c > 0} \mathrm{Pois}\big\{y_c \mid \lambda_c(\boldsymbol{\beta})\big\}\, \mathcal{N}\big(\boldsymbol{\beta};\, \mathbf{0}, \mathbf{P}_{\beta}^{-1}\big), \qquad \lambda_c(\boldsymbol{\beta}) = O_c \exp\{\tilde{\mathbf{x}}_c^\top \boldsymbol{\beta} + e_c\},
\end{equation}
The proposal is the IWLS proposal, that is, \eqref{eq:asisprop} with $a = 0$, with precision $\mathbf{P}(\boldsymbol{\beta}) = \tilde{\mathbf{X}}^\top \boldsymbol{\Lambda} \tilde{\mathbf{X}} + \mathbf{P}_{\beta}$, $\boldsymbol{\Lambda} = \mathrm{diag}\{\lambda_c(\boldsymbol{\beta})\}$, and Newton point $\hat{\boldsymbol{\beta}} = \boldsymbol{\beta} + \mathbf{P}(\boldsymbol{\beta})^{-1}\{\tilde{\mathbf{X}}^\top(\mathbf{y} - \boldsymbol{\lambda}) - \mathbf{P}_{\beta}\boldsymbol{\beta}\}$. It is accepted with the ratio \eqref{eq:asisratio}, and on acceptance $\mathbf{z} \leftarrow \mathbf{z} + \tilde{\mathbf{X}}(\boldsymbol{\beta}^* - \boldsymbol{\beta})$.

\subsection{Error variances}
\label{sec:S1s2}

For level $j$ of mode $s$, with $n_j = N/n_s$ cells and residual sum of squares $\mathrm{rss}_j = \sum_{c \,:\, c_s = j} (z_c - \eta_c)^2$,
\begin{equation}
\sigma^2_j \mid \cdot \sim \mathcal{IG}\!\left( c_0 + \tfrac{n_j}{2},\; C_0 + \tfrac{\mathrm{rss}_j}{2} \right),
\end{equation}
with $c_0 = 1$ and $C_0 = 0.01$ as in the main text.
The interweaving step holds the standardized residuals $\epsilon_c = (z_c - \eta_c)/\sigma_{c_s}$ fixed, so that $z_c = \eta_c + \sigma_{c_s}\epsilon_c$, and updates every $\sigma^2_j$ by a random-walk Metropolis step on $\log \sigma^2_j$. A proposal $\sigma^{2*}$ is accepted with probability
\begin{equation}
\min\left\{1,\ \prod_{c \,:\, c_s = j,\, O_c > 0} \frac{\mathrm{Pois}\{y_c \mid O_c \exp(\eta_c + \sigma^* \epsilon_c)\}}{\mathrm{Pois}\{y_c \mid O_c \exp(\eta_c + \sigma_j \epsilon_c)\}} \times \frac{\mathcal{IG}(\sigma^{2*};\, c_0, C_0)\; \sigma^{2*}}{\mathcal{IG}(\sigma^2_j;\, c_0, C_0)\; \sigma^2_j}\right\},
\end{equation}
where $\mathcal{IG}(\cdot\,; c_0, C_0)$ denotes the inverse gamma density and the factor $\sigma^2$ is the Jacobian of the log transformation. On acceptance, $z_c = \eta_c + \sigma^* \epsilon_c$ for the cells of level $j$.

\subsection{Computation time}
\label{sec:S1time}

Table~\ref{tab:timing} reports times measured on an AMD Ryzen 5 5500U at 2.1 GHz, one thread, OpenBLAS, and R 4.4.0 under Windows. For the Bayesian models, the cost per sweep is the median over three repetitions of 50 sampling sweeps after 50 warm-up sweeps, and the total is extrapolated to the budget of $20{,}000$ warm-up and $20{,}000$ sampling sweeps. This is done as the full fits are run on a computing cluster where runtimes differ substantially based on current load and available CPUs. The point estimators are fitted completely. On the Swedish array, BATS costs about as much per sweep as the Poisson tensor factorization with $R = 20$, roughly twice as much as CP-PLN, which has no kernel parameters and draws its mode vectors level by level, and about four times as much as the additive model. The full Swedish fit takes about eleven hours and the Austrian fit, on an array with $1.66$ times as many cells, about fifteen. 

\subsection{Derivation of the induced covariance}
\label{sec:S1cov}

This subsection derives the second-order structure stated in equation~(8) of the main text. Conditional on the kernel parameters and the weight scales, the weights and the mode vectors are independent with mean zero. Every term $g_r \prod_d m_{d,r}(c_d)$ is a product of independent zero-mean factors, so $\mathbb{E}(\eta_c) = \mu$. For two cells $c$ and $c'$, products of terms from different components have expectation zero for the same reason. Within a component, the expectation of the product factorizes over the independent factors, which gives
\begin{align}
\mathrm{Cov}(\eta_c, \eta_{c'})
&= \sum_{r=1}^{R} \mathbb{E}(g_r^2) \prod_{d=1}^{D} \mathbb{E}\{ m_{d,r}(c_d)\, m_{d,r}(c_d') \} \nonumber \\
&= \sum_{r=1}^{R} s_r^2 \prod_{d=1}^{D} \mathbf{K}_{d,r}(c_d, c_d').
\end{align}
Every factor of the product is a kernel entry. All kernels have unit diagonal, so the prior variance of every cell equals $\sum_r s_r^2$. The error layer adds an independent diagonal term, which yields the covariance of $\mathbf{z}$ stated in the main text. The entries of $\mathbf{K}_{d,r}$ are the correlations discussed in Section~4.2 of the main text: $1/(1+\kappa^2)$ for two distinct levels of an exchangeable mode, and $\{1 + \kappa^2 c_\phi(\delta)\}/(1+\kappa^2)$ for two levels at distance $\delta$ of a smooth mode.

\section{Gram Matrices and Directions}
\label{sec:S2}

All summaries reported in the main text and in this supplement are functions of the surface $\boldsymbol{\eta}$ and are computed per stored draw. The Gram matrix $\boldsymbol{\Gamma}_d$ of mode $d$ and its eigendecomposition are defined in Section~4.4 of the main text. Because the surface is a CP decomposition, $\boldsymbol{\Gamma}_d$ can be computed from the factors without forming the surface, as $\boldsymbol{\Gamma}_d = N^{-1} \mathbf{H} \mathbf{M}_d \mathrm{diag}(\mathbf{g}) (\circ_{d' \neq d} \mathbf{M}_{d'}^\top \mathbf{M}_{d'}) \mathrm{diag}(\mathbf{g}) \mathbf{M}_d^\top \mathbf{H}$, where $\mathbf{H} = \mathbf{I} - \mathbf{1}\mathbf{1}^\top / n_d$ centers over the levels of mode $d$.

The directions are reported as $\sqrt{n_d \lambda_j}\,\mathbf{u}_j$. Because $\mathbf{u}_j$ has unit length, the scaled direction has mean square $\lambda_j$ over the $n_d$ levels, so its entries are typical log-rate deviations of the levels along direction $j$ and are comparable across modes of different sizes. The sign of $\mathbf{u}_j$ is aligned to the $j$-th eigenvector $\mathbf{v}_j$ of the posterior mean Gram matrix, and no rotation is applied. When two eigenvalues are nearly equal, every rotation of the two eigenvectors within the space they span describes the surface almost equally well, so that across draws the two directions can swap their order or mix. Direction $j$ is flagged as not individually identified when $\mathbf{u}_j$ lies closer to another eigenvector of the posterior mean Gram matrix than to $\mathbf{v}_j$, that is, $|\langle \mathbf{u}_j, \mathbf{v}_j \rangle| \leq \max_{k \neq j} |\langle \mathbf{u}_j, \mathbf{v}_k \rangle|$, in more than five percent of the draws.

\section{Simulation Experiments}
\label{sec:S3}

\subsection{Recovery of a four-way truth}
\label{sec:S3demo}

The first experiment simulates data from a CP-PLN truth on a four-way array and asks whether the sampler recovers the identified quantities of the noise-free latent surface $\eta$, namely the surface itself, the leading directions of every mode, and the age-indexed error variances. The array has dimensions $30 \times 40 \times 2 \times 12$ and emulates the structure of the application, with a smooth time mode of 30 years, a smooth age mode of 40 age groups of 2.5 years covering ages 0 to 97.5, a two-level sex mode, and a twelve-level unit mode. The truth consists of four components with amplitudes $2.0$, $0.8$, $0.4$, and $0.2$ around a level of $-5$. Their time and age vectors are draws from Mat\'ern-$3/2$ processes with component-specific length scales, the sex and unit vectors are exchangeable draws with component-specific variation around a common level, and all vectors are scaled to unit root mean square. Exposures are drawn cell-wise between $200$ and $20{,}000$, and the age-indexed error standard deviations between $0.05$ and $0.15$. The model is fitted with an upper bound of $R = 20$ components and the sampling settings of the application.

\begin{figure}[h]
    \centering
    \begin{subfigure}[t]{0.36\textwidth}
        \includegraphics[width=\textwidth]{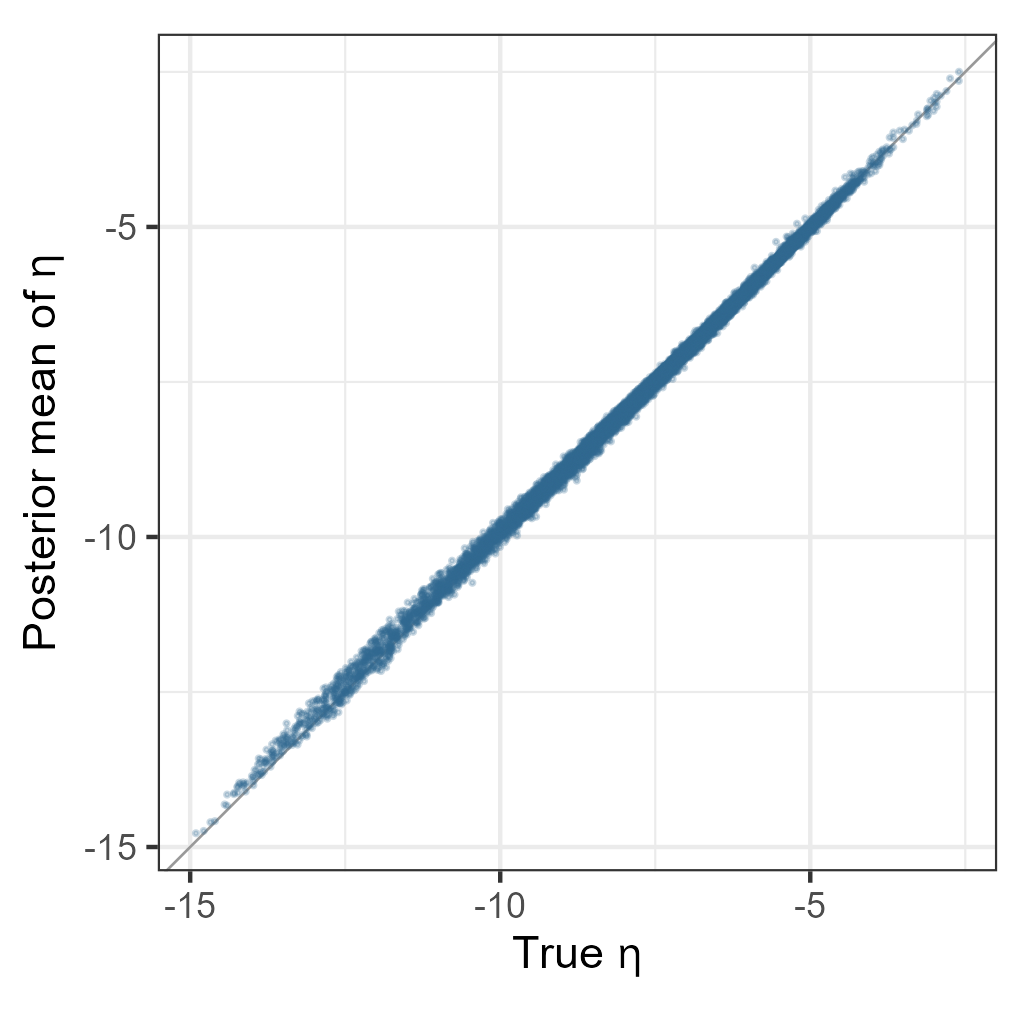}
    \end{subfigure}
    \hfill
    \begin{subfigure}[t]{0.55\textwidth}
        \includegraphics[width=\textwidth]{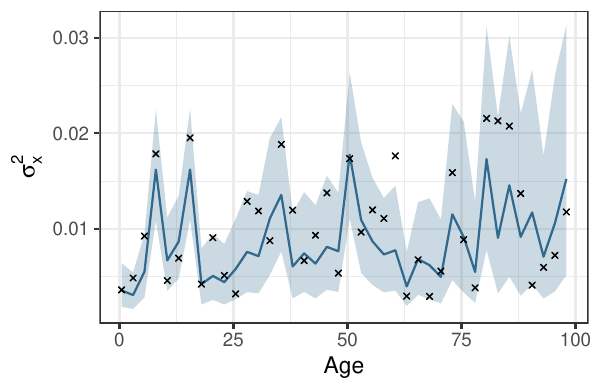}
    \end{subfigure}

    \begin{subfigure}[t]{\textwidth}
        \includegraphics[width=\textwidth]{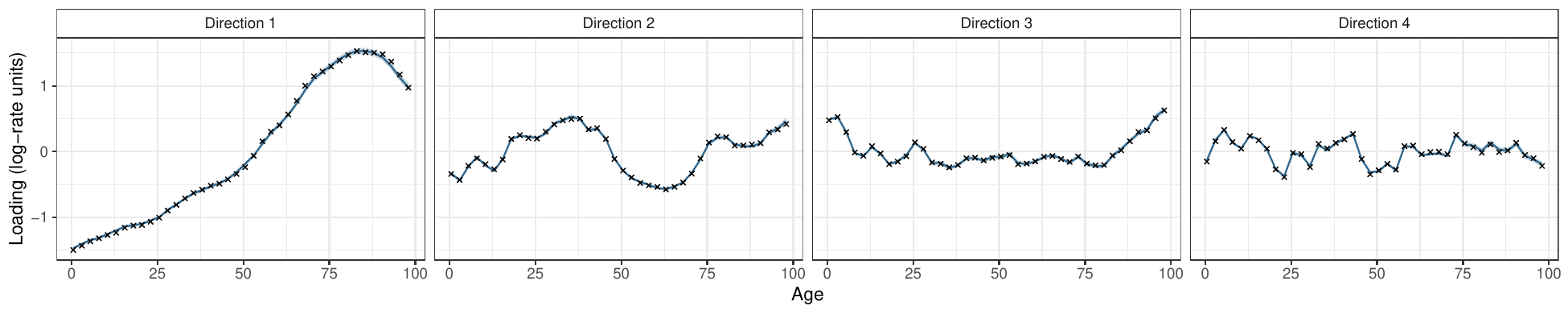}
    \end{subfigure}

    \begin{subfigure}[t]{\textwidth}
        \includegraphics[width=\textwidth]{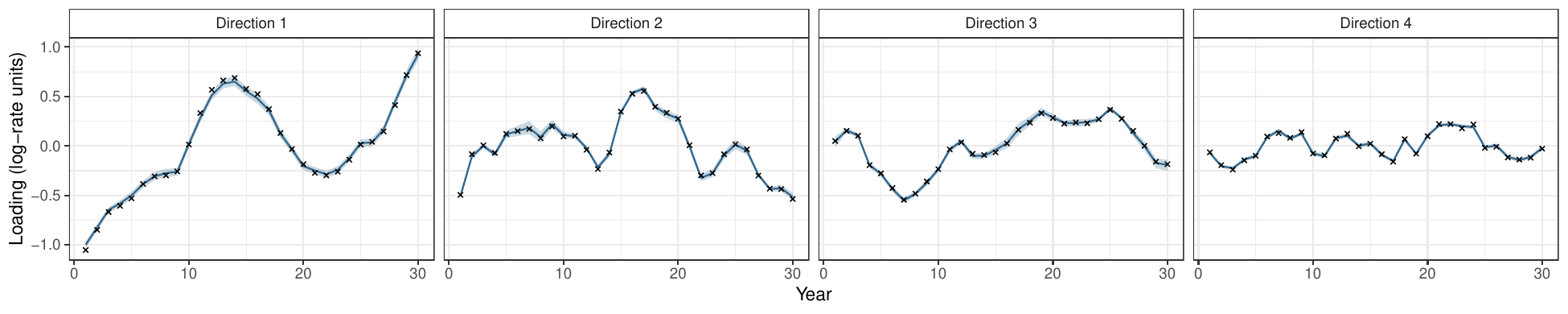}
    \end{subfigure}
    \caption{Four-way simulation. Top left: posterior mean of $\boldsymbol{\eta}$ against the true $\boldsymbol{\eta}$ for all cells. Top right: posterior medians and 90 percent intervals of the age-indexed error variances against the truth (crosses). Middle and bottom: the leading directions of the age and time modes in log-rate units, posterior medians with 90 percent bands against the truth (crosses).}
    \label{fig:simdemo}
\end{figure}

Figure~\ref{fig:simdemo} shows that the posterior mean surface lies close to the true surface over its whole range, that the leading directions of the age and time modes are recovered with posterior medians on top of the truth, and that the posterior of the error variances tracks their age pattern.

\subsection{Factorial experiment}
\label{sec:S3fact}

The second experiment examines the two sparsity mechanisms and the smoothing kernels in a $2 \times 2 \times 2$ factorial design that crosses the rank of the truth, the noise level, and the exposure. The array has dimensions $20 \times 41 \times 15$ (year, age, region), with log rates between about $-8.5$ and $-0.5$. Every component of the truth is constant in some modes. The low-rank truth has four components: an age schedule that varies in age only, a mortality improvement that varies in year and age, regional differences that vary in age and region, and a regional level that varies in region only. The high-rank truth adds four smaller components: regional trends (year and region), a three-way pattern (year, age, and region), regional differences at the oldest ages (age and region), and a second, slower age--time pattern (year and age). The error standard deviation is $0.05$ or $0.25$, and the exposure per cell below age 50 is $20{,}000$ or $200$ person-years, lower at older ages. Every setting is replicated five times, and three models are fitted to the same data with $R = 20$ and the sampling settings of the application: BATS, whose age kernel separates age 0 from the ages above as in the application, a model with both shrinkage mechanisms but exchangeable kernels in every mode (no Mat\'ern), and CP-PLN.

Three identified quantities are compared with the truth. First, the spurious share measures the structure a model fits beyond what the truth contains. For every mode, it is the share of the variation along that mode, that is, of the trace of the Gram matrix, that falls on the eigenvalues beyond the true rank of the mode, summarized by its posterior median and averaged over the three modes. A value of zero means that the model uses exactly as many directions as the truth. Second, the direction RMSE is the root mean squared error of the posterior medians of the scaled directions $\sqrt{n_d \lambda_j}\,\mathbf{u}_j$, aligned in sign to the truth, against the true directions, pooled over the three modes and over the directions that carry at least one percent of the variation of their mode in the truth. Third, the surface RMSE is the root mean squared error of the posterior mean of $\boldsymbol{\eta}$ against the true noise-free $\boldsymbol{\eta}$, which excludes the cell-level error and Poisson variation. Table~\ref{tab:simfact} reports the three quantities.

BATS has the smallest spurious share in every setting. It stays at or below $0.002$ at high exposure and reaches $0.016$ only for the low-rank truth at high noise and low exposure, whereas CP-PLN fits between $0.3$ and $28$ percent of the variation of a mode to directions the truth does not contain, most at low exposure. The model without Mat\'ern kernels removes most of this spurious structure, so the shrinkage toward constant mode vectors does the larger part of the work. The smoothing kernels add accuracy. BATS has the smallest surface RMSE in every setting, between $34$ and $47$ percent below CP-PLN, and the smallest direction RMSE in seven of the eight settings, the exception being the low-rank truth at low noise and low exposure, where the model without Mat\'ern kernels is marginally ahead within one standard deviation. In absolute terms the gains are largest at low exposure, where the information per cell is scarce.

\section{Prior Demonstration: Full Comparison}
\label{sec:S4}

This section complements Section~5 of the main text with the full comparison on the subset of the four least populous counties in 2021--2024. In addition to BATS and CP-PLN, it includes a model that keeps both shrinkage mechanisms but replaces the Mat\'ern kernels of the time and age modes by exchangeable kernels (no Mat\'ern), which isolates the contribution of smoothing. All three models use $R = 20$ and the same sampling settings as throughout in the main text.

\begin{figure}[p]
    \centering
    \includegraphics[width=\textwidth]{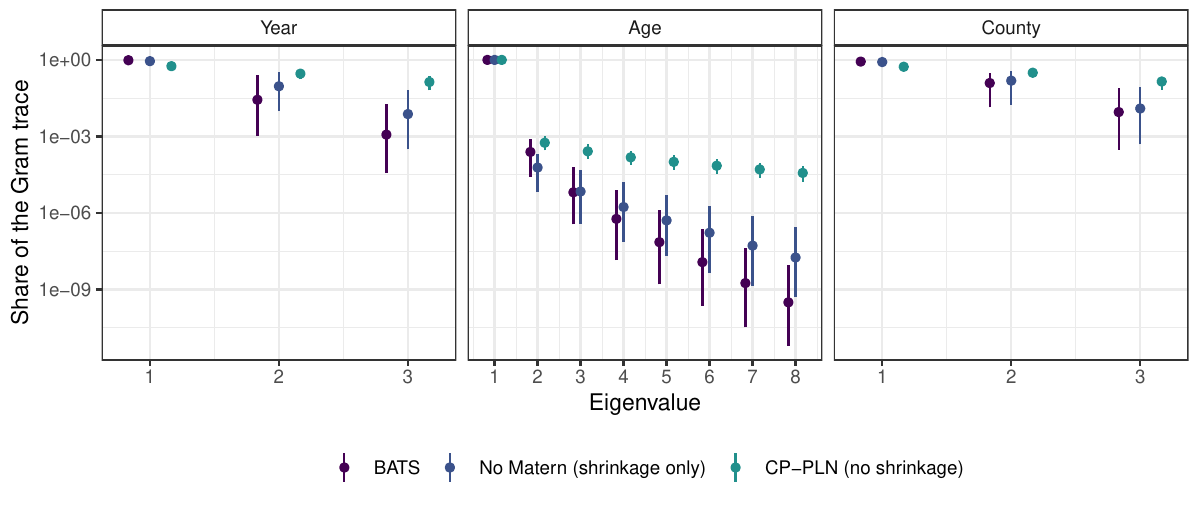}
    \caption{Prior demonstration. Posterior medians and 90 percent intervals of the Gram eigenvalues of the time, age, and county modes as shares of the trace, on a log scale.}
    \label{fig:subset_spectra}
\end{figure}

\begin{figure}[p]
    \centering
    \includegraphics[width=\textwidth]{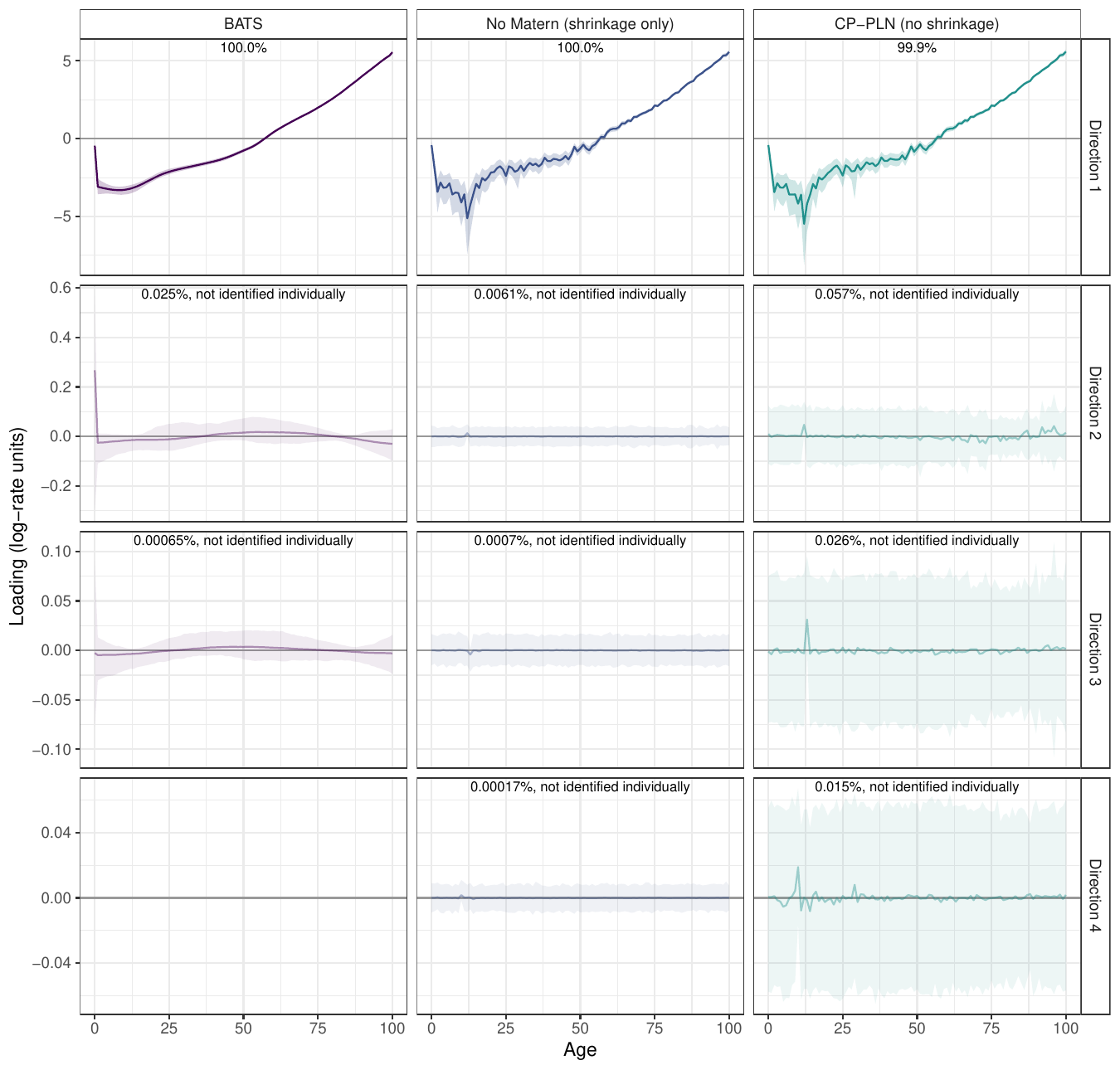}
    \caption{Prior demonstration. Leading directions of the age mode in log-rate units, posterior medians with 90 percent bands, for the three models (columns). Panel labels give the median share of the variation along age. Directions that are not individually identified are drawn faded.}
    \label{fig:subset_dir1}
\end{figure}

\begin{figure}[p]
    \centering
    \includegraphics[width=0.9\textwidth]{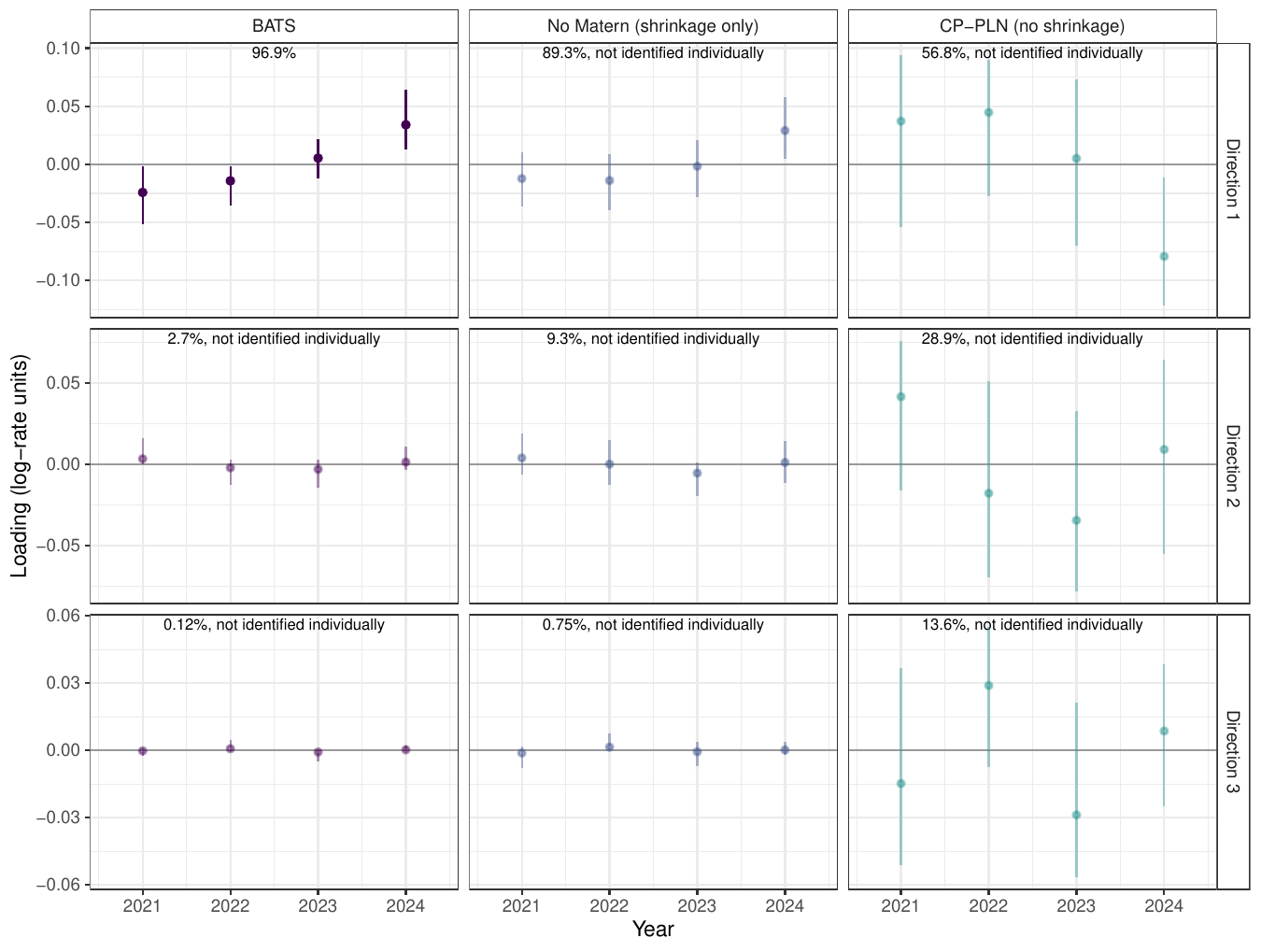}

    \includegraphics[width=0.9\textwidth]{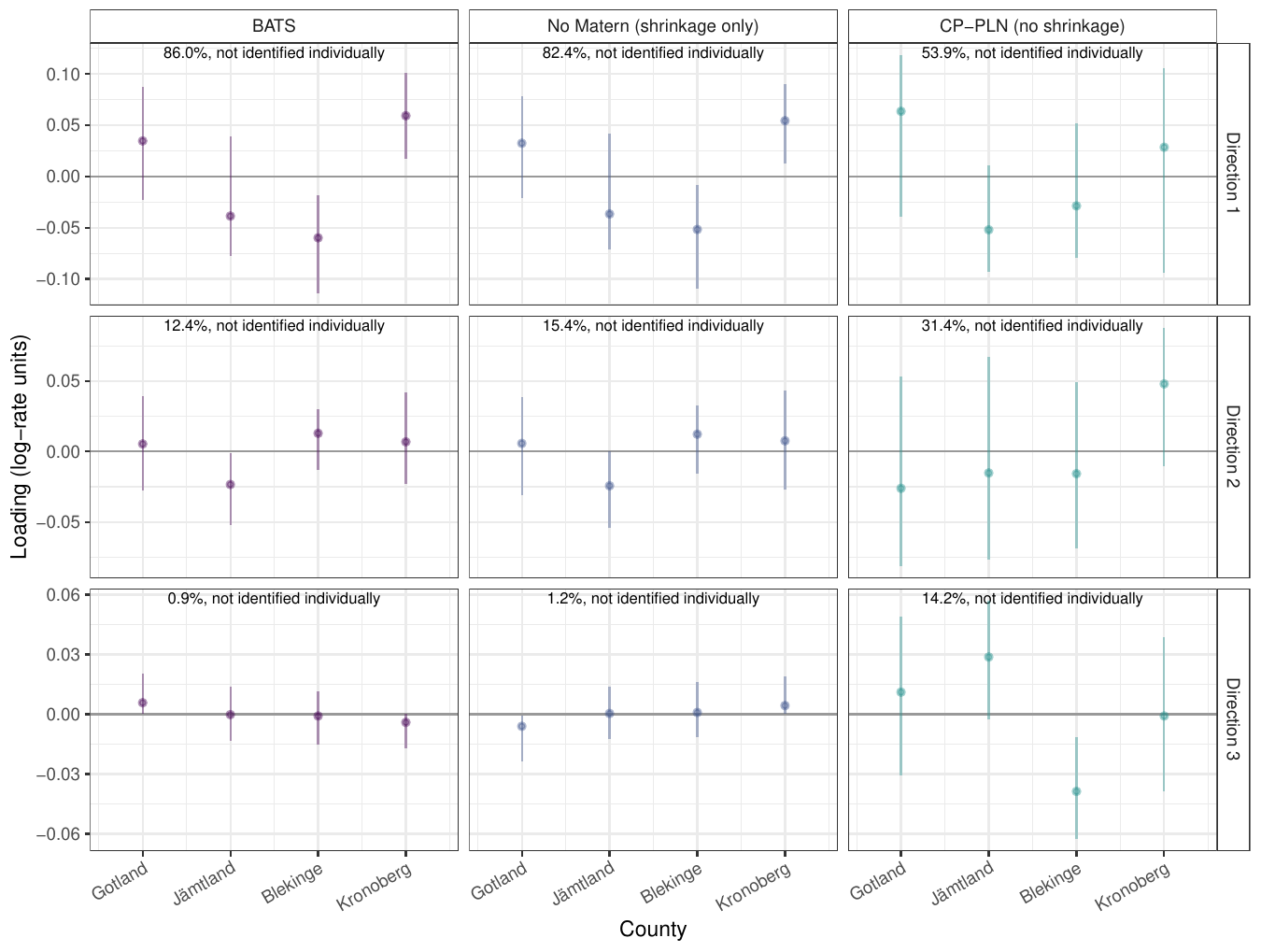}
    \caption{Prior demonstration. Leading directions of the time (top) and county (bottom) modes in log-rate units, posterior medians with 90 percent bands, for the three models (columns). Panel labels give the median share of the variation along the mode. Directions that are not individually identified are drawn faded.}
    \label{fig:subset_dir2}
\end{figure}

\begin{figure}[p]
    \centering
    \includegraphics[width=0.85\textwidth]{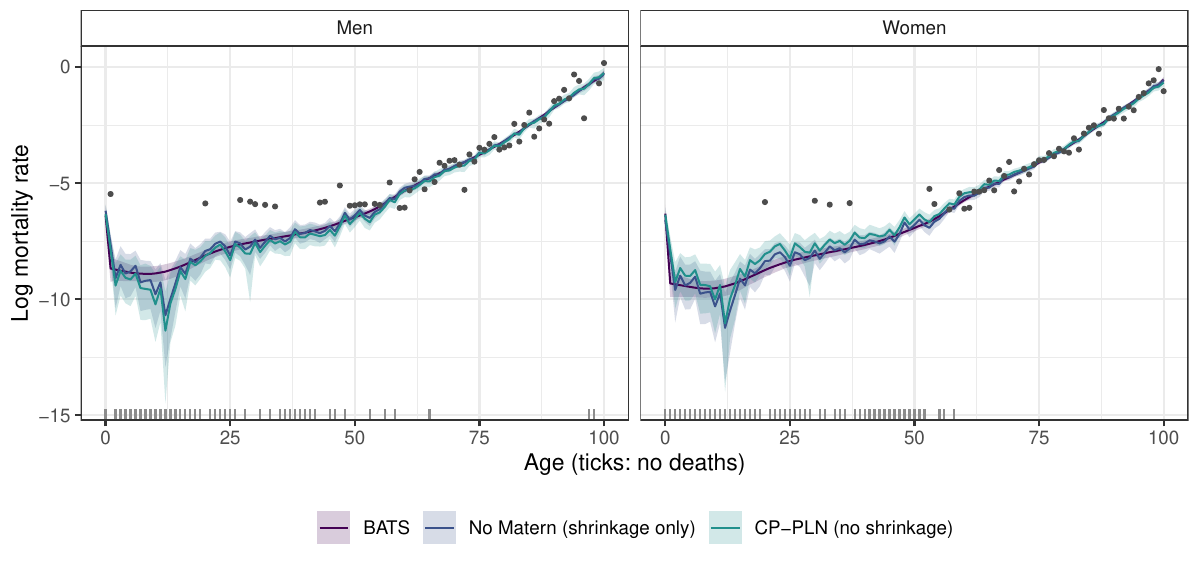}
    \caption{Prior demonstration. Posterior medians and 90 percent intervals of the smoothed log mortality rates $\boldsymbol{\eta}$ of Gotland in 2024 under the three models, by sex, which exclude the error layer and Poisson variation, and the raw log rates (points). Rug ticks mark ages with zero observed deaths.}
    \label{fig:subset_curves}
\end{figure}

Figure~\ref{fig:subset_spectra} shows the Gram eigenvalue spectra of the time, age, and county modes. Under CP-PLN, the spectra of all modes level off at a floor, which is the spurious structure fitted to sampling noise, whereas under the two shrinkage models they keep decaying. The Mat\'ern kernels mainly change the shape of the age and time directions. Figures~\ref{fig:subset_dir1} and~\ref{fig:subset_dir2} show the leading directions of all four modes. Along age, the leading direction is smooth under BATS and jagged under both models without smoothing, and the further directions are smooth and close to zero under BATS, flat but noisy under the model without Mat\'ern kernels, and pure noise under CP-PLN. Along time and county, where four levels carry the information of many cells each, the two shrinkage models agree on the leading direction, whereas CP-PLN spreads the variation over all directions, so that its leading direction differs and is poorly determined. The sex mode has a single direction, on which the two shrinkage models agree, whereas CP-PLN places a smaller part of the difference between men and women in it. Figure~\ref{fig:subset_curves} shows the fitted age schedules of Gotland in 2024.

\section{Additional Application Results}
\label{sec:S5}

\subsection{Life table computations}
\label{sec:S5lt}

All life table quantities are computed from single-age period life tables for ages $x = 0, \dots, 99$ and the open age group $100+$, following the methods protocol of the Human Mortality Database \citep{wilmoth2007methods}. Given an age schedule of death rates $m_x$, the average number of years lived in the year of death is $a_x = 1/2$ for $x \geq 1$, and $a_0$ follows the sex-specific formulas of \citet{andreev2015average} as a piecewise linear function of $m_0$. Death probabilities are $q_x = m_x / \{1 + (1 - a_x)\, m_x\}$ for the closed ages and $q_{100+} = 1$. With radix $l_0 = 1$, survivorship is $l_{x+1} = l_x (1 - q_x)$, the life table distribution of deaths by age is $d_x = l_x q_x$, person-years lived are $L_x = l_{x+1} + a_x d_x$ for the closed ages and $L_{100+} = l_{100+}/m_{100+}$ for the open age group, and remaining life expectancy is $e_x = \sum_{y \geq x} L_y / l_x$, so that life expectancy at birth is $e_0 = \sum_x L_x$. Life disparity is approximated by
$e^{\dagger} =
\sum_{x=0}^{99} d_x
\big[e_x + a_x(e_{x+1}-e_x)\big]
+ d_{100+}e_{100+}$,
the average remaining life expectancy at the age
of death \citep{vaupel2003decomposing}.

The raw quantities are computed from the observed rates $m_x = y_x / O_x$. Where an area records no deaths in the open age group in a year, its rate is replaced by the national rate of the same year and sex, because a zero rate in the open age group implies an infinite life expectancy. The posterior quantities are computed per stored draw from the smoothed rates $\exp(\boldsymbol{\eta})$, so that every draw of the surface yields one draw of every life table quantity. For Sweden as a whole and for groups of counties, the rates are the exposure-weighted averages of the smoothed county rates, $\sum_i O_{i}\exp(\eta_{i}) / \sum_i O_{i}$ over the counties $i$ of the group, and for the raw quantities the pooled counts divided by the pooled exposures. The standard deviation of life expectancy across counties is the unweighted standard deviation of the twenty county values, computed per draw for the posterior and once for the raw rates.

\subsection{Convergence}

Convergence is assessed on quantities that are invariant to the non-identified features of the decomposition, namely the Poisson data-layer log likelihood of the observed cells, the mean and root mean square of the latent surface, fixed random projections of the latent surface, the two leading Gram eigenvalues $\lambda_1$ and $\lambda_2$ of the time, age, and county modes, the age-indexed error variances $\sigma^2_x$, indexed by age $x$, and predicted expected death counts $O_c \exp(\eta_c + \sigma^2_{c_s}/2)$ at six individual cells that span the range from sparse to dense. These are the cells of men in Gotland at ages 5, 30, and 85, of women in Stockholm at ages 40 and 85, and of men in Halland at age 0, all in 2024.

\begin{figure}[p]
    \centering
    \begin{subfigure}[t]{0.325\textwidth}
        \includegraphics[width=\textwidth]{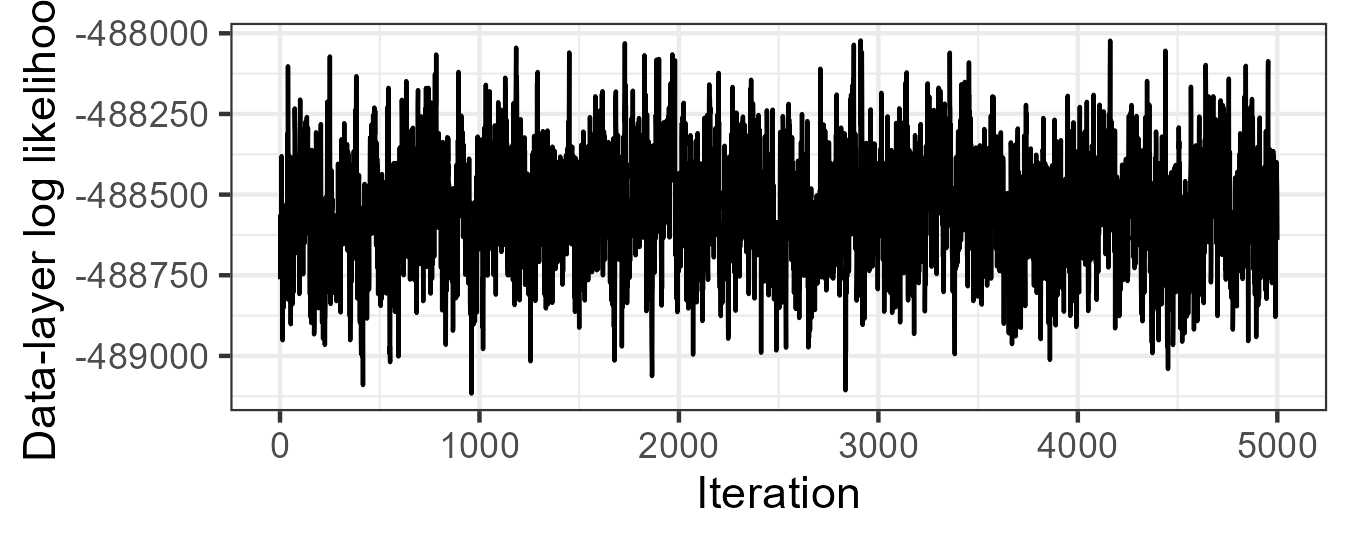}
    \end{subfigure}
    \begin{subfigure}[t]{0.325\textwidth}
        \includegraphics[width=\textwidth]{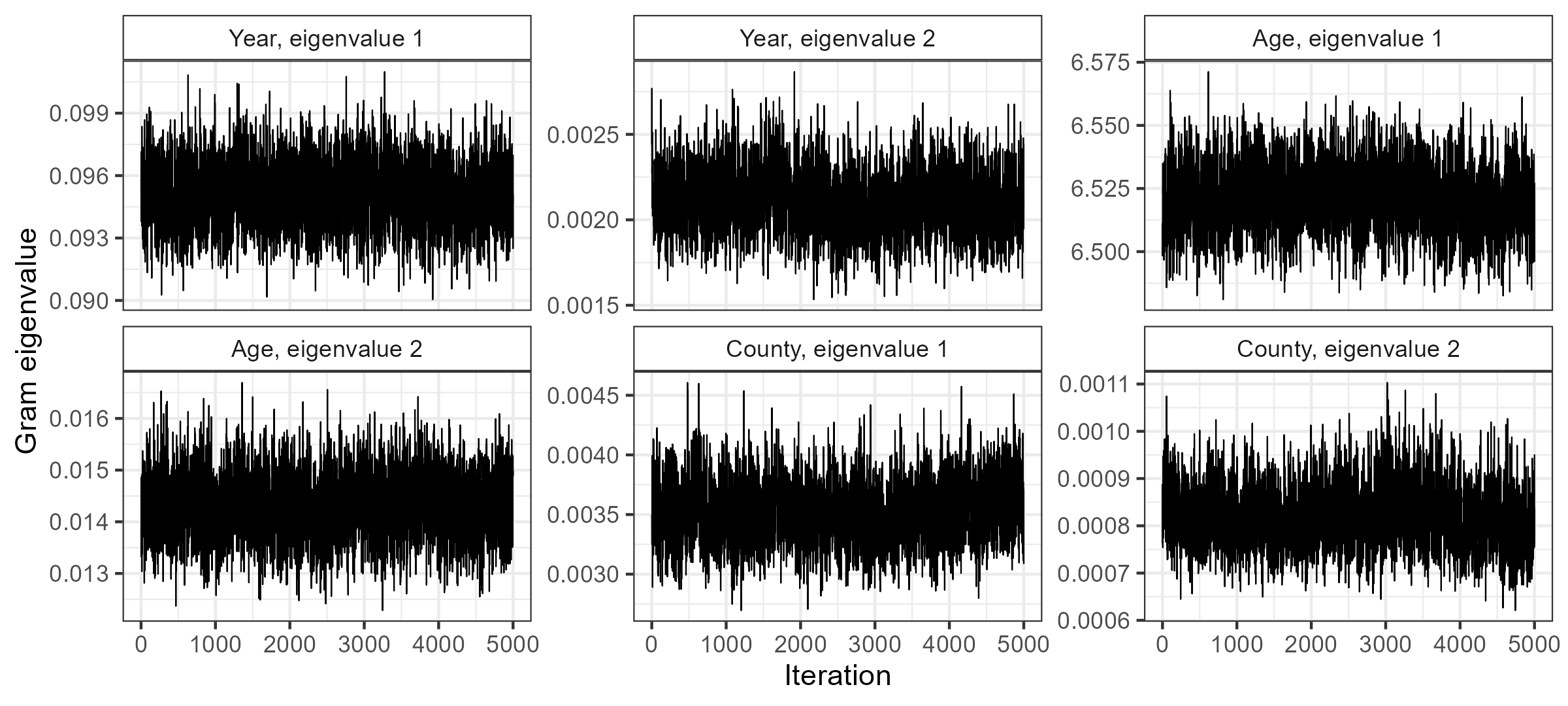}
    \end{subfigure}
    \begin{subfigure}[t]{0.325\textwidth}
        \includegraphics[width=\textwidth]{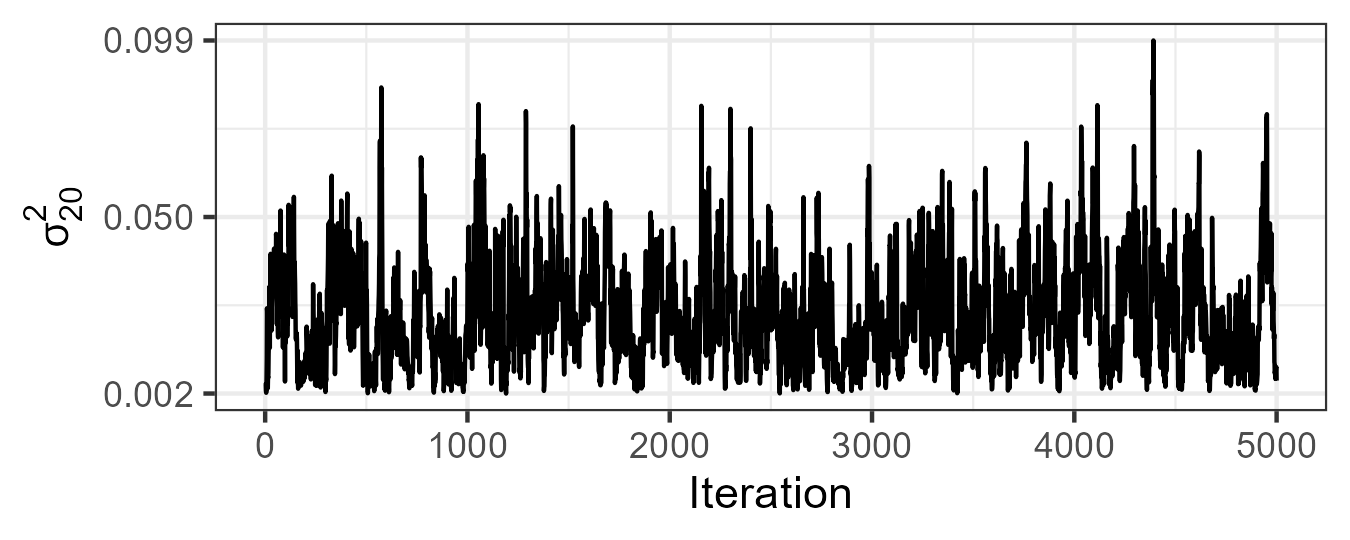}
    \end{subfigure}
    \caption{Convergence diagnostics for tensor-level series. Trace of the Poisson data-layer log likelihood of the observed cells (left), traces of the two leading Gram eigenvalues of the time, age, and county modes (middle), and trace of the observation variance at age 20 (right).}
    \label{fig:conv}
\end{figure}

\begin{figure}[p]
    \centering
    \includegraphics[width=\textwidth]{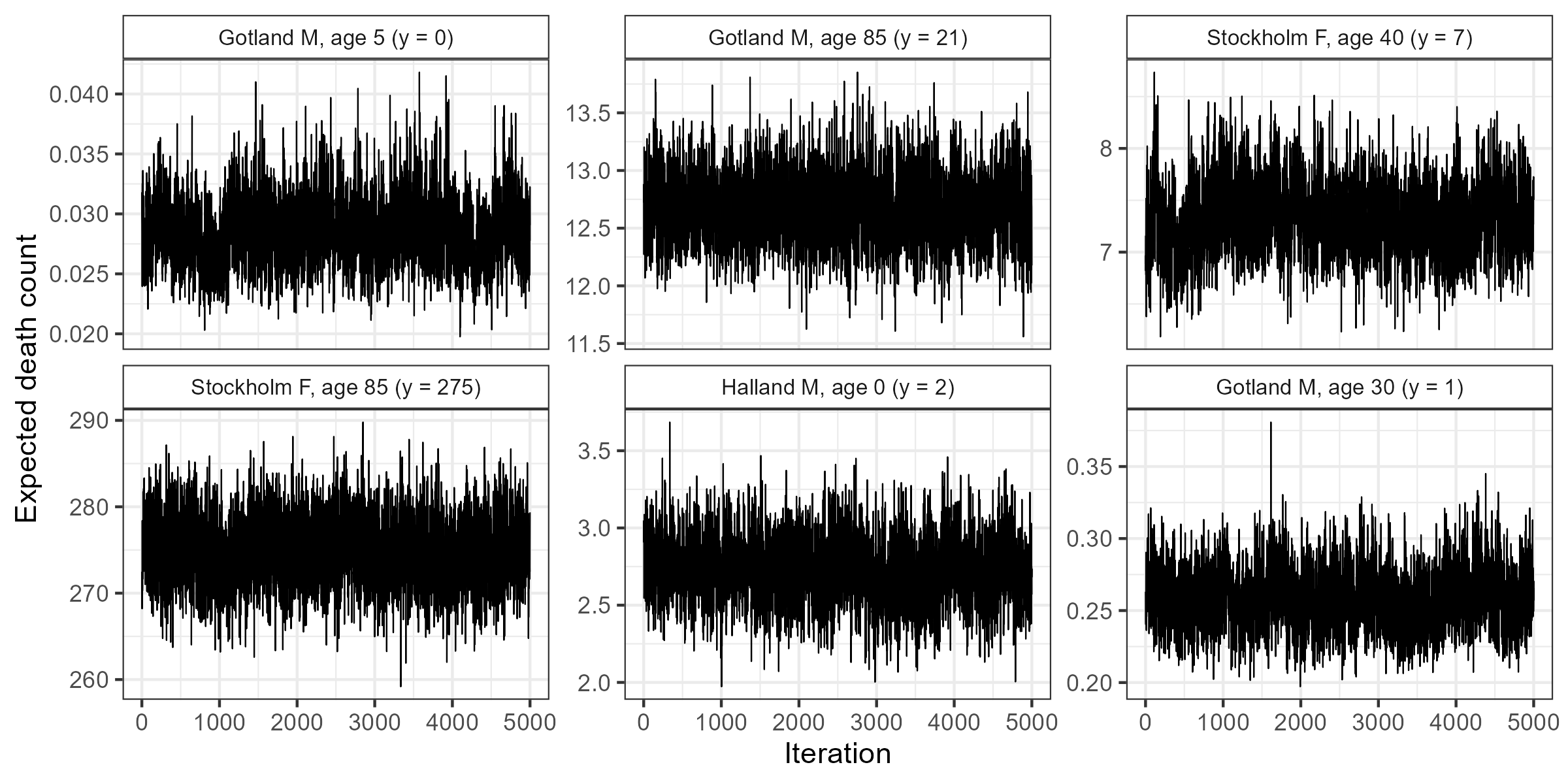}
    \caption{Convergence diagnostics for predicted expected death counts at six cells ranging from sparse to dense, with the observed count of the cell given in the panel label. Their effective sample sizes and split-$\hat{R}$ values are given in Table~\ref{tab:ess}.}
    \label{fig:convpred}
\end{figure}

Table~\ref{tab:ess} reports effective sample sizes and split-$\hat{R}$ values for all monitored series, computed from the stored draws of the main fit, and Figures~\ref{fig:conv} and~\ref{fig:convpred} show representative traces. Of the $5{,}000$ stored draws, the effective sample sizes lie between $210$ and $1{,}710$ for the log likelihood, the surface summaries, and the predicted counts, between $330$ and $2{,}690$ for the Gram eigenvalues, and between $260$ and $910$ for the error variances, and all split-$\hat{R}$ values are at or below $1.02$. The slowest series are the predicted count of women in Stockholm at age 40 and the error variance at age 20, where few deaths inform it.

\subsection{Age-indexed error variances}

\begin{figure}[p]
    \centering
    \includegraphics[width=0.7\textwidth]{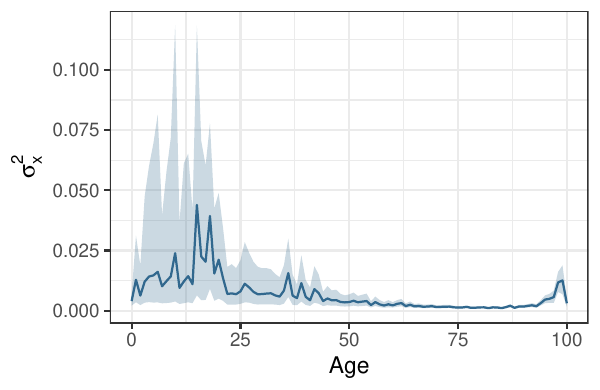}
    \caption{Posterior medians with 90 percent credible bands of the age-indexed error variances $\sigma^2_x$.}
    \label{fig:sigma2}
\end{figure}

Figure~\ref{fig:sigma2} shows the posterior of the age-indexed error variances $\sigma^2_x$. The error variances are largest in childhood and adolescence, where deaths are rare and dominated by external causes and the posterior is correspondingly wide, decline to about $0.002$ between ages 50 and 90, where chronic disease dominates and counts are large, and rise again toward age 100, where exposures become small. The separate age-0 level of the age kernel absorbs the infant drop into the structured surface, so the error variance at age 0 is small. This pattern is consistent with indexing the dispersion by age.

\subsection{Systematic fit assessment}
\label{sec:S5fit}

To assess the fit beyond the units selected for display in the main text, all $2{,}240$ age schedules, that is, all year, sex, and county combinations, are ranked by the held-out log predictive score they received in the cross-validation exercise. However, a raw mean log predictive score would rank schedules by their count level rather than by their fit. The largest log density that a Poisson-type predictive can attain at an observed count $y$, $\log \mathrm{Pois}(y \mid y)$, decreases with $y$, from about zero at an empty cell to about $-3.8$ at a count of 300, even when the model is perfect. The ranking statistic is therefore the excess log predictive score, defined per cell as the log predictive score minus $\log \mathrm{Pois}(y_c \mid y_c)$ and averaged over the cells of the schedule. This is essentially the predictive analogue of the deviance. It equals zero for a perfect prediction and is negative otherwise.

\begin{figure}[p]
    \centering
    \begin{subfigure}[t]{\textwidth}
        \includegraphics[width=\textwidth]{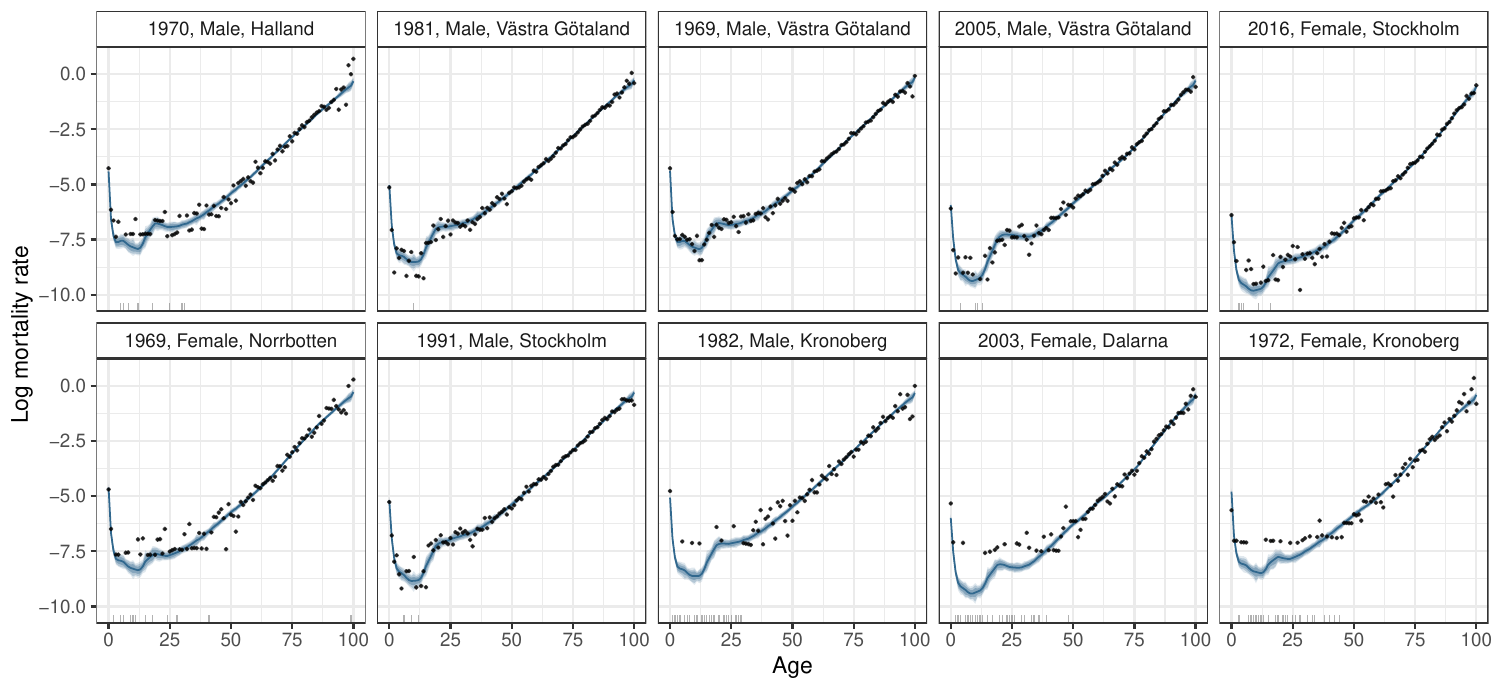}
    \end{subfigure}

    \begin{subfigure}[t]{\textwidth}
        \includegraphics[width=\textwidth]{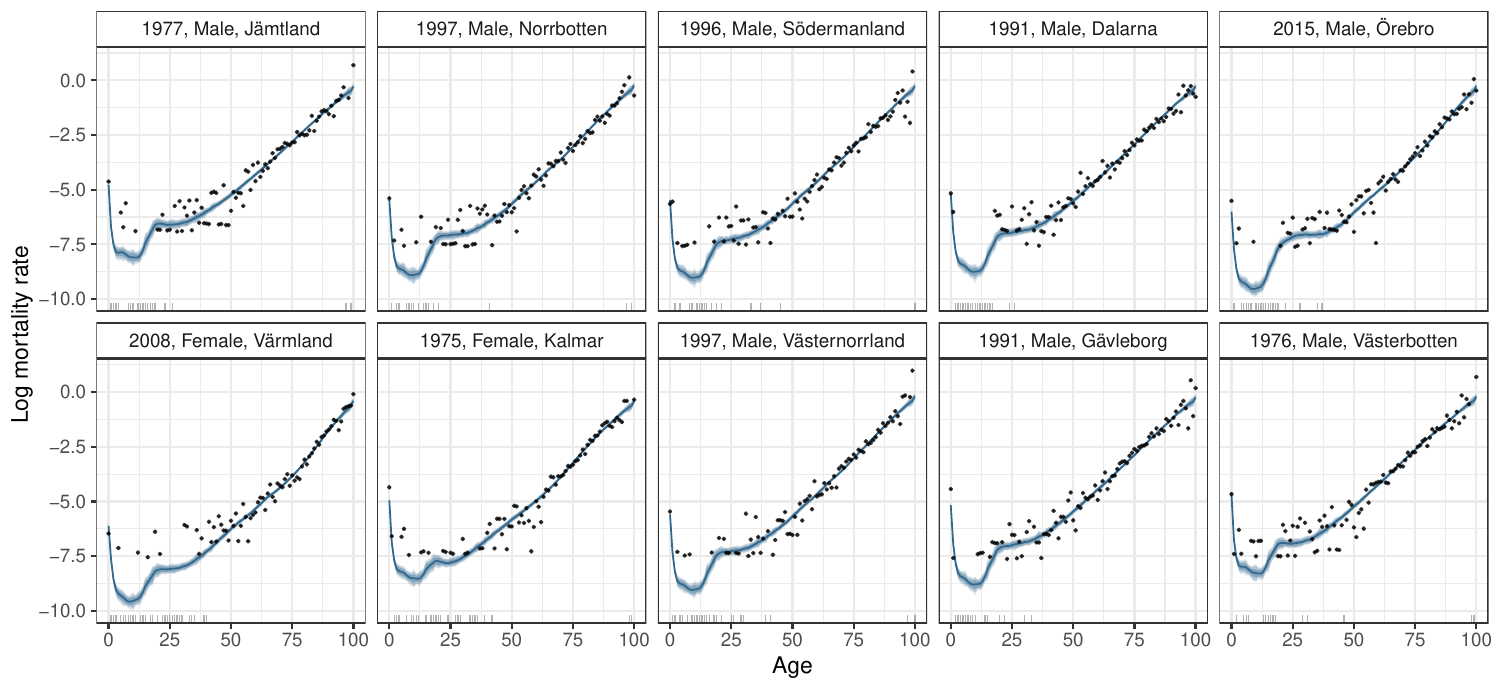}
    \end{subfigure}

    \begin{subfigure}[t]{\textwidth}
        \includegraphics[width=\textwidth]{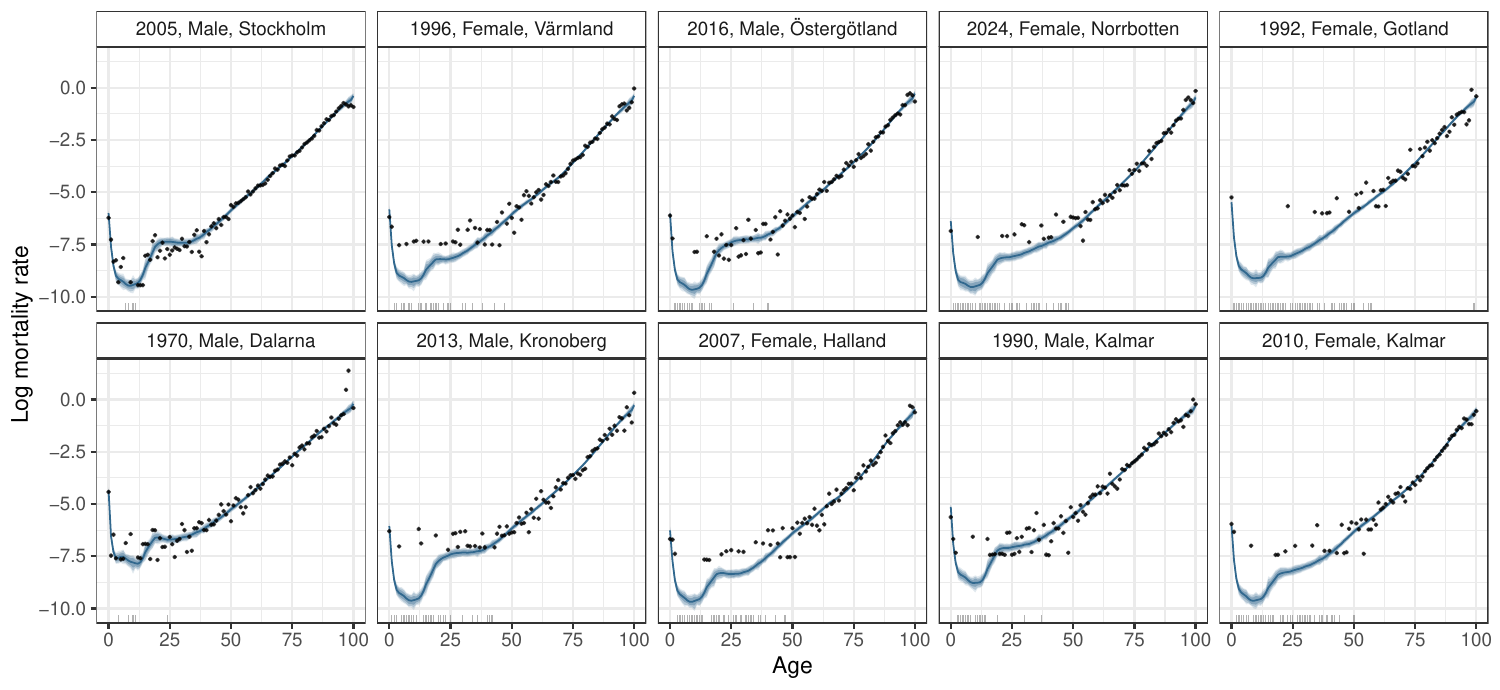}
    \end{subfigure}
    \caption{Best (top), worst (middle), and randomly selected (bottom) schedules by the held-out excess log predictive score. Lines are posterior means of the smoothed log rate $\boldsymbol{\eta}$, and bands are 75, 90, and 95 percent intervals of the latent rate $\exp(\boldsymbol{\eta} + \sigma_x \varepsilon)$ with $\varepsilon$ drawn anew for every posterior draw, which exclude Poisson variation. Points are the raw log rates, and rug ticks mark ages with zero observed deaths.}
    \label{fig:insample}
\end{figure}

Figure~\ref{fig:insample} shows the ten best, the ten worst, and ten randomly selected schedules. The line is the posterior mean of the smoothed log rate $\boldsymbol{\eta}$. The bands are the 75, 90, and 95 percent intervals of $\exp(\boldsymbol{\eta}^{(s)} + \sigma^{(s)}_x \varepsilon)$ over the stored draws $s$ of the main fit, where $\varepsilon \sim \mathcal{N}(0,1)$ is drawn anew for every draw, that is, of the latent rate of a new cell with the same smoothed rate and error variance. They include the error layer but exclude Poisson variation. The best-ranked schedules belong to small populations, most of them women in Gotland, J\"amtland, and Norrbotten, whereas the worst-ranked schedules belong predominantly to men in medium-sized and large counties, such as V\"astra G\"otaland, Sk\aa ne, \"Osterg\"otland, and V\"asterbotten, and to the 1970s to 1990s. This ordering partially reflects what the statistic can detect rather than a failure of the fit in large populations. In a small population, the counts carry little information beyond the low-rank smooth structure, so the fitted schedule and the predictive spread are consistent with the data. In larger units, the counts are informative enough to reveal local structure that a smooth surface of at most twenty components does not capture, such as year-specific irregularities in the age profile, and these departures are penalized because the predictive distribution is narrow relative to them. Even in the worst-ranked schedules, however, the fitted curves remain smooth and demographically plausible, and their lower scores stem from local departures of this kind rather than from a systematic misfit.

\subsection{Regional inequality, women}

\begin{figure}[p]
    \centering
    \begin{subfigure}[b]{0.49\textwidth}
        \includegraphics[width=\textwidth]{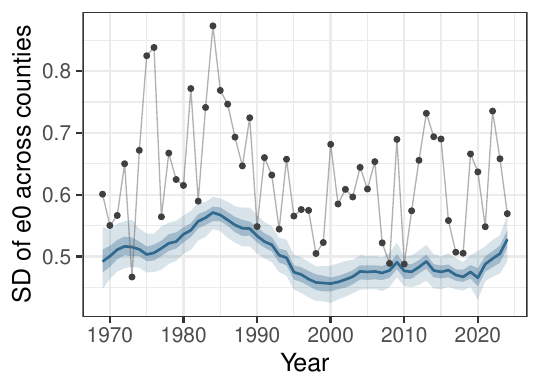}
    \end{subfigure}
    \hfill
    \begin{subfigure}[b]{0.49\textwidth}
        \includegraphics[width=\textwidth]{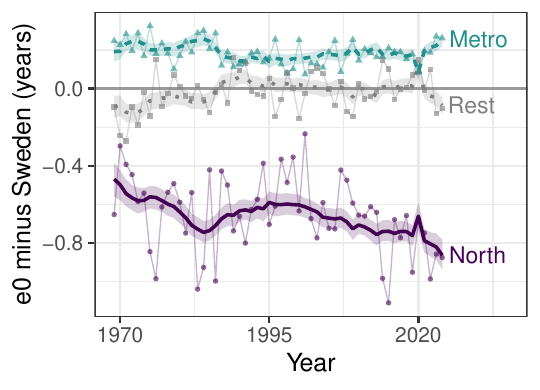}
    \end{subfigure}
    \caption{Regional inequality in life expectancy at birth, women, as in Figure~6 of the main text.}
    \label{fig:regional_women}
\end{figure}

Figure~\ref{fig:regional_women} repeats Figure~6 of the main text for women. The pattern is similar to that of men but weaker. The posterior standard deviation of life expectancy across counties rises from $0.50$ years in 1970 to $0.56$ years in the mid-1980s, declines to a minimum of $0.46$ years in 2000, and rises again to $0.53$ years in 2024. The northern counties fall from about $0.6$ years below the national level in the 1990s to almost $0.9$ years below it in 2024.

\subsection{Cross-validation design and scoring}
\label{sec:S5scoring}

Paired differences between BATS and a competitor are computed per cell. Standard errors are clustered by county--sex unit. The misfit of a model in one unit can persist across the ages and years of that unit, so the held-out cells of a unit are not independent. Different units are, however, treated as independent clusters. Size stratification uses the mean annual person-years of the county and sex unit. 

The log predictive score of a held-out cell is computed per posterior draw as the integral of the Poisson likelihood against the Gaussian state layer by adaptive Gauss--Hermite quadrature. For BPTF the log predictive score is the exact Poisson mixture over the rate draws. Point predictions are scored by the Poisson deviance $2\{y_c \log(y_c/\hat{y}_c) - (y_c - \hat{y}_c)\}$ of the held-out count against the posterior mean expected count $\hat{y}_c$, or against the plug-in Poisson mean of the point estimators. Randomized probability integral transform ranks are computed as $u = \{\#(\text{draws} < y) + U \cdot (\#(\text{draws} = y) + 1)\}/(S + 1)$ with $U$ uniform, which is exactly uniform under calibration at any finite number of stored draws $S$ and is unaffected by the discreteness of counts, and their Kolmogorov--Smirnov distance from the uniform distribution is reported. The empirical coverage of the central 90 (95) percent predictive interval is the share of held-out cells whose PIT value lies in $[0.05, 0.95]$ ($[0.025, 0.975]$). 

\subsection{Cross-validation results by population size and paired differences}
\label{sec:S5cv}

Table~\ref{tab:cvpaired} reports the paired differences between BATS and every competitor over all held-out cells, for both panels, and Table~\ref{tab:cvsize} the paired improvement by quartile of county--sex unit size, measured by mean annual person-years. Raw score differences on the count scale grow mechanically with count magnitude, so the gain in log predictive score is expressed in nats per $1{,}000$ observed deaths of the age schedule. The reduction in mean Poisson deviance per held-out cell is reported for all arms.

The gains over CP-PLN and BPTF are positive in every quartile, both in log predictive score and in deviance. The gain in log predictive score over BPTF is largest in the smallest quartile, where the predictive spread is dominated by the likelihood and the missing overdispersion layer of BPTF matters most. The gain over the additive model declines mildly toward larger units in log predictive score and grows toward larger units in deviance, because the deviance of a point prediction is dominated by the largest cells. Among the point estimators, the per-unit P-spline surface is closest to BATS in every quartile, the per-unit Lee--Carter model follows, and the schedule-wise P-spline and TOPALS, which pool little information across years, lose substantially in every quartile.

\section{Competing Methods}
\label{sec:S6}

\subsection{Schedule-wise P-spline smoother}

The first competitor fits an independent one-dimensional Poisson P-spline to every age schedule, that is, to every year, sex, and county combination, with a cubic B-spline basis on equally spaced knots over the age range extended by one percent on either side, with $\lfloor A/5 \rfloor = 20$ intervals, that is, a knot about every five years of age, a second-order difference penalty, and the penalty selected by the Bayesian information criterion per schedule, following the univariate smoother of \citet{camarda2012mortalitysmooth}. It pools no information across years or units and is included as the pooling-free end of the benchmark spectrum.

\subsection{Per-unit P-spline surface}

The second competitor is a two-dimensional Poisson P-spline surface fitted independently to each sex and county unit, following \citet{currie2004smoothing} as operationalized in the MortalitySmooth package \citep{camarda2012mortalitysmooth}, which has since been archived on CRAN. The implementation is a faithful reimplementation of that package. It uses tensor-product cubic B-splines with equally spaced knots on both axes, $\lfloor T/5 \rfloor$ intervals in time and $\lfloor A/5 \rfloor$ intervals in age (11 and 20 for Sweden), second-order difference penalties on both axes, and the two penalties selected by the Bayesian information criterion per unit. 

\subsection{Per-unit Poisson Lee--Carter model}

The third competitor fits the Poisson log-bilinear Lee--Carter model of \citet{brouhns2002poisson} separately to each sex and county unit by maximum likelihood, with the usual identification constraints on the age loadings and the period index. The model imposes a rank-one age--time structure within a unit and no smoothness.

\subsection{TOPALS relational model}

The fourth competitor is a two-stage TOPALS model \citep{debeer2012smoothing}. A sex-specific standard age schedule is first pooled from the training cells of the fold across all years and counties. Every age schedule is then fitted as the standard schedule plus a deviation on the log scale, represented by a linear B-spline basis with knots at ages 0, 1, 10, 20, 40, 70, and 100, with a first-order difference penalty on the spline coefficients and one global penalty selected by the Bayesian information criterion. Information is pooled through the standard schedule only.

\subsection{Additive main-effects model}

The fifth competitor specifies $\eta_c = \mu + \sum_d a_d(c_d)$ with zero-sum main effects, exchangeable Gaussian priors on the basis coefficients, and conjugate variance updates. It shares the observation layer, the latent-rate machinery, and the sampling budget with BATS, but removes all interactions.

\subsection{CP Poisson lognormal model}

The sixth competitor, CP-PLN, keeps the count layer, the upper bound on components, and the sampler of BATS, but replaces the priors of Section~4.1 of the main text by independent standard Gaussian priors on all mode vector entries and $\mathcal{N}(0, 10)$ priors on the weights. It has neither the shrinkage of the weights, nor the shrinkage of the mode vectors toward constants, nor smoothing kernels, and corresponds to the customary priors of probabilistic CP factorization.

\subsection{Poisson--gamma CP factorization}
\label{sec:S6bptf}

The seventh and eighth competitors are the Bayesian Poisson tensor factorization of \citet{schein2015bayesian} with $R = 10$ and $R = 20$ components, the CP analogue of Poisson--gamma nonnegative matrix factorization and a representative of the count tensor factorization tradition reviewed in Section~3 of the main text. It differs from BATS in every ingredient except the multiway low-rank structure: it has a pure Poisson likelihood without an overdispersion layer, nonnegative factors without weights, and no smoothness. The specification is
\begin{equation}
y_c \sim \mathrm{Poisson}\Big( O_c \sum_{r=1}^{R} \prod_{d=1}^{D} \theta_{d,r}(c_d) \Big),
\qquad
\theta_{d,r}(j) \sim \mathcal{G}(\alpha, \alpha \beta_d),
\qquad
\beta_d \sim \mathcal{G}(a_0, b_0),
\label{eq:bptf}
\end{equation}
where $\mathcal{G}(a, b)$ denotes the gamma distribution with shape $a$ and rate $b$, so that every factor entry has prior mean $1/\beta_d$. The exposure offset enters the rate multiplicatively, as in BATS. The shape $\alpha$ is fixed at $0.1$, below one, so that the prior has a spike at zero and the entries of components the data do not need collapse toward zero. This is the only complexity control of the model. The per-mode rates $\beta_d$ receive $a_0 = b_0 = 0.1$ and are learned. The sampling budget is the same as for the other Bayesian arms.

Inference is by an exact Gibbs sampler based on the Poisson--multinomial augmentation. For every observed cell with a positive count, the count is allocated to the $R$ components,
\begin{equation}
(y_{c,1}, \dots, y_{c,R}) \mid \cdot \sim \mathrm{Multinomial}\big(y_c, \, p_{c,1}, \dots, p_{c,R}\big),
\qquad p_{c,r} \propto \prod_{d} \theta_{d,r}(c_d),
\end{equation}
which is drawn vectorized over cells as a sequence of $R - 1$ conditional binomials. Conditional on the allocations, the factor entries are independent with gamma full conditionals,
\begin{gather}
\theta_{d,r}(j) \mid \cdot \sim \mathcal{G}\Big( \alpha + \sum_{c : c_d = j} y_{c,r}, \;
\alpha \beta_d + \sum_{c : c_d = j} O_c \prod_{d' \neq d} \theta_{d',r}(c_{d'}) \Big),
\\
\beta_d \mid \cdot \sim \mathcal{G}\Big( a_0 + \alpha n_d R, \; b_0 + \alpha \sum_{r, j} \theta_{d,r}(j) \Big).
\end{gather}
The sums are accumulated over observed cells by mode level, retaining cell-specific exposure weights. Held-out and zero-exposure cells contribute no likelihood terms.

\subsection{Fitted schedules of all methods for the least populous unit}
\label{sec:S6smallest}

\begin{figure}[p]
    \centering
    \includegraphics[width=\textwidth,page=5]{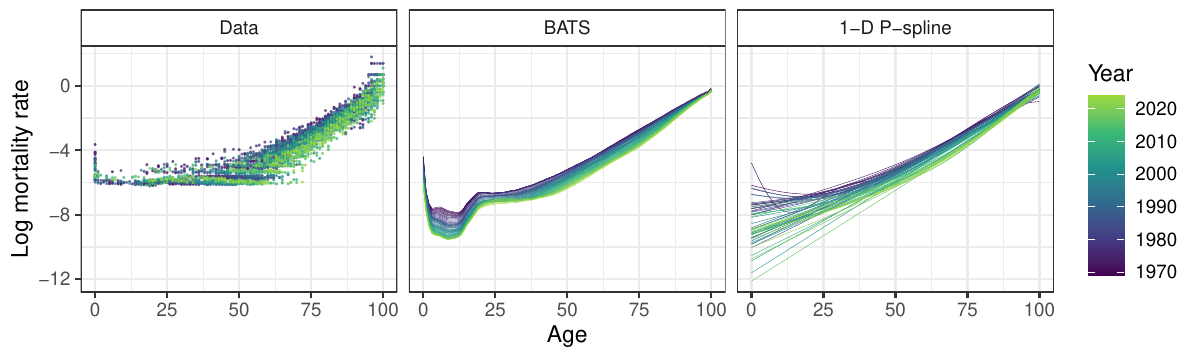}

    \includegraphics[width=\textwidth,page=6]{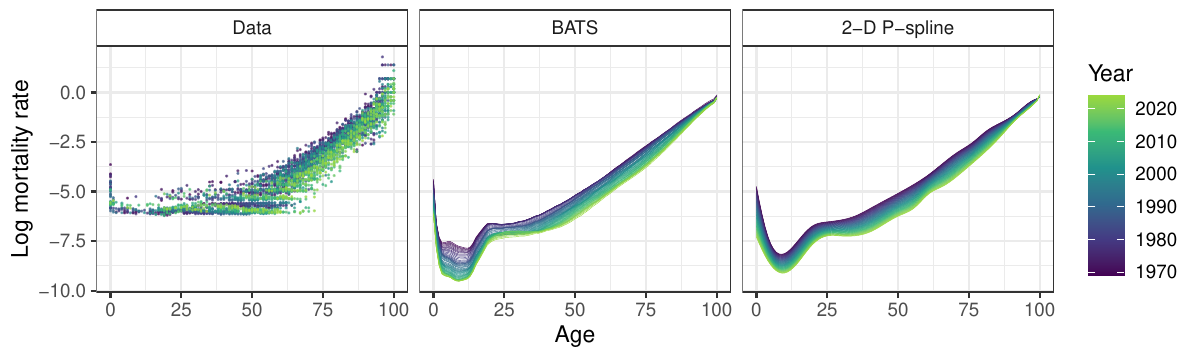}

    \includegraphics[width=\textwidth,page=7]{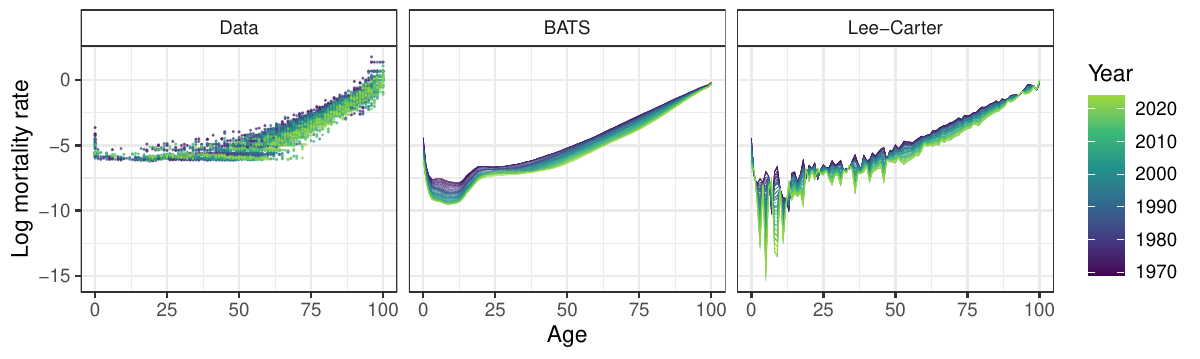}

    \includegraphics[width=\textwidth,page=8]{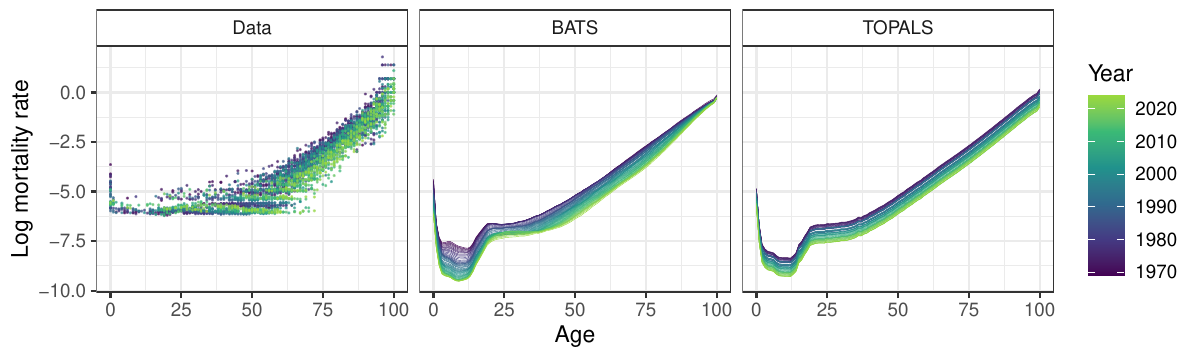}
    \caption{Fitted log mortality schedules for men in Gotland, as age schedules colored by year, for the four demographic point estimators. Each row shows the raw rates (left), the posterior mean of the smoothed log rate \( \boldsymbol{\eta} \) of BATS (middle), and the fitted schedules of one competitor (right): the schedule-wise P-spline, the per-unit P-spline surface, the per-unit Lee--Carter model, and TOPALS, from top to bottom. The vertical scale is shared within a row only.}
    \label{fig:smallest_point}
\end{figure}

\begin{figure}[p]
    \centering
    \includegraphics[width=\textwidth,page=1]{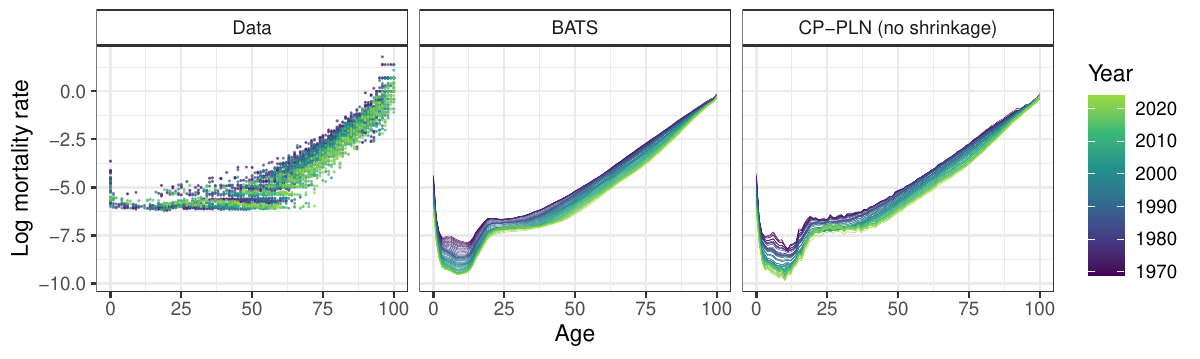}

    \includegraphics[width=\textwidth,page=2]{compare_surface_gotland_men_curves_pages.pdf}

    \includegraphics[width=\textwidth,page=3]{compare_surface_gotland_men_curves_pages.pdf}

    \includegraphics[width=\textwidth,page=4]{compare_surface_gotland_men_curves_pages.pdf}
    \caption{As Figure~\ref{fig:smallest_point}, for the four Bayesian competitors CP-PLN, BPTF with $R = 10$ and $R = 20$, and the additive main-effects model, from top to bottom.}
    \label{fig:smallest_bayes}
\end{figure}

The cross-validation scores of Section~6 of the main text summarize predictive performance over the entire panel. To illustrate how the methods differ where the differences are most visible, this section compares the fitted schedules of every method for the least populous unit in the panel, men in Gotland, whose posterior surface under BATS is shown in Figure~4 of the main text. Every competitor is fitted once to the full panel with the settings of the cross-validation study, and the point estimate of each method, that is, the fitted surface of the point estimators and the posterior mean of the smoothed log rate of the Bayesian methods, is displayed against the raw rates as age schedules colored by year. Figures~\ref{fig:smallest_point} and \ref{fig:smallest_bayes} show the result, one row per competitor, with the raw rates and BATS repeated in every row so that each competitor can be read against both. The vertical scale is shared within a row but not across rows, because the schedule-wise P-spline and the per-unit Lee--Carter model produce rates far below the range of the other methods.

The two figures make the mechanisms behind the cross-validation scores visible. The schedule-wise P-spline sees one year of Gotland's counts at a time, about a hundred cells most of which are empty below age 50, and its information criterion responds by selecting very heavy penalties. The fitted schedules degenerate to nearly log-linear lines in age that carry neither the infant drop nor the young-adult hump, while their levels at the youngest ages fan out from one year to the next. The method thus underfits the shape of the schedule and overfits its year-specific level at the same time. The per-unit Lee--Carter model pools across years within Gotland but has no smoothness. It is rough across ages in every year and inherits the runs of zero counts at young ages as spurious dips of several log units. The per-unit P-spline surface is smooth in both directions and reproduces the overall shape of the schedule, but with only Gotland's own counts to inform it, its age profile is less sharply resolved at young ages and slightly overfitting at older ages. TOPALS is close in shape, because the pooled standard schedule supplies the shape that Gotland's counts cannot. What remains visible is that the schedules are essentially shifted copies of the standard, so that the secular decline is compressed to a nearly uniform shift across ages. The additive model borrows from the entire panel and is smooth, but on the log scale it imposes one common age profile shifted by year and county, so the fitted schedules of different years are parallel by construction and the model cannot represent that the decline over the study period is much stronger in childhood than at the oldest ages. CP-PLN and BPTF come closest to BATS. Both borrow across all modes through the count layer and reproduce the level and shape of the schedule at the well-determined ages above 50, but without shrinkage and smoothness priors their profiles are visibly rough across adjacent ages and years in childhood and early adulthood, where deaths are rare. BATS yields the smoothest surface across both ordered modes while retaining sufficient detail, especially at the infant drop and the young-adult hump.

\section{Second Application: Austrian District-Level Outmigration}
\label{sec:S7}

The cross-validation study of Section~6 of the main text concerns one demographic component, mortality, in one country. To check that the findings are not specific to that setting, this section repeats the identical cross-validation design on a panel that differs in the component, the country, the granularity of the unit mode, and the temporal regime.

\subsection{Data}
\label{sec:S7data}

The panel comprises district-specific outmigrant counts by single year of age ($x = 0, \dots, 90$, $A = 91$), sex, and year across the 94 political districts of Austria from 2002 to 2023 ($T = 22$), together with the corresponding resident populations as exposures, so that the modeled quantity is an age-specific outmigration rate. The data are sourced from the Austrian national statistical office, \textit{Statistik Austria}, via its public database \textit{StatCube} (\textit{Wanderungsstatistik}). The data are based on the Austrian residence register and combine registered moves between Austrian districts with international emigration. The four-way array has $22 \times 91 \times 2 \times 94 = 376{,}376$ cells and records $6.9$ million outmigrants. The median cell contains seven outmigrants at a rate of about two percent of the resident population, $9.3$ percent of the cells are empty, and $60$ cells in the smallest district, Rust, at ages above $75$ carry zero exposure and hence no likelihood term. District populations range from about two million residents in Vienna to fewer than two thousand in Rust, so the unit mode spans three orders of magnitude of population size, one more than the Swedish county panel, and the array is $1.7$ times larger.

The panel is a demanding test for a smoothing model in two respects that the Swedish panel does not share. First, migration schedules over age are far from monotone. They show a sharp peak between ages 18 and 30 that reflects moves for education, military service, and early careers, a secondary rise at retirement ages in some districts, and a spike at ages 18 and 19 when young adults often move for education. Hence, the age mode combines smooth and locally rough structure. Second, the study period contains several disruptions of the Austrian migration system, namely the 2015--2016 refugee inflows, the COVID-19 pandemic of 2020, and the inflows from Ukraine in 2022, which affected internal migration patterns, so the time mode is far less regular than the secular decline of mortality.

\subsection{Results}
\label{sec:S7results}

The same cross-validation design as in the main text is used. Table~\ref{tab:cvat} reports the same scores as Table~1 of the main text, Table~\ref{tab:cvpaired} the paired differences, and Table~\ref{tab:cvatsize} the paired improvement of BATS over every competitor by quartile of district--sex unit size. Every entry of Table~\ref{tab:cvat} is a mean over $376{,}316$ scored cells, belonging to $4{,}136$ age schedules.

The conclusions of Section~6 of the main text are generally reproduced. First, the count tensor models are again far ahead of every other method. However, the benchmarks are ordered differently than on the Swedish panel. The per-unit P-spline surface is the best point estimator among the point estimators, but its deviance is $14$ percent above that of BATS. The schedule-wise P-spline and the per-unit Lee--Carter model share the last place among the point estimators. The per-unit Lee--Carter model loses most in the smallest districts. The additive model, which cannot represent that the height of the young-adult peak of migration varies across districts, has by far the largest deviance, driven by the largest cells.

Second, within the group of count tensor models, BATS again has the best log predictive score and deviance. The two count models with the overdispersion layer are calibrated to within a tenth of a percentage point at both nominal levels, as in Sweden, whereas BPTF undercovers by $1.5$ to $3.1$ percentage points, at least twice its shortfall on the Swedish panel. In log predictive score, BATS leads CP-PLN by $0.0027$ nats per cell, $5.9$ standard errors from zero, and BPTF by $0.021$ and $0.070$ nats per cell with $R = 20$ and $R = 10$. 

In deviance, BATS leads CP-PLN by $2.9$ standard errors, whereas its lead over BPTF with $R = 20$ in Table~\ref{tab:cvat} is within the sampling uncertainty, at $0.5$ standard errors. BATS improves on or matches CP-PLN in every quartile of unit size, and improves on BPTF with $R = 20$ in the three smaller quartiles, whereas BPTF with $R = 20$ has a slightly smaller deviance in the largest quartile, which contains Vienna and the other most populous districts, where the data leave little to smooth. Importantly, the smoothing kernels apparently do not oversmooth strongly at the rough features of these schedules. The spike of outmigration at ages 18 and 19 in some districts and the shock years are exactly where a smoothness prior could be expected to lose against CP-PLN, yet BATS improves on it in log predictive score in every quartile and in deviance in all but one, where the two are tied, which shows that the learned kernel parameters accommodate the local roughness of the age mode and the irregularity of the time mode rather than smoothing across them.

\newpage

\section{Supplementary Tables}
\label{sec:S8}

This section collects all tables of the supplement, in the order in which they are referenced in the text.
\vspace{3.65cm}
\begin{table}[h]
  \centering
  \footnotesize
  \setlength{\tabcolsep}{5pt}
  \begin{threeparttable}
    \caption{Computation time}
    \label{tab:timing}
    \begin{tabular}{llccc}
      \toprule
      Data & Model & Init (s) & s / sweep & Total \\
      \midrule
      Sweden & 1-D P-spline & -- & -- & 21 s \\
       & 2-D P-spline & -- & -- & 51 s \\
       & Lee--Carter & -- & -- & 1 s \\
       & TOPALS & -- & -- & 29 s \\
       & Additive main effects & 0 & 0.227 & 2.5 h \\
       & BPTF, $R = 10$ & 9 & 0.543 & 6.0 h \\
       & BPTF, $R = 20$ & 16 & 0.978 & 10.9 h \\
       & CP-PLN & 2 & 0.528 & 5.9 h \\
       & BATS & 2 & 0.966 & 10.7 h \\
      \addlinespace
      Austria & 1-D P-spline & -- & -- & 27 s \\
       & 2-D P-spline & -- & -- & 44 s \\
       & Lee--Carter & -- & -- & 2 s \\
       & TOPALS & -- & -- & 51 s \\
       & Additive main effects & 0 & 0.368 & 4.1 h \\
       & BPTF, $R = 10$ & 15 & 0.915 & 10.2 h \\
       & BPTF, $R = 20$ & 26 & 1.689 & 18.8 h \\
       & CP-PLN & 4 & 0.841 & 9.3 h \\
       & BATS & 4 & 1.372 & 15.2 h \\
      \bottomrule
    \end{tabular}
    \begin{tablenotes}[flushleft]
      \footnotesize
      \item\textit{Notes}: Elapsed time on one core of an AMD Ryzen 5 5500U (2.1 GHz) with single-threaded OpenBLAS. Init is the time of the warm start, s / sweep the median cost of a sampling sweep over three repetitions of 50 timed sweeps, and Total the initialization plus $40{,}000$ sweeps at that cost. The point estimators are fitted completely on the full array.
    \end{tablenotes}
  \end{threeparttable}
\end{table}

\begin{table}[t]
  \centering
  \footnotesize
  \setlength{\tabcolsep}{4pt}
  \begin{threeparttable}
    \caption{Factorial experiment}
    \label{tab:simfact}
    \begin{tabular}{llllccc}
    \toprule
    Rank & Noise & Exposure & Model & Spurious share & Direction RMSE & Surface RMSE \\
    \midrule
      Low & Low & High & BATS & \textbf{0.001} (0.000) & \textbf{0.009} (0.002) & \textbf{0.016} (0.002) \\
       &  &  & No Mat\'ern & \textbf{0.001} (0.000) & \textbf{0.009} (0.001) & 0.022 (0.002) \\
       &  &  & CP-PLN & 0.009 (0.002) & 0.011 (0.001) & 0.030 (0.003) \\
    \addlinespace
      High & Low & High & BATS & \textbf{0.000} (0.000) & \textbf{0.008} (0.001) & \textbf{0.023} (0.002) \\
       &  &  & No Mat\'ern & 0.001 (0.000) & 0.010 (0.001) & 0.035 (0.003) \\
       &  &  & CP-PLN & 0.003 (0.001) & 0.012 (0.000) & 0.042 (0.003) \\
    \addlinespace
      Low & High & High & BATS & \textbf{0.002} (0.001) & \textbf{0.014} (0.003) & \textbf{0.029} (0.004) \\
       &  &  & No Mat\'ern & 0.004 (0.001) & 0.016 (0.003) & 0.038 (0.003) \\
       &  &  & CP-PLN & 0.081 (0.011) & 0.018 (0.002) & 0.053 (0.003) \\
    \addlinespace
      High & High & High & BATS & \textbf{0.002} (0.000) & \textbf{0.014} (0.001) & \textbf{0.038} (0.001) \\
       &  &  & No Mat\'ern & 0.004 (0.001) & 0.015 (0.001) & 0.053 (0.002) \\
       &  &  & CP-PLN & 0.027 (0.003) & 0.017 (0.001) & 0.064 (0.002) \\
    \addlinespace
      Low & Low & Low & BATS & \textbf{0.006} (0.003) & 0.072 (0.008) & \textbf{0.124} (0.016) \\
       &  &  & No Mat\'ern & 0.008 (0.003) & \textbf{0.067} (0.007) & 0.129 (0.009) \\
       &  &  & CP-PLN & 0.162 (0.060) & 0.080 (0.009) & 0.214 (0.037) \\
    \addlinespace
      High & Low & Low & BATS & \textbf{0.001} (0.001) & \textbf{0.057} (0.004) & \textbf{0.134} (0.008) \\
       &  &  & No Mat\'ern & 0.002 (0.001) & 0.070 (0.014) & 0.177 (0.019) \\
       &  &  & CP-PLN & 0.042 (0.009) & 0.074 (0.015) & 0.212 (0.011) \\
    \addlinespace
      Low & High & Low & BATS & \textbf{0.016} (0.002) & \textbf{0.064} (0.006) & \textbf{0.115} (0.011) \\
       &  &  & No Mat\'ern & 0.035 (0.016) & 0.078 (0.010) & 0.147 (0.018) \\
       &  &  & CP-PLN & 0.281 (0.070) & 0.087 (0.012) & 0.219 (0.042) \\
    \addlinespace
      High & High & Low & BATS & \textbf{0.002} (0.002) & \textbf{0.065} (0.005) & \textbf{0.148} (0.008) \\
       &  &  & No Mat\'ern & 0.008 (0.006) & 0.071 (0.007) & 0.183 (0.015) \\
       &  &  & CP-PLN & 0.086 (0.006) & 0.079 (0.005) & 0.223 (0.011) \\
    \bottomrule
    \end{tabular}
    \begin{tablenotes}[flushleft]
      \footnotesize
      \item\textit{Notes}: Means (standard deviations) over five replications; the best value per setting and column is in bold. Spurious share is the share of the variation along a mode that falls on the eigenvalues beyond the true rank of that mode, posterior median, averaged over the three modes. Direction RMSE is the root mean squared error of the leading scaled directions against the truth, pooled over modes. Surface RMSE is the root mean squared error of the posterior mean of $\boldsymbol{\eta}$ on the log scale.
    \end{tablenotes}
  \end{threeparttable}
\end{table}

\begin{table}[t]
  \centering
  \footnotesize
  \begin{threeparttable}
    \caption{Convergence diagnostics of the main fit}
    \label{tab:ess}
    \begin{tabular}{lcc@{\hspace{2em}}lcc}
      \toprule
      Series & ESS & split-$\hat{R}$ & Series & ESS & split-$\hat{R}$ \\
      \midrule
      Data-layer log likelihood & 586 & 1.00 & RMS of latent surface & 1193 & 1.00 \\
      Mean of latent surface & 1358 & 1.00 & $\lambda_1$, time & 2438 & 1.00 \\
      Random projection 1 & 498 & 1.00 & $\lambda_2$, time & 332 & 1.02 \\
      Random projection 2 & 299 & 1.01 & $\lambda_1$, age & 1020 & 1.00 \\
      $\sigma^2_x$, age 0 & 557 & 1.00 & $\lambda_2$, age & 2689 & 1.00 \\
      $\sigma^2_x$, age 20 & 257 & 1.00 & $\lambda_1$, county & 461 & 1.00 \\
      $\sigma^2_x$, age 40 & 521 & 1.00 & $\lambda_2$, county & 333 & 1.00 \\
      $\sigma^2_x$, age 70 & 907 & 1.00 & Count, Halland M, 0 ($y = 2$) & 812 & 1.01 \\
      $\sigma^2_x$, age 100+ & 593 & 1.00 & Count, Gotland M, 5 ($y = 0$) & 290 & 1.00 \\
      Count, Stockholm W, 40 ($y = 7$) & 210 & 1.00 & Count, Gotland M, 30 ($y = 1$) & 383 & 1.00 \\
      Count, Stockholm W, 85 ($y = 275$) & 1708 & 1.00 & Count, Gotland M, 85 ($y = 21$) & 1493 & 1.00 \\
      \bottomrule
    \end{tabular}
    \begin{tablenotes}[flushleft]
      \footnotesize
      \item\textit{Notes}: Effective sample sizes (ESS) and split-$\hat{R}$ from the stored draws of the main fit: $20{,}000$ warm-up sweeps are followed by $20{,}000$ sampling sweeps, of which every fourth is stored, so that $5{,}000$ draws enter. $\lambda_1$ and $\lambda_2$ are the two leading eigenvalues of the Gram matrix of a mode. Counts are predicted expected death counts in 2024 for men (M) or women (W) at the given age, with the observed count of the cell in parentheses.
    \end{tablenotes}
  \end{threeparttable}
\end{table}

\vspace{3cm}

\begin{table}[h]
  \centering
  \small
  \begin{threeparttable}
    \caption{Paired differences between BATS and each competitor over all held-out cells}
    \label{tab:cvpaired}
    \begin{tabular}{lcccc@{\hspace{2em}}cccc}
      \toprule
      & \multicolumn{4}{c}{Sweden} & \multicolumn{4}{c}{Austria} \\
      \cmidrule(lr){2-5} \cmidrule(lr){6-9}
      Competitor & $\Delta$LPS & $z$ & $\Delta$Dev. & $z$ & $\Delta$LPS & $z$ & $\Delta$Dev. & $z$ \\
      \midrule
      CP-PLN                  & 0.0045 & 5.9  & 0.010 & 5.5  & 0.0027 & 5.9  & 0.008 & 2.9  \\
      BPTF, $R = 10$          & 0.0063 & 6.9  & 0.012 & 5.7  & 0.0696 & 9.5  & 0.102 & 11.9 \\
      BPTF, $R = 20$          & 0.0057 & 6.4  & 0.011 & 5.2  & 0.0207 & 9.2  & 0.005 & 0.5  \\
      Additive main effects   & 0.0925 & 7.2  & 0.287 & 5.5  & 0.1751 & 17.1 & 1.038 & 3.8  \\
      1-D P-spline            & --     & --   & 0.755 & 8.5  & --     & --   & 0.268 & 27.0 \\
      2-D P-spline            & --     & --   & 0.029 & 10.3 & --     & --   & 0.164 & 12.0 \\
      Lee--Carter             & --     & --   & 0.072 & 21.3 & --     & --   & 0.268 & 5.2  \\
      TOPALS                  & --     & --   & 0.138 & 12.5 & --     & --   & 0.248 & 11.2 \\
      \bottomrule
    \end{tabular}
    \begin{tablenotes}[flushleft]
      \footnotesize
      \item\textit{Notes}: $\Delta$LPS is the log predictive score of BATS minus that of the competitor and $\Delta$Dev.\ the mean Poisson deviance of the competitor minus that of BATS, both per held-out cell, so positive values favor BATS. $z$ is the mean difference divided by its standard error, clustered by county--sex unit (Sweden) or district--sex unit (Austria).
    \end{tablenotes}
  \end{threeparttable}
\end{table}

\begin{table}[h]
  \centering
  \small
  \begin{threeparttable}
    \caption{Paired improvement of BATS over each competitor by quartile of county--sex unit size}
    \label{tab:cvsize}
    \begin{tabular}{lcccc}
      \toprule
      Competitor & Q1 (smallest) & Q2 & Q3 & Q4 (largest) \\
      \midrule
      \multicolumn{5}{l}{\textit{Gain in log predictive score, nats per 1,000 observed deaths}} \\
      CP-PLN                       & 0.20 & 0.25 & 0.16 & 0.20 \\
      BPTF, $R = 10$               & 0.36 & 0.34 & 0.28 & 0.24 \\
      BPTF, $R = 20$               & 0.32 & 0.33 & 0.24 & 0.22 \\
      Additive main effects        & 4.80 & 4.96 & 4.27 & 3.67 \\
      \midrule
      \multicolumn{5}{l}{\textit{Reduction in mean Poisson deviance per held-out cell}} \\
      CP-PLN                       & 0.004 & 0.007 & 0.006 & 0.023 \\
      BPTF, $R = 10$               & 0.005 & 0.008 & 0.007 & 0.027 \\
      BPTF, $R = 20$               & 0.005 & 0.008 & 0.006 & 0.026 \\
      Additive main effects        & 0.108 & 0.179 & 0.193 & 0.667 \\
      1-D P-spline (per age schedule) & 0.361 & 0.539 & 0.583 & 1.535 \\
      2-D P-spline (per unit)      & 0.021 & 0.023 & 0.023 & 0.049 \\
      Lee--Carter (per unit)       & 0.060 & 0.060 & 0.066 & 0.101 \\
      TOPALS                       & 0.078 & 0.126 & 0.125 & 0.224 \\
      \bottomrule
    \end{tabular}
    \begin{tablenotes}[flushleft]
      \footnotesize
      \item\textit{Notes}: Each quartile contains ten county--sex units and $560$ age schedules. Entries are means of the paired per-schedule improvement of BATS over the competitor. Positive values favor BATS.
    \end{tablenotes}
  \end{threeparttable}
\end{table}

\begin{table}[t]
  \centering
  \small
  \begin{threeparttable}
    \caption{Cross-validation results on the Austrian outmigration panel}
    \label{tab:cvat}
    \begin{tabular}{lccccc}
      \toprule
      Model & LPS & Deviance & Cov.\ 90 & Cov.\ 95 & PIT KS \\
      \midrule
      1-D P-spline (per age schedule)   & --       & 1.467 & --    & --    & --    \\
      2-D P-spline (per unit)           & --       & 1.362 & --    & --    & --    \\
      Lee--Carter (per unit)            & --       & 1.467 & --    & --    & --    \\
      TOPALS                            & --       & 1.447 & --    & --    & --    \\
      Additive main effects             & $-2.554$ & 2.237 & 0.903 & 0.951 & 0.007 \\
      BPTF, $R = 10$                    & $-2.449$ & 1.301 & 0.869 & 0.926 & 0.021 \\
      BPTF, $R = 20$                    & $-2.400$ & 1.204 & 0.881 & 0.935 & 0.013 \\
      CP-PLN                            & $-2.382$ & 1.207 & 0.899 & 0.949 & $\mathbf{0.002}$ \\
      BATS                              & $\mathbf{-2.379}$ & $\mathbf{1.199}$ & 0.899 & 0.949 & $\mathbf{0.002}$ \\
      \bottomrule
    \end{tabular}
    \begin{tablenotes}[flushleft]
      \footnotesize
      \item\textit{Notes}: Same design and scoring as Table~1 of the main text, on the Austrian panel of $376{,}376$ cells ($22$ years, $91$ single ages, $2$ sexes, $94$ districts) with resident populations as exposures. Every cell is held out exactly once across the twenty folds, and results are averaged across the $376{,}316$ scored cells with positive exposure, belonging to $4{,}136$ age schedules. LPS is the mean log predictive score per held-out cell (larger is better), Deviance the mean Poisson deviance per held-out cell, Cov.\ 90 and Cov.\ 95 the empirical coverages of central 90 and 95 percent predictive intervals based on randomized PIT ranks, and PIT KS the Kolmogorov--Smirnov distance of the PIT values from uniformity. Column-wise best performance is highlighted in \textbf{bold}. The first four methods are point estimators, so probabilistic scores are not defined for them.
    \end{tablenotes}
  \end{threeparttable}
\end{table}

\begin{table}[t]
  \centering
  \small
  \begin{threeparttable}
    \caption{Paired improvement of BATS by quartile of district--sex unit size, Austrian panel}
    \label{tab:cvatsize}
    \begin{tabular}{lcccc}
      \toprule
      Competitor & Q1 (smallest) & Q2 & Q3 & Q4 (largest) \\
      \midrule
      \multicolumn{5}{l}{\textit{Gain in log predictive score, nats per 1,000 observed outmigrants}} \\
      CP-PLN                       & 0.42  & 0.11  & 0.15  & 0.12 \\
      BPTF, $R = 10$               & 5.68  & 5.43  & 4.46  & 3.03 \\
      BPTF, $R = 20$               & 2.24  & 1.58  & 1.35  & 0.83 \\
      Additive main effects        & 18.30 & 15.52 & 12.44 & 6.35 \\
      \midrule
      \multicolumn{5}{l}{\textit{Reduction in mean Poisson deviance per held-out cell}} \\
      CP-PLN                       & 0.005 & 0.000 & 0.006 & 0.019 \\
      BPTF, $R = 10$               & 0.059 & 0.093 & 0.090 & 0.164 \\
      BPTF, $R = 20$               & 0.021 & 0.018 & 0.011 & $-0.031$ \\
      Additive main effects        & 0.316 & 0.531 & 0.568 & 2.737 \\
      1-D P-spline (per age schedule) & 0.204 & 0.265 & 0.300 & 0.304 \\
      2-D P-spline (per unit)      & 0.081 & 0.143 & 0.164 & 0.265 \\
      Lee--Carter (per unit)       & 0.388 & 0.236 & 0.202 & 0.247 \\
      TOPALS                       & 0.118 & 0.196 & 0.270 & 0.407 \\
      \bottomrule
    \end{tabular}
    \begin{tablenotes}[flushleft]
      \footnotesize
      \item\textit{Notes}: Same construction as Table~\ref{tab:cvsize}, with district--sex units grouped into quartiles of mean annual resident population ($47$ units and $1{,}034$ age schedules per quartile) and the gain in log predictive score expressed per $1{,}000$ observed outmigrants. Positive values favor BATS.
    \end{tablenotes}
  \end{threeparttable}
\end{table}

\clearpage
\section{Supplementary Figures}
\label{sec:S9}

This section collects all figures of the supplement, in the order in which they are referenced in the text.

\vspace{3cm}

\end{document}